\documentclass[smallextended,numbook]{svjour3}    
\usepackage[numbers,sort&compress]{natbib}

\usepackage{mathptmx}

\usepackage{amsmath}
\usepackage{amsfonts}
\usepackage{amssymb}
\usepackage{bm}
\usepackage{mathtools}
\usepackage{hyperref}

\providecommand{\Ent}[1]{\lfloor #1 \rfloor}

\numberwithin{equation}{section}

\journalname{Journal of Mathematical Chemistry}

\begin{document}

\title{On the Mathematical Nature of Guseinov's Rearranged One-Range
  Addition Theorems for Slater-Type Functions}

\titlerunning{Guseinov's Rearranged One-Range Addition Theorems}

\author{Ernst Joachim Weniger}

\institute{Institut f\"{u}r Physikalische und Theoretische Chemie,
  Universit\"{a}t Regensburg, D-93040 Regensburg, Germany,
  \email{joachim.weniger@chemie.uni-regensburg.de}}

\date{Submitted: \today / Received: date / Accepted: date}

\maketitle

\begin{abstract}
  Starting from one-range addition theorems for Slater-type functions,
  which are expansion in terms of complete and orthonormal functions
  based on the generalized Laguerre polynomials, Guseinov constructed
  addition theorems that are expansions in terms of Slater-type functions
  with a common scaling parameter and integral principal quantum
  numbers. This was accomplished by expressing the complete and
  orthonormal Laguerre-type functions as finite linear combinations of
  Slater-type functions and by rearranging the order of the nested
  summations. Essentially, this corresponds to the transformation of a
  Laguerre expansion, which in general only converges in the mean, to a
  power series, which converges pointwise. Such a transformation is not
  necessarily legitimate, and this contribution discusses in detail the
  difference between truncated expansions and the infinite series that
  result in the absence of truncation.
\end{abstract}

\newpage

\tableofcontents

\newpage


\typeout{==> Section: Introduction}
\section{Introduction}
\label{Sec:Introduction}

Electronic structure theory is a highly interdisciplinary research topic
which benefited greatly from interactions with related scientific
disciplines.  It is generally agreed that molecular electronic structure
calculations only became feasible because of the spectacular advances in
computer hard and software. But we must not forget that advances in pure
and applied mathematics also played a crucial role.

While mathematicians and theoretical physicists have a venerable
tradition of a close and mutually beneficial collaboration, the same
cannot be said about the interaction of mathematicians and theoretical
chemists (my personal views on this topic are explained in
\cite{Weniger/2009a}). This is highly deplorable. Theoretical chemists
could -- and should -- learn more about new mathematical concepts or
powerful numerical techniques. On the other hand, electronic structure
theory offers quite a few challenging mathematical and computational
problems which could serve as a valuable source of inspiration for
mathematicians. A more extensive collaboration would certainly be
mutually beneficial (see for example \cite[Section 8]{Weniger/2009a} and
references therein).

Advances in mathematics are particularly important for molecular
multicenter integrals, which occur in molecular calculations on the basis
of the Hartree-Fock-Roothaan equations
\cite{Hartree/1928,Fock/1930,Roothaan/1951} and also in other
approximation schemes. The efficient and reliable evaluation of the
three- and six-dimensional molecular integrals, which are notoriously
difficult in the case of the physically better motivated exponentially
decaying basis functions, is one of the oldest mathematical and
computational problems of molecular electronic structure theory. In spite
of heroic efforts of numerous researchers, no completely satisfactory
solution has been found yet, and we have to continue searching for more
powerful mathematical and computational techniques. A review of the older
literature can be found in articles by Browne \cite{Browne/1971},
Dalgarno \cite{Dalgarno/1954}, Harris and Michels
\cite{Harris/Michels/1967}, and Huzinaga \cite{Huzinaga/1967}.

Quantum mechanics only determines which types of molecular integrals we
have to evaluate, but how we manipulate and ultimately evaluate them is a
mathematical problem. Thus, for a researcher working on multicenter
integrals, physical insight and a good knowledge of quantum mechanics is
actually less important than the ability of skillfully manipulating
complicated expressions involving special functions and a profound
knowledge of advanced mathematical techniques with a special emphasis on
numerics.

The evaluation of multicenter integrals is difficult because of the
different centers occurring in their integrands. This effectively
prevents the straightforward separation of the three- and six-dimensional
integrals into products of simpler integrals. A promising computational
strategy requires that we find a way of separating the integration
variables at tolerable computational costs.

Principal tools, which can accomplish a separation of integration
variables, are so-called addition theorems. These are special series
expansions of a function $f (\bm{r} \pm \bm{r}')$ with $\bm{r}, \bm{r}'
\in \mathbb{R}^{3}$ in terms of other functions that only depend on
either $\bm{r}$ or $\bm{r}'$.  The basic features of these fairly
complicated series expansions are reviewed in Section
\ref{Sec:BasicFeaturesOfAdditionTheorems}. A much more detailed treatment
will be given in my forthcoming review \cite{Weniger/2011a*}.

Infinite series expansions are among the most fundamental mathematical
tools with countless applications not only in mathematics, but also in
science and engineering. Nevertheless, a mathematically rigorous use of
infinite series is not necessarily an easy thing. Many scientists have a
highly cavalier attitude when it comes to questions of existence and
convergence. Since mathematics is ultimately used to describe natural
phenomena, it is tempting to believe that nature guarantees that all
intermediate mathematical manipulations are legitimate. Needless to say
that such an attitude is overly optimistic and can easily have
catastrophic consequences.

Strictly speaking, an infinite series is a meaningless object unless we
specify its convergence type -- for example pointwise convergence,
convergence in the mean, or even distributional or weak convergence --
and provide convincing evidence that this series converges according to
its specified type of convergence. It is important to take into account
that different convergence types of series expansions imply different
mathematical properties with different advantages and disadvantages. This
applies also to addition theorems. As discussed in my forthcoming review
\cite{Weniger/2011a*}, the different convergence types of addition
theorems may well provide the most useful characterization of their
properties.

Slater-type functions were originally introduced by Slater
\cite{Slater/1930,Slater/1932} to provide computationally convenient
analytical approximations to numerically determined solutions of
effective one-particle Schr\"{o}dinger equations. In unnormalized form,
they can be expressed as follows:
\begin{equation}
  \label{Def_STF}
  \chi_{N, L}^{M} (\beta, \bm{r}) \; = \;
  (\beta r)^{N-L-1} \, \mathrm{e}^{- \beta r} \, 
  \mathcal{Y}_{L}^{M} (\beta \bm{r})  \; = \; 
  (\beta r)^{N-1} \, \mathrm{e}^{- \beta r} \, Y_{L}^{M} (\bm{r}/r) \, .  
\end{equation}
Here, $\beta > 0$ is a scaling parameter, $\mathcal{Y}_{L}^{M} (\beta
\bm{r}) = (\beta r)^{L} Y_{L}^{M} (\bm{r}/r)$ is a regular solid harmonic
and $Y_{L}^{M} (\bm{r}/r)$ is a surface spherical harmonic (in my own
work, I have always used the phase condition of Condon and Shortley
\cite[Eqs.\ (6) and (9) on p.\ 115]{Condon/Shortley/1970}). By a slight
abuse of language, the index $N$ is frequently called principal quantum
number. In most, but not in all cases $N$ is a positive integer
satisfying $N \ge L + 1$.
 
Because of the importance of Slater-type functions as basis functions in
atomic and molecular electronic structure calculations, it is not
surprising that there is an extensive literature on their notoriously
difficult multicenter integrals in general as well as on their addition
theorems in special. A reasonably complete bibliography on their addition
theorems would be beyond the scope of this article. Let me just mention
two classic articles by Barnett and Coulson \cite{Barnett/Coulson/1951}
and by L\"{o}wdin \cite{Loewdin/1956}, respectively, which have inspired
many other researchers.

As discussed in more detail in Section
\ref{Sec:GuseinovsRearrangementsOfAdditionTheorems}, Guseinov
\cite{Guseinov/1978,Guseinov/1980a,Guseinov/1985a,Guseinov/2001a,%
  Guseinov/2002c} derived so-called one-range addition theorems for
Slater-type functions with integral and nonintegral principal quantum
numbers. He expanded Slater-type functions $\chi_{N, L}^{M} (\beta,
\bm{r} \pm \bm{r}')$ in terms of the functions $\bigl\{
\prescript{}{k}{\Psi}_{n, \ell}^{m} (\gamma, \bm{r}) \bigr\}_{n, \ell,
  m}$, which are complete and orthonormal in certain Hilbert spaces and
which are defined in (\ref{Def_Psi_Guseinov}). The radial parts of these
functions are based on the generalized Laguerre polynomials whose most
inportant properties are reviewed in Appendix
\ref{App:GeneralizedLaguerrePolynomials}.

On the basis of the approach developed in
\cite{Guseinov/1978,Guseinov/1980a,Guseinov/1985a,Guseinov/2001a,%
  Guseinov/2002c}, Guseinov and coworkers produced an amazing number of
articles on addition theorems and related topics
\cite{Guseinov/2002b,Guseinov/2002d,Guseinov/2003b,Guseinov/2003c,%
  Guseinov/2003d,Guseinov/2003e,Guseinov/2004a,Guseinov/2004b,%
  Guseinov/2004c,Guseinov/2004d,Guseinov/2004e,Guseinov/2004f,%
  Guseinov/2004i,Guseinov/2004k,Guseinov/2005a,Guseinov/2005b,%
  Guseinov/2005c,Guseinov/2005d,Guseinov/2005e,Guseinov/2005f,%
  Guseinov/2005g,Guseinov/2006a,Guseinov/2006b,Guseinov/2007e,%
  Guseinov/2007f,Guseinov/2007g,Guseinov/2008a,Guseinov/2008b,%
  Guseinov/2008c,Guseinov/2008d,Guseinov/2008f,Guseinov/2008h,%
  Guseinov/2009a,Guseinov/2009b,Guseinov/2009c,Guseinov/2009d,%
  Guseinov/2009e,Guseinov/2009f,Guseinov/2009h,Guseinov/2009i,%
  Guseinov/2009j,Guseinov/2009k,Guseinov/2010a,Guseinov/2010b,%
  Guseinov/2010c,Guseinov/2010d,Guseinov/2011a,Guseinov/2011b,%
  Guseinov/Aksu/2008,Guseinov/Aydin/Bagci/2008,Guseinov/Aydin/Mamedov/2003,%
  Guseinov/Ertuerk/2008,Guseinov/Ertuerk/2009a,Guseinov/Ertuerk/2009b,%
  Guseinov/Goerguen/2011,Guseinov/Gorgun/Zaim/2010,Guseinov/Mamedov/2002d,%
  Guseinov/Mamedov/2003,Guseinov/Mamedov/2004b,Guseinov/Mamedov/2004d,%
  Guseinov/Mamedov/2004e,Guseinov/Mamedov/2004h,Guseinov/Mamedov/2005c,%
  Guseinov/Mamedov/2005d,Guseinov/Mamedov/2005g,Guseinov/Mamedov/2007e,%
  Guseinov/Mamedov/2008a,Guseinov/Mamedov/2010b,Guseinov/Mamedov/2011a,%
  Guseinov/Mamedov/2011b,Guseinov/Mamedov/Andic/2010,%
  Guseinov/Mamedov/Andic/Cicek/2009,Guseinov/Mamedov/Oezdogan/Orbay/1999,%
  Guseinov/Mamedov/Orbay/2000,Guseinov/Mamedov/Rzaeva/2001,%
  Guseinov/Mamedov/Suenel/2002,%
  Guseinov/Rzaeva/Mamedov/Orbay/Oezdogan/Oener/1999,Guseinov/Sahin/2010,%
  Guseinov/Sahin/2011,Guseinov/Sahin/Aydin/Bagci/2008}.

The immediate reason for writing this article is a one-center expansion
constructed and applied by Guseinov and Mamedov
\cite{Guseinov/Mamedov/2008a}. They expanded a Slater-type function
$\chi_{N, L}^{M} (\beta, \bm{r})$ with an in general nonintegral
principal quantum number $N \in \mathbb{R} \setminus \mathbb{N}$ in terms
of Slater-type functions $\bigl\{ \chi_{n, L}^{M} (\gamma, \bm{r})
\bigr\}_{n=L+1}^{\infty}$ with integral principal quantum numbers $n \in
\mathbb{N}$ and an in general different common scaling parameter $\gamma
\neq \beta > 0$ \cite[Eq.\ (4)]{Guseinov/Mamedov/2008a}. This expansion
was used by Guseinov and Mamedov for the construction of what they called
series expansions for overlap integrals of Slater-type functions with
nonintegral principal quantum numbers in terms of overlap integrals of
Slater-type functions with integral principal quantum numbers.

For Guseinov and coworkers, who had done a lot of work on Slater-type
functions with nonintegral principal quantum numbers
\cite{Guseinov/2002b,Guseinov/2003a,Guseinov/2003b,Guseinov/2004c,%
  Guseinov/2004g,Guseinov/2004j,Guseinov/2005g,Guseinov/2007e,%
  Guseinov/2008c,Guseinov/2008g,Guseinov/2008h,Guseinov/2009a,%
  Guseinov/2009b,Guseinov/2009e,Guseinov/2009i,Guseinov/2009k,%
  Guseinov/2010a,Guseinov/2010c,Guseinov/2010d,Guseinov/2011a,%
  Guseinov/Aksu/2008,Guseinov/Ertuerk/2009a,Guseinov/Ertuerk/2009b,%
  Guseinov/Ertuerk/Sahin/2011,Guseinov/Ertuerk/Sahin/Aksu/2008,%
  Guseinov/Ertuerk/Sahin/Aksu/Bagci/2008,Guseinov/Mamedov/2002a,%
  Guseinov/Mamedov/2002d,Guseinov/Mamedov/2002e,Guseinov/Mamedov/2003,%
  Guseinov/Mamedov/2004b,Guseinov/Mamedov/2004d,Guseinov/Mamedov/2004e,%
  Guseinov/Mamedov/2004h,Guseinov/Mamedov/2005c,Guseinov/Mamedov/2005d,%
  Guseinov/Mamedov/2005h,Guseinov/Mamedov/2007e,Guseinov/Mamedov/2008a,%
  Guseinov/Mamedov/2010b,Guseinov/Mamedov/Suenel/2002}, such an expansion
would be extremely useful: Slater-type functions with nonintegral
principal quantum numbers could be replaced by Slater-type functions with
integral principal quantum numbers whose multicenter integrals can be
evaluated (much) more easily.

It is, however, trivially simple to show that such a one-center
expansion, which in my notation can be written as follows,
\begin{equation}
  \label{GuExp_1}
  \chi_{N, L}^{M} (\beta, \bm{r}) \; = \; 
  \sum_{n=L+1}^{\infty} \, \mathbb{X}_{n}^{(N, L)} 
  (\beta, \gamma) \, \chi_{n, L}^{M} (\gamma, \bm{r}) \, ,
  \qquad N \in \mathbb{R} \, ,
\end{equation}
only exists if the principal quantum number $N$ is a positive integer
satisfying $N \ge L+1$. If $N$ is nonintegral or zero, i.e., if $N \in
\mathbb{R} \setminus \mathbb{N}$, this expansion does not exist. The
nonexistence of this expansion also played a major role in my discussion
\cite[Section 7]{Weniger/2008} of Guseinov's treatment of one-range
addition theorems for Slater-type functions. However, my discussion in
\cite[Section 7]{Weniger/2008} was incomplete. Its scope is extended
considerably by this article.

From a methodological point of view, we only have to utilize the obvious
fact that an expansion of $\chi_{N, L}^{M} (\beta, \bm{r})$ in terms of
Slater-type functions $\bigl\{ \chi_{n, L}^{M} (\gamma, \bm{r})
\bigr\}_{n=L+1}^{\infty}$ with integral principal quantum numbers $n$ and
a common scaling parameter $\gamma > 0$ is nothing but a power series
expansion of $\exp (\gamma r) \chi_{N, L}^{M} (\beta, \bm{r})$ about
$r=0$ in disguise. This interpretation of the expansion (\ref{GuExp_1})
offers some valuable insight. Relatively little is known about the
convergence and existence of expansions in terms of Slater-type functions
$\bigl\{ \chi_{n, L}^{M} (\gamma, \bm{r}) \bigr\}_{n}$, but it is
normally comparatively easy to decide whether such a corresponding power
series expansion about $r=0$ exists and for which values of $r$ it
converges.

We can convert the expansion (\ref{GuExp_1}) to a more transparent
one-dimensional radial problem by canceling the spherical harmonics.
Then, the right-hand side of (\ref{GuExp_1}) can be converted to a power
series in $\gamma r$ by multiplying either side of (\ref{GuExp_1}) by
$\exp (\gamma r)$:
\begin{equation}
  \label{GuExp_2}
  \mathrm{e}^{{(\gamma-\beta)r}} \, (\beta r)^{N-1} \; = \;  
  \sum_{n=L+1}^{\infty} \, \mathbb{X}_{n}^{(N, L)} (\beta, \gamma) 
  \, (\gamma r)^{n-1} \, . 
\end{equation}  

Let us now assume that the principal quantum number $N$ is a positive
integer satisfying $N \ge L+1$. Then we only have to replace the
exponential on the left-hand side of (\ref{GuExp_2}) by its power series
to obtain:
\begin{equation}
  \label{GuExp_3}
  (\beta/\gamma)^{N-1} \, \sum_{\nu=0}^{\infty} \, 
  \frac{[1-(\beta/\gamma)]^{\nu}}{\nu!} \, (\gamma r)^{N+\nu-1} \; = \;  
  \sum_{n=L+1}^{\infty} \, 
  \mathbb{X}_{n}^{(N, L)} (\beta, \gamma) \, (\gamma r)^{n-1} \, . 
\end{equation}
Comparison of the coefficients of equal powers yields explicit
expressions for the coefficients $\mathbb{X}_{n}^{(N, L)} (\beta,
\gamma)$.

But if the principal quantum number $N$ is not a positive integer, the
matching of the coefficients of equal powers does not work: On the
left-hand side of (\ref{GuExp_3}) there are either only nonintegral
powers or some negative powers, and on the right-hand side there are only
integral and nonnegative powers. Accordingly, a power series expansion
for $\mathrm{e}^{{(\gamma-\beta)r}} \, (\beta r)^{N-1}$ can only exist if
$N$ is a positive integer satisfying $N \ge L+1$.

Since every power series is also a Taylor series for some function (see
for example \cite{Meyerson/1981}), the expansion (\ref{GuExp_1}) can also
be derived by doing a Taylor expansion of $\exp \bigl((\gamma-\beta)r)
\bigr) (\beta r)^{N-1}$ about $r=0$, yielding the same
conclusions. Fractional or nonintegral powers $r^{\alpha}$ with $\alpha
\in \mathbb{R} \setminus \mathbb{N}_{0}$ and thus also $\exp
([\gamma-\beta]r) r^{N-1}$ with $N \in \mathbb{R} \setminus \mathbb{N}$
do not have continuous derivatives of arbitrary order at $r=0$. Thus, $N
\in \mathbb{R} \setminus \mathbb{N}$ implies that the leading expansion
coefficients $\mathbb{X}_{n}^{(N, L)} (\beta, \gamma)$ with indices $n
\le \Ent {N}$ are zero, but all the remaining coefficients with indices
$n > \Ent {N}$ are infinite in magnitude. Here, $\Ent {N}$ stands for the
integral part of $N$.

The nonexistence problems mentioned above are a direct consequence of the
obvious fact that the radial part of a Slater-type function with a
nonintegral or negative principal quantum number is not analytic in the
sense of complex analysis at $r=0$. This is so elementary that it is hard
to understand why nobody had noticed the nonexistence of the expansion
(\ref{GuExp_1}) for $N \in \mathbb{R} \setminus \mathbb{N}$ before.

In spite of the nonexistence of (\ref{GuExp_1}) for $N \in \mathbb{R}
\setminus \mathbb{N}$, Guseinov and Mamedov \cite{Guseinov/Mamedov/2008a}
had used their version of this one-center expansion for numerical
purposes and presented apparently meaningful numerical results for
overlap and other, closely related integrals of Slater-type functions
with nonintegral principal quantum numbers in \cite[Tables 1 and
2]{Guseinov/Mamedov/2008a}.

If these numerical results are genuine -- which I assume -- then there is
only one logically satisfactory conclusion: Contrary to their claim (see
the text before Eq.\ (4) of \cite{Guseinov/Mamedov/2008a}), Guseinov and
Mamedov did not use \emph{expansions} involving an \emph{infinite} number
of Slater-type functions with integral principal quantum numbers, which
-- as discussed above -- do not exist for $N \in \mathbb{R} \setminus
\mathbb{N}$. Instead, they only employed some \emph{approximations}
consisting of a \emph{finite} number of terms and extrapolated -- albeit
incorrectly -- that these finite approximations remain meaningful in the
limit of an infinite number of terms.

As is well known from the mathematical literature, truncated expansions,
which are approximations consisting of a finite number of terms, may have
mathematical properties that are not at all related to those of a
complete expansion consisting of an infinite number of terms. In
particular, it can happen that a truncated expansion can be
mathematically meaningful as well as numerically useful, although its
limit as an expansion of infinite length does not exist (see also
Appendix \ref{App:Semiconvergence}, where the semiconvergence of certain
infinite series is reviewed briefly). It is the purpose of this article
to investigate this as well as some closely related question.

The one-center expansion used by Guseinov and Mamedov \cite[Eq.\
(4)]{Guseinov/Mamedov/2008a} was originally derived by Guseinov
\cite[Eq.\ (21)]{Guseinov/2002c} as the one-center limit of a class of
addition theorems for Slater-type functions \cite[Eq.\
(15)]{Guseinov/2002c}. Guseinov's approach is based on one-range addition
theorems for Slater-type functions, which are expansions in terms of
Guseinov's Laguerre-type functions $\bigl\{ \prescript{}{k}{\Psi}_{n,
  \ell}^{m} (\gamma, \bm{r}) \bigr\}_{n, \ell, m}$. These functions are
then replaced by Slater-type functions according to
(\ref{GusFun2STF}). Ultimately, this yields addition theorems that are
expansions in terms of Slater-type functions with integral principal
quantum numbers. Unfortunately, the derivation of these expansions is
fairly difficult and involves manipulations whose validity is not at all
obvious \cite{Weniger/2008,Weniger/2007b,Weniger/2007c}.

In view of the amazing number of articles on addition theorems and
related topics published by Guseinov and coworkers
\cite{Guseinov/1978,Guseinov/1980a,Guseinov/1985a,Guseinov/2001a,%
  Guseinov/2002b,Guseinov/2002c,Guseinov/2002d,Guseinov/2003b,Guseinov/2003c,%
  Guseinov/2003d,Guseinov/2003e,Guseinov/2004a,Guseinov/2004b,%
  Guseinov/2004c,Guseinov/2004d,Guseinov/2004e,Guseinov/2004f,%
  Guseinov/2004i,Guseinov/2004k,Guseinov/2005a,Guseinov/2005b,%
  Guseinov/2005c,Guseinov/2005d,Guseinov/2005e,Guseinov/2005f,%
  Guseinov/2005g,Guseinov/2006a,Guseinov/2006b,Guseinov/2007e,%
  Guseinov/2007f,Guseinov/2007g,Guseinov/2008a,Guseinov/2008b,%
  Guseinov/2008c,Guseinov/2008d,Guseinov/2008f,Guseinov/2008h,%
  Guseinov/2009a,Guseinov/2009b,Guseinov/2009c,Guseinov/2009d,%
  Guseinov/2009e,Guseinov/2009f,Guseinov/2009h,Guseinov/2009i,%
  Guseinov/2009j,Guseinov/2009k,Guseinov/2010a,Guseinov/2010b,%
  Guseinov/2010c,Guseinov/2010d,Guseinov/2011a,Guseinov/2011b,%
  Guseinov/Aksu/2008,Guseinov/Aydin/Bagci/2008,Guseinov/Aydin/Mamedov/2003,%
  Guseinov/Ertuerk/2008,Guseinov/Ertuerk/2009a,Guseinov/Ertuerk/2009b,%
  Guseinov/Goerguen/2011,Guseinov/Gorgun/Zaim/2010,Guseinov/Mamedov/2002d,%
  Guseinov/Mamedov/2003,Guseinov/Mamedov/2004b,Guseinov/Mamedov/2004d,%
  Guseinov/Mamedov/2004e,Guseinov/Mamedov/2004h,Guseinov/Mamedov/2005c,%
  Guseinov/Mamedov/2005d,Guseinov/Mamedov/2005g,Guseinov/Mamedov/2007e,%
  Guseinov/Mamedov/2008a,Guseinov/Mamedov/2010b,Guseinov/Mamedov/2011a,%
  Guseinov/Mamedov/2011b,Guseinov/Mamedov/Andic/2010,%
  Guseinov/Mamedov/Andic/Cicek/2009,Guseinov/Mamedov/Oezdogan/Orbay/1999,%
  Guseinov/Mamedov/Orbay/2000,Guseinov/Mamedov/Rzaeva/2001,%
  Guseinov/Mamedov/Suenel/2002,%
  Guseinov/Rzaeva/Mamedov/Orbay/Oezdogan/Oener/1999,Guseinov/Sahin/2010,%
  Guseinov/Sahin/2011,Guseinov/Sahin/Aydin/Bagci/2008}, such an
investigation should be of some principal interest. 

This article sheds some light on certain more subtle differences of
orthogonal and nonorthogonal expansions which -- although in principle
known -- are often not sufficiently appreciated and taken into account in
the literature on electronic structure calculations.

In Section \ref{Sec:BasicFeaturesOfAdditionTheorems}, basic features of
both one-range and two-range addition theorems are reviewed. Section
\ref{Sec:GuseinovsRearrangementsOfAdditionTheorems} discusses Guseinov's
derivation of expansions in terms of Slater-type functions by rearranging
his one-range addition theorems for Slater-type functions with a special
emphasis on numerical stability issues. From a mathematical point of
view, Guseinov's treatment of addition theorems ultimately boils down to
the transformation of a Laguerre series to a power series. The most
important properties of this transformation, whose theory had recently
been formulated in \cite{Weniger/2008}, are reviewed in Section
\ref{Sec:LagSer->PowSer}.

The central results of this manuscript are presented in Sections
\ref{Sec:OneCenterExpansionSTF} -- \ref{Sec:NumericalImplications}. In
Section \ref{Sec:OneCenterExpansionSTF} it is shown that the
transformation formulas described in Section \ref{Sec:LagSer->PowSer}
fully explain the nonexistence of Guseinov's rearrangements of one-center
expansions in the case of nonintegral principal quantum numbers. 

In the case of the simpler one-center expansions, which correspond to the
Laguerre series (\ref{GenPow2GLag}) for $z^{\rho}$ with $\rho \in
\mathbb{R} \setminus \mathbb{N}_{0}$, this analysis had already been done
in \cite[Section 3]{Weniger/2008}. But my results for the more
complicated one-center expansions, which correspond to the Laguerre
series (\ref{ExpoPow2GLag}) for $z^{\rho} \mathrm{e}^{u z}$ with $\rho
\in \mathbb{R} \setminus \mathbb{N}_{0}$, are new. In \cite[Section
7]{Weniger/2008} it was only stated on the basis of general analyticity
principles that the Laguerre series (\ref{ExpoPow2GLag}) cannot exist if
$\rho \in \mathbb{R} \setminus \mathbb{N}_{0}$, but a detailed analysis
of the transformation formulas could not be done. At that time I did not
know yet the large index asymptotics of some special Gaussian
hypergeometric series ${}_{2} F_{1}$ which were derived in Section
\ref{Sec:OneCenterExpansionSTF}. These asymptotic expressions make an
asymptotic analysis of the transformation formulas possible.

As discussed in more details in Section
\ref{Sec:BasicFeaturesOfAdditionTheorems}, a one-range addition theorem
for a function $f (\bm{r} \pm \bm{r}')$ provides for all arguments
$\bm{r}, \bm{r}' \in \mathbb{R}^{3}$ a \emph{unique} representation of
$f$ in \emph{separated} form that is valid for the whole argument set
$\mathbb{R}^{3} \times \mathbb{R}^{3}$. Since the one-center limits of
Guseinov's rearranged one-range addition theorems do not exist in the
important special case of nonintegral principal quantum numbers, it was
argued in \cite[Section 7]{Weniger/2008} that at least some of Guseinov's
rearrangements cannot be one-range addition theorems. However, the exact
nature of Guseinov's rearrangements was not yet understood in
\cite{Weniger/2008}.

Because of technical problems, a straightforward analysis of the
transformation formulas described in Section \ref{Sec:LagSer->PowSer} is
not possible in the case of Guseinov's one-range addition theorems,
However, in Section \ref{Sec:OneRangeAdd->TwoRangeAdd} it is shown that
the rearrangement of Guseinov's one-range addition theorems produces
two-range addition theorems. This follows at once from the fact that a
Slater-type function $\chi_{N, L}^{M} (\beta, \bm{r} \pm \bm{r}')$ or
equivalently the function $\mathrm{e}^{\gamma r} \chi_{N, L}^{M} (\beta,
\bm{r} \pm \bm{r}')$ is singular for $\bm{r} \pm \bm{r}' = \bm{0}$.

Section \ref{Sec:NumericalImplications} discusses some numerical
implications of the results derived in Sections
\ref{Sec:OneCenterExpansionSTF} and
\ref{Sec:OneRangeAdd->TwoRangeAdd}. The most important result is that
rearrangements of Guseinov's truncated one-center expansions are
semiconvergent with respect to a variation of the truncation order if the
principal quantum number is nonintegral. This follows from the leading
order asymptotic approximation (\ref{ZetaTypeTail}).

This article is concluded by a summary and an outlook in Section
\ref{Sec:SummaryOutlook}. Appendices
\ref{App:GeneralAspectsOfSeriesExpansions} -- \ref{App:Semiconvergence}
review general aspects of series expansions, the largely complementary
features of power series and orthogonal expansions, the for our purposes
most important properties of generalized Laguerre polynomials, and
semiconvergent expansions.

\typeout{==> Section: Basic Features of Addition Theorems}
\section{Basic Features of Addition Theorems}
\label{Sec:BasicFeaturesOfAdditionTheorems}

Addition theorems are series expansions of a function $f (\bm{r} \pm
\bm{r}')$ with $\bm{r}, \bm{r}' \in \mathbb{R}^{3}$ in terms of other
functions that depend only on either $\bm{r}$ or $\bm{r}'$. Because of
the completeness and the convenient orthogonality properties of the
(surface) spherical harmonics $Y_{\ell}^{m} (\theta, \phi)$, it is
customary to express the angular parts of addition theorems in terms of
spherical harmonics whose arguments correspond to the solid angles
$\bm{r}/r$ and $\bm{r}'/r'$, respectively.

In atomic and molecular electronic structure theory, we are predominantly
interested in addition theorems for functions
\begin{equation}
  \label{Def_IrrSphericalTensor}
  F_{\ell}^{m} (\bm{r}) \; = \;
  f_{\ell} (r) \, Y_{\ell}^{m} (\bm{r}/r) \, ,
\end{equation}
which can be factored into a radial part $f_{\ell} (r)$ multiplied by a
surface spherical harmonic. In the language of angular momentum theory,
such a function $F_{\ell}^{m} (\bm{r})$ is an irreducible spherical
tensor with a fixed rank $\ell \in \mathbb{N}_{0}$.

The best known and probably also the most simple addition theorem of such
an irreducible spherical tensor is the Laplace expansion of the Coulomb
or Newton potential $1/\vert {\bm{r}} \pm {\bm{r}}' \vert$. As discussed
in Hobson's classic book \cite[\S 11 on pp.\ 16 - 17]{Hobson/1965}, the
mathematical tools needed for the construction of this addition theorem
were developed by Laplace and Legendre already in the late 18th
century. In modern notation, the Laplace expansion can be expressed as
follows:
\begin{align}
  \label{LapExp}
  \frac{1}{\vert {\bm{r}} \pm {\bm{r}}' \vert} & \; = \; 4\pi \,
  \sum_{\lambda=0}^{\infty} \, \frac{(\mp 1)^{\lambda}}{2\lambda+1} \,
  \frac{r_{<}^{\lambda}}{r_{>}^{\lambda+1}} \, 
  \sum_{\mu=-\lambda}^{\lambda} \,
  \bigl[ Y_{\lambda}^{\mu} ({\bm{r}}_{<}/r_{<}) \bigr]^{*} \, 
  Y_{\lambda}^{\mu} ({\bm{r}}_{>}/r_{>}) \, ,
  \notag \\
  & \qquad \vert \bm{r}_{<} \vert = \min (r,r') \, , \qquad
           \vert \bm{r}_{>} \vert = \max (r,r') \, .
\end{align}

The Laplace expansion is the simplest prototype for a large class of
addition theorems that possess a characteristic two-range form. Numerous
techniques for the derivation of two-range addition theorems are
described in the literature. With the possible exception of the so-called
Fourier transform method advocated almost simultaneously but
independently by Ruedenberg \cite{Ruedenberg/1967} and Silverstone
\cite{Silverstone/1967d}, they are all related to power series expansions
in one way or the other.

For example, the Laplace expansion (\ref{LapExp}) can be derived easily
with the help of the generating function of the Legendre polynomials
$P_{n} (x)$ \cite[p.\ 232]{Magnus/Oberhettinger/Soni/1966}:
\begin{equation}
  \label{GenFun_LegPol}
  \bigl[ 1 - 2 x t + t^2 \bigr]^{-1/2} \; = \;
  \sum_{n=0}^{\infty} \, P_{n} (x) \, t^{n} \, ,
  \qquad -1 < x < 1 \, , \quad \vert t \vert < 1 \, .
\end{equation}
Obviously, this generating function is nothing but a power series
expansion about $t=0$. For the derivation of the Laplace expansion
(\ref{LapExp}), we only need the so-called spherical harmonic addition
theorem (see for example \cite[Eq.\ (1.2-3a)]{Weissbluth/1978})
\begin{equation}
  \label{Ylm_AddThm}
  P_{\ell} (\cos \theta) \; = \; \frac{4\pi}{2\ell+1} \,
  \sum_{m=-\ell}^{\ell} \, \bigl[ Y_{\ell}^{m} (\bm{u}/u) \bigr]^{*} \,
  Y_{\ell}^{m} (\bm{v}/v) \, , \qquad
  \cos \theta \; = \; \frac{\bm{u} \cdot \bm{v}}{u v} \, .
\end{equation}
The two-range form of the Laplace expansion (\ref{LapExp}) is necessary
to ensure convergence, because the generating function
(\ref{GenFun_LegPol}) converges only for $\vert t \vert < 1$. Thus, the
Laplace expansion converges \emph{pointwise} for $r_{<}/r_{>} < 1$ and
diverges for $r_{<}/r_{>} > 1$.

The use of generating functions of the Gegenbauer polynomials, which
generalize the generating function (\ref{GenFun_LegPol}) of the Legendre
polynomials, for the construction of addition theorems of spherically
symmetric functions will be discussed briefly in Section
\ref{Sec:OneRangeAdd->TwoRangeAdd}. The two-range forms of these addition
theorems follow from the convergence conditions of the corresponding
generating functions which are nothing but special power series
expansions, usually in the variable $r/r'$ or more precisely in
$r_{<}/r_{>}$.

The most principal approach for the construction of pointwise convergent
addition theorems consists in interpreting such an addition theorem as a
three-dimensional Taylor expansion (see for example \cite[p.\
181]{Biedenharn/Louck/1981a}):
\begin{equation}
  \label{ExpDifOp}
  f (\bm{r} \pm \bm{r}') \; = \; \sum_{n=0}^{\infty} \,
  \frac{(\bm{r} \cdot \nabla')^n}{n!} \, f (\pm \bm{r}') \; = \;
  \mathrm{e}^{\bm{r} \cdot \nabla'} \, f (\pm \bm{r}') \, .
\end{equation}
Thus, the translation operator $\mathrm{e}^{\bm{r} \cdot \nabla'} =
\mathrm{e}^{x \partial /\partial x'} \mathrm{e}^{y \partial /\partial y'}
\mathrm{e}^{z \partial /\partial z'}$ generates $f (\bm{r} \pm \bm{r}')$
by constructing a three-dimensional Taylor expansion of $f$ about $\pm
\bm{r}'$ with shift vector $\bm{r}$. Since the variables $\bm{r}$ and
$\bm{r}'$ are separated, the series expansion (\ref{ExpDifOp}) is indeed
an addition theorem.

Of course, the series expansion (\ref{ExpDifOp}) tacitly assumes that $f
(\pm \bm{r}')$ possesses continuous derivatives of arbitrary order with
respect to the Cartesian components of its argument $\pm \bm{r}' = \pm
(x', y', z')$. Thus, $f$ has to be \emph{analytic} at $\pm \bm{r}'$,
which guarantees that the series expansion (\ref{ExpDifOp}) converges at
least for sufficiently small $\vert \bm{r} \vert > 0$. Eventual
singularities of $f$ determine for which values of the shift vector
$\bm{r}$ the three-dimensional Taylor expansion (\ref{ExpDifOp})
converges.

We could also expand $f$ about $\bm{r}$ and use $\pm \bm{r}'$ as the
shift vector. This would produce an addition theorem for $f (\bm{r} \pm
\bm{r}')$ in which the roles of $\bm{r}$ and $\bm{r}'$ are interchanged.
Both approaches are mathematically legitimate and equivalent if $f$ is
analytic at $\bm{r}$, $\bm{r}'$, and $\bm{r} \pm \bm{r}'$ for essentially
arbitrary vectors $\bm{r}, \bm{r}' \in \mathbb{R}^3$.  This is normally
not true. Many of the functions, that are of interest in the context of
molecular electronic structure calculations, have a singularity at the
origin. Obvious examples are the Coulomb potential, which has a pole at
the origin, or the commonly used exponentially decaying functions as for
example Slater-type functions, which have a branch point singularity at
the origin. 

Accordingly, for the addition theorem of a function $f (\bm{r} \pm
\bm{r}')$, which is singular for $\bm{r} \pm \bm{r}' = \bm{0}$, the
natural variables are not $\bm{r}$ and $\bm{r}'$, but the vectors
$\bm{r}_{<}$ and $\bm{r}_{>}$ satisfying $\vert \bm{r}_{<} \vert = \min
(r,r')$ and $\bm{r}_{>} \vert = \max (r,r')$. This implies that the
expansion formula (\ref{ExpDifOp}) should be rewritten as follows:
\begin{equation}
  \label{TransOp<>}
  f (\bm{r}_{<} \pm \bm{r}_{>}) \; = \;
  \sum_{n=0}^{\infty} \,
  \frac{(\bm{r}_{<} \cdot \nabla_{>})^n}{n!} \, f (\pm \bm{r}_{>}) \; = \;
  \mathrm{e}^{\bm{r}_{<} \cdot \nabla_{>}} \, f (\pm \bm{r}_{>}) \, .
\end{equation}
In this way, the convergence of the three-dimensional Taylor expansion is
guaranteed provided that $f$ is analytic everywhere with the possible
exception of the origin.

From a practical point of view, the translation operator
$\mathrm{e}^{\bm{r}_{<} \cdot \nabla_{>}}$ does not seem to be a
particularly useful analytical tool. In electronic structure theory, we
are usually interested in addition theorems of irreducible spherical
tensors of the type of (\ref{Def_IrrSphericalTensor}), which are defined
in terms of the spherical polar coordinates $r$, $\theta$, and
$\phi$. Differentiating such an irreducible spherical tensor with respect
to the Cartesian components of $\bm{r}_{>} = (x_{>}, y_{>}, z_{>})$ would
lead to extremely messy expressions and to difficult technical
problems. Thus, it is a seemingly obvious conclusion that the translation
operator $\mathrm{e}^{\bm{r}_{<} \cdot \nabla+>}$ only provides a formal
solution to the problem of separating the variables $\bm{r}_{<}$ and
$\bm{r}_{>}$ of a function $f (\bm{r}_{>} \pm \bm{r}_{>})$. Nevertheless,
this conclusion is wrong.

The crucial step, which ultimately makes the Taylor expansion method
practically useful, is the expansion of the translation operator
$\mathrm{e}^{\bm{r}_{<} \cdot \nabla_{>}}$ in terms of differential
operators that are irreducible spherical tensors with a fixed rank $\ell
\in \mathbb{N}_{0}$:
\begin{equation}
  \label{ST_TransOp}
  \mathrm{e}^{\bm{r}_{<} \cdot \nabla_{>}} \; = \; 
  2 \pi \, \sum_{\ell=0}^{\infty} \, \sum_{m=-\ell}^{\ell} \,
  \left[ \mathcal{Y}_{\ell}^{m} (\bm{r}_{<}) \right]^{*} \,
  \mathcal{Y}_{\ell}^{m} (\nabla_{>}) \, \sum_{k=0}^{\infty} \,
  \frac{r_{<}^{2k} \, \nabla_{>}^{2k}}
  {2^{\ell+2k} k! (1/2)_{\ell+k+1}} \, .
\end{equation}
Here, $\mathcal{Y}_{\ell}^{m} (\nabla)$ is the so-called spherical tensor
gradient operator which is obtained by replacing the Cartesian components
of $\bm{r} = (x, y, z)$ by the Cartesian components of $\nabla =
(\partial/\partial x \partial/\partial y, \partial/\partial z)$ in the
explicit expression of the solid harmonic $\mathcal{Y}_{\ell}^{m}
(\bm{r}) = r^{\ell} Y_{\ell}^{m} (\bm{r}/r)$ which is a homogeneous
polynomial of degree $\ell$. More details as well as numerous references
can be found in \cite[Section 2]{Weniger/2005}.

It seems that the expansion (\ref{ST_TransOp}) was first published by
Santos \cite[Eq.\ (A.6)]{Santos/1973}, who emphasized that this expansion
should be useful for the derivation of addition theorems, but he
apparently never used it for that purpose.

In \cite{Weniger/2000a,Weniger/2002} or in \cite[Section 7]{Weniger/2005}
it was shown that two-range addition theorems of irreducible spherical
tensors are nothing but rearranged three-dimensional Taylor expansions
about $\pm \bm{r}_{>}$ with shift vector $\bm{r}_{<}$. Such an expansion
converges pointwise and uniformly in the interior of suitable subsets of
$\mathbb{R}^{3} \times \mathbb{R}^{3}$ whose boundaries are defined by
the singularities of the function which is to be expanded. If the
function under consideration has a singularity for $\bm{r} \pm \bm{r}' =
\bm{0}$, the two-range form of its addition theorem is necessary to
guarantee pointwise or even uniform convergence in a suitable
neighborhood of the expansion point.

There is a different class of addition theorems based on Hilbert space
theory. Let us assume that $\{ \varphi_{n, \ell}^{m} (\bm{r}) \}_{n,
  \ell, m}$ is a \emph{complete} and \emph{orthonormal} function set in
the Hilbert space
\begin{equation}
  \label{HilbertL^2}
  L^{2} (\mathbb{R}^3) \; = \; \Bigl\{ f \colon \mathbb{R}^3 \to
  \mathbb{C} \Bigm\vert \, \int \, \vert f (\bm{r}) \vert^2 \,
  \mathrm{d}^3 \bm{r} < \infty \Bigr\}
\end{equation}
of functions that are square integrable with respect to an integration
over the whole $\mathbb{R}^{3}$. Since any $f \in L^{2} (\mathbb{R}^{3})$
can be expanded in terms of the complete and orthonormal functions $\{
\varphi_{n, \ell}^{m} (\bm{r}) \}_{n, \ell, m}$, a one-range addition
theorem for $f (\bm{r} \pm \bm{r}')$ can be formulated as follows:
\begin{subequations}
  \label{OneRangeAddTheor}
  \begin{align}
     \label{OneRangeAddTheor_a}
    f (\bm{r} \pm \bm{r}') & \; = \; \sum_{n \ell m} \, C_{n, \ell}^{m}
    (f; \pm \bm{r}') \, \varphi_{n, \ell}^{m} (\bm{r}) \, ,
    \\
     \label{OneRangeAddTheor_b}
    C_{n, \ell}^{m} (f; \pm \bm{r}') & \; = \; \int \, \bigl[
    \varphi_{n, \ell}^{m} (\bm{r}) \bigr]^{*} \, f (\bm{r} \pm \bm{r}')
    \, \mathrm{d}^3 \bm{r} \, .
  \end{align}
\end{subequations}
The expansion (\ref{OneRangeAddTheor}) is a \emph{one-range} addition
theorem since the variables ${\bm{r}}$ and ${\bm{r}}'$ are completely
separated in a unique way independent of the lengths of $\bm{r}$ and
$\bm{r}'$: The dependence on $\bm{r}$ is contained in the functions
$\varphi_{n, \ell}^{m} (\bm{r})$, whereas $\bm{r}'$ occurs only in the
expansion coefficients $C_{n, \ell}^{m} (f; \pm \bm{r}')$ which are
overlap or convolution-type integrals.

Since the functions $\{ \varphi_{n, \ell}^{m} (\bm{r}) \}_{n, \ell, m}$
are by assumption complete and orthonormal in the Hilbert space $L^{2}
(\mathbb{R}^{3})$, the existence of the one-range addition theorem
(\ref{OneRangeAddTheor}) as well as its convergence in the mean is
guaranteed if $f \in L^{2} (\mathbb{R}^{3})$.

The summation limits in (\ref{OneRangeAddTheor}) depend on the exact
definition of the function set $\{ \varphi_{n, \ell}^{m} (\bm{r}) \}_{n,
  \ell, m}$. Unless explicitly specified, this article tacitly uses the
convention
\begin{equation}
  \sum_{n \ell m} \; = \;
  \sum_{n=1}^{\infty} \, \sum_{\ell=0}^{n-1} \, \sum_{m=-\ell}^{\ell} \, ,
\end{equation}
which is in agreement with the usual convention for the bound state
hydrogen eigenfunctions.

It is a typical feature of \emph{all} expansions of Hilbert space
elements in terms of function sets, which are complete and orthonormal in
this Hilbert space, that they do not necessarily converge pointwise but
only in the mean with respect to the norm of the corresponding Hilbert
space. As is well known from the mathematical literature, convergence in
the mean is weaker than pointwise convergence. Accordingly, expansions of
that kind are not necessarily suited for a pointwise representation of a
function. However, as discussed in more detail in Section
\ref{Sec:OneRangeAdd->TwoRangeAdd}, we must use a weaker form of
convergence if we want construct one-range addition theorems for
functions that are singular for $\bm{r} \pm \bm{r}' = \bm{0}$. Moreover,
we do not really need the stronger pointwise convergence if we only want
to evaluate multicenter integrals with the help of addition theorems.

One-range addition theorems of the kind of (\ref{OneRangeAddTheor}) were
constructed by Filter and Steinborn \cite[Eqs.\ (5.11) and
(5.12)]{Filter/Steinborn/1980} and later applied by Kranz and Steinborn
\cite{Kranz/Steinborn/1982} and by Trivedi and Steinborn
\cite{Trivedi/Steinborn/1982}. An alternative derivation of these
addition theorems based on Fourier transformation combined with weakly
convergent expansions of the plane wave $\exp (\pm \mathrm{i} \bm{p}
\cdot \bm{r})$ with $\bm{p}, \bm{r} \in \mathbb{R}^{3}$ was presented in
\cite[Section VII]{Weniger/1985} and in
\cite{Homeier/Weniger/Steinborn/1992a}. A similar approach based on the
work of Shibuya and Wulfman \cite{Shibuya/Wulfman/1965} was pursued by
Novosadov \cite[Section 3]{Novosadov/1983}.

As discussed in \cite[Section 3]{Weniger/2007b}), it is also possible to
formulate one-range addition theorems that converge with respect to the
norm of a \emph{weighted} Hilbert space
\begin{equation}
  \label{HilbertL_w^2}
  L_{w}^{2} (\mathbb{R}^3) \; = \; \Bigl\{ f \colon \mathbb{R}^3 \to
  \mathbb{C} \Bigm\vert \, \int \, w (\bm{r}) \, \vert f (\bm{r}) \vert^2
  \, \mathrm{d}^3 \bm{r} < \infty \Bigr\} \, ,
\end{equation}
where $w (\bm{r}) \ne 1$ is a suitable positive weight function.  If we
assume that $f \in L_{w}^{2} (\mathbb{R}^3)$ and that the functions $\{
\psi_{n, \ell}^{m} (\bm{r}) \}_{n, \ell, m}$ are complete and orthonormal
in $L_{w}^{2} (\mathbb{R}^3)$, then we obtain the following one-range
addition theorem \cite[Eq.\ (3.6)]{Weniger/2007b}):
\begin{subequations}
  \label{OneRangeAddTheor_w}
  \begin{align}
    \label{OneRangeAddTheor_w_a}
    f (\bm{r} \pm \bm{r}') & \; = \; \sum_{n \ell m} \, \mathbf{C}_{n,
      \ell}^{m} (f, w; \pm \bm{r}') \, \psi_{n, \ell}^{m} (\bm{r}) \, ,
    \\
    \label{OneRangeAddTheor_w_b}
    \mathbf{C}_{n, \ell}^{m} (f, w; \pm \bm{r}') & \; = \; \int \, \bigl[
    \psi_{n, \ell}^{m} (\bm{r}) \bigr]^{*} \, w (\bm{r}) \, f (\bm{r} \pm
    \bm{r}') \, \mathrm{d}^3 \bm{r} \, .
  \end{align}
\end{subequations}

A one-range addition theorem of the type of either
(\ref{OneRangeAddTheor}) or (\ref{OneRangeAddTheor_w}) for a function $f
\colon \mathbb{R}^{3} \to \mathbb{C}$ is a mapping $\mathbb{R}^3 \times
\mathbb{R}^3 \to \mathbb{C}$. Compared to the better known two-range
addition theorems like the Laplace expansion (\ref{LapExp}), which depend
on $\bm{r}$ and $\bm{r}'$ only indirectly via $\bm{r}_{<}$ and
$\bm{r}_{>}$, one-range addition theorems have the highly advantageous
feature that they provide unique infinite series representations of
functions $f (\bm{r} \pm \bm{r}')$ with separated variables $\bm{r}$ and
$\bm{r}'$ that are valid for the whole argument set $\mathbb{R}^3 \times
\mathbb{R}^3$.

In his $k$-dependent one-range addition theorems for Slater-type
functions and related functions, Guseinov used as a complete and
orthonormal set the following class of functions \cite[Eq.\
(1)]{Guseinov/2002c}, which -- if the mathematical notation
(\ref{GLag_1F1}) for the generalized Laguerre polynomials is used -- can
be expressed as follows \cite[Eq.\ (4.16)]{Weniger/2007b}:
\begin{align}
  \label{Def_Psi_Guseinov}
  \prescript{}{k}{\Psi}_{n, \ell}^{m} (\gamma, \bm{r}) & \; = \; \left[
  \frac{(2\gamma)^{k+3} (n-\ell-1)!}{(n+\ell+k+1)!} \right]^{1/2} \,
  \mathrm{e}^{-\gamma r} \, L_{n-\ell-1}^{(2\ell+k+2)} (2 \gamma r) \,
  \mathcal{Y}_{\ell}^{m} (2 \gamma \bm{r}) \, ,
  \notag \\
  & \qquad n \in \mathbb{N} \, , \quad n \ge \ell+1 \, , 
  \quad k = -1, 0, 1, 2, \dots \, ,  \quad \gamma > 0 \, .
\end{align}
The restriction to integral values of $k = -1, 0, 1, 2, \dots$ is
unnecessary. The mathematical definition (\ref{GLag_1F1}) of the
generalized Laguerre polynomials $U_{n}^{{(\alpha)}} (z)$ permits
nonintegral superscripts $\alpha$. Thus, the condition $k = -1, 0, 1, 2,
\dots$ in (\ref{Def_Psi_Guseinov}) can be replaced by $k \in [-1,
\infty)$. If this is done, one only has to replace the factorial
$(n+\ell+k+1)!$ in (\ref{Def_Psi_Guseinov}) by the gamma function $\Gamma
(n+\ell+k+2)$.

These functions are orthonormal with respect to the weight function $w
(\bm{r}) = r^{k}$ (compare also \cite[Eq.\ (4)]{Guseinov/2002c}):
\begin{equation}
  \label{Psi_Guseinov_OrthoNor}
  \int \, \bigl[ \prescript{}{k}{\Psi}_{n, \ell}^{m} (\gamma, \bm{r})
  \bigr]^{*} \, r^k \, \prescript{}{k}{\Psi}_{n', \ell'}^{m'} (\gamma,
  \bm{r}) \, \mathrm{d}^3 \bm{r} \; = \;
  \delta_{n n'} \, \delta_{\ell \ell'} \, \delta_{m m'} \, .
\end{equation}
Accordingly, these functions are complete and orthonormal in the weighted
Hilbert space
\begin{equation}
  \label{HilbertL_r^k^2}
  L_{r^k}^{2} (\mathbb{R}^3) \; = \; \Bigl\{ f \colon \mathbb{R}^3 \to
  \mathbb{C} \Bigm\vert \, \int \, r^k \, \vert f (\bm{r}) \vert^2 \,
  \mathrm{d}^3 \bm{r} < \infty \Bigr\} \, ,
  \qquad k = -1, 0, 1, 2, \dots \; .
\end{equation}

As discussed in the text following \cite[Eq.\ (4.20)]{Weniger/2007b}, the
functions $\prescript{}{k}{\Psi}_{n, \ell}^{m} (\gamma, \bm{r})$ can --
depending on the value of the free parameter $k = -1, 0, 1, 2, \dots$ --
reproduce several other physically relevant complete and orthonormal
function sets.

If we set $k=0$ in (\ref{HilbertL_r^k^2}), we retrieve the Hilbert space
$L^{2} (\mathbb{R}^{3})$ of square integrable functions defined by
(\ref{HilbertL^2}) which because of the Born interpretation of bound
state wave functions is the most natural choice for the representation of
effective one-particle wave functions in electronic structure
calculations.

For $k \ne 0$, there is, however, a problem since we neither have $L^{2}
(\mathbb{R}^3) \subset L_{r^{k}}^{2} (\mathbb{R}^3)$ nor $L_{r^{k}}^{2}
(\mathbb{R}^3) \subset L^{2} (\mathbb{R}^3)$. Thus, the Hilbert spaces
$L^{2} (\mathbb{R}^3)$ and $L_{r^{k}}^{2} (\mathbb{R}^3)$ are for $k \ne
0$ inequivalent. This applies also to approximation processes which
(should) converge with respect to the norms of these Hilbert spaces (see
for example \cite[Appendix E]{Weniger/2007c}).

Conceptually, the derivation of a one-range addition is a
triviality. This follows from the obvious fact that an arbitrary Hilbert
space element can be expanded in terms of a function set that is complete
and orthonormal in this Hilbert space.

Unfortunately, this does not imply that the derivation of a one-range
addition theorem is necessarily a simple task. The challenging part is
the construction of computationally convenient expressions for the
overlap integrals (\ref{OneRangeAddTheor_b}) or
(\ref{OneRangeAddTheor_w_b}), respectively. In realistic applications, we
cannot not tacitly assume that the use of one-range addition theorems in
multicenter integrals necessarily leads to rapidly convergent expansions
(compare for instance the convergence rates reported by Trivedi and
Steinborn \cite{Trivedi/Steinborn/1982}). Therefore, we must be able to
compute the overlap integrals (\ref{OneRangeAddTheor_b}) or
(\ref{OneRangeAddTheor_w_b}) both efficiently and reliably even for
possibly very large indices.

Let me conclude this Section with some short remarks on the relative
advantages and disadvantages of some of the various sets of exponentially
decaying function sets that are described in the literature.

The best known and most often used exponentially decaying functions are
undoubtedly Slater-type functions (\ref{Def_STF}) introduced in
\cite{Slater/1930,Slater/1932}. Slater himself had argued that
sufficiently accurate approximations to numerically determined solutions
of effective one-particle Schr\"{o}dinger equations can be obtained even
if the radial nodes of these solutions are completely ignored \cite[p.\
57]{Slater/1930}. Slater's prime concern was not accuracy but analytic
simplicity \cite[p.\ 42]{Slater/1932}. In view of the limited
computational resources at that time, Slater's pragmatic attitude
certainly made sense.

Because of their remarkably simple structure, Slater-type functions are
often considered to be the most basic prototypes of all exponentially
decaying functions. However, the simplicity of Slater-type functions in
the coordinate representation is deceptive. Slater-type functions have
been used with considerable success in the case of atomic electronic
structure calculations, where multicenter integrals do not occur. 

But there is considerable evidence that in the case of multicenter
integrals it is the simplicity of a function in the momentum
representation that really matters. As for example discussed in
\cite{Weniger/Steinborn/1983a}, the Fourier transform of a Slater-type
function is a comparatively complicated object having the same level of
complexity as the Fourier transform of a bound state hydrogen
eigenfunction (see for example \cite[Section
IV]{Weniger/1985}). Therefore, it is certainly worth while to look for
alternative exponentially decaying function sets with simpler Fourier
transforms and also more convenient properties in multicenter problems.

Inspired by previous work of Shavitt \cite[Eq.\ (55) on p.\
15]{Shavitt/1963}, Steinborn and Filter \cite[Eqs.\ (3.1) and
(3.2)]{Steinborn/Filter/1975c} introduced the so-called reduced Bessel
function
\begin{equation}
   \label{Def:RBF}
\hat{k}_{\nu} (z) \; = \; (2/\pi)^{1/2} \, z^{\nu} \, K_{\nu} (z) \, .
\end{equation}
Here, $K_{\nu} (z)$ is a modified Bessel function of the second kind
\cite[p.\ 66]{Magnus/Oberhettinger/Soni/1966}. If the order $\nu$ is
half-integral, $\nu = m + 1/2$ with $m \in \mathbb{N}_0$, the reduced
Bessel function is an exponential multiplied by a terminating confluent
hypergeometric series ${}_1 F_1$ (see for example \cite[Eq.\
(3.7)]{Weniger/Steinborn/1983b}):
\begin{equation}
  \label{RBF_HalfInt}
  \hat{k}_{m+1/2} (z) \; = \; 2^m \, (1/2)_m \,
  \mathrm{e}^{-z} \, {}_1 F_1 (-m; -2m; 2z) \, . 
\end{equation}
As discussed in more detail in Section \ref{Sec:OneRangeAdd->TwoRangeAdd}
or in \cite[Section 4]{Weniger/2009a}, Steinborn and Filter became
interested in reduced Bessel functions because of a known Gegenbauer-type
addition theorem which allowed a simple derivation a two-range addition
theorem for reduced Bessel functions with half-integral orders (see for
example \cite[Eq.\ (3.4)]{Steinborn/Filter/1975c} or, as an improved
version \cite[Eq.\ (5.5)]{Weniger/Steinborn/1989b}).

In connection with convolution and Coulomb integrals, Filter and
Steinborn later introduced the so called $B$ functions as an anisotropic
generalization of the reduced Bessel functions with half-integral orders
\cite[Eq.\ (2.14)]{Filter/Steinborn/1978b}:
\begin{equation}
  \label{Def:B_Fun}
  B_{n,\ell}^{m} (\beta, \hm{r}) \; = \; 
   [2^{n+\ell} (n+\ell)!]^{-1} \, \hat{k}_{n-1/2} (\beta r) \, 
    \mathcal{Y}_{\ell}^{m} (\beta \bm{r}) \, ,
   \qquad n \in \mathbb{Z} \, .
\end{equation}

$B$ functions are fairly complicated mathematical objects, and
(\ref{Def:B_Fun}) and (\ref{RBF_HalfInt}) imply that a $B$ function can
be expressed as a linear combination of Slater-type functions with
integral principal quantum numbers. Hence, it is not at all clear why the
use of the comparatively complicated $B$ functions should offer any
advantages over Slater-type functions which possess an exceptionally
simple explicit expression in the coordinate representation. 

Let us for the moment assume that we form some finite linear combinations
of Slater-type functions and that we do some mathematical manipulations
with this linear combination. Normally, the complexity of the resulting
expression increases, depending on the number of Slater-type functions
occurring in the linear combination. In fortunate cases, however, it may
happen that most terms of the resulting expression cancel exactly. Thus,
a significant reduction of complexity is also possible if we form
appropriate linear combinations.

The Fourier transform of a $B$ function seems to be such a fortunate case
since it is of exceptional simplicity among exponentially decaying
functions:
\begin{align}
  \label{FT_B_Fun}
  \bar{B}_{n,\ell}^{m} (\alpha, \bm{p}) & \; = \; (2\pi)^{-3/2} \, \int
  \, \mathrm{e}^{- \mathrm{i} \bm{p} \cdot \bm{r}} \, B_{n,\ell}^{m}
  (\alpha, \bm{r}) \, \mathrm{d}^3 \bm{r}
  \notag \\
  & \; = \; 
    (2/\pi)^{1/2} \, \frac{\alpha^{2n+\ell-1}}{[\alpha^2 +
    p^2]^{n+\ell+1}} \, \mathcal{Y}_{\ell}^{m} (- \mathrm{i} \bm{p}) \, .
\end{align}
This is the most consequential and also the most often cited result of my
PhD thesis \cite[Eq.\ (7.1-6) on p.\ 160]{Weniger/1982}. Later,
(\ref{FT_B_Fun}) was published in \cite[Eq.\
(3.7)]{Weniger/Steinborn/1983a}. Independently and almost simultaneously,
(\ref{FT_B_Fun}) was also derived by Niukkanen \cite[Eqs.\ (57) -
(58)]{Niukkanen/1984c}.

The exceptionally simple Fourier transform (\ref{FT_B_Fun}) explains why
multicenter integrals of $B$ functions are often simpler than the
corresponding integrals of other exponentially decaying functions (see
for example \cite{Safouhi/2010a,Safouhi/2010b,Weniger/2009a} and
references therein). It also explains why it was comparatively easy to
derive two-range \cite{Weniger/2002,Weniger/Steinborn/1989b} and
one-range \cite{Filter/Steinborn/1980} addition theorems for $B$
functions.

The exceptionally simple Fourier transform (\ref{FT_B_Fun}) also explains
why other exponentially decaying functions can be expressed in terms of
$B$ functions. For example, a Slater-type function with an integral
principal quantum number can be expressed by the following finite sum of
$B$ functions \cite[Eqs.\ (3.3) and (3.4)]{Filter/Steinborn/1978b}:
\begin{align}
  \label{STF->Bfun}
  \chi_{n, \ell}^{m} (\beta, \bm{r}) & \; = \; 
  2^n \, \sum_{\sigma \ge 0} \, (-1)^{\sigma} \, 
  \frac{(-[n-\ell-1]/2)_{\sigma} \, (-[n-\ell]/2)_{\sigma}}{{\sigma}!}
  \notag \\
  & \qquad \qquad \times \, 
  (n-\sigma)! \, B_{n-\ell-\sigma, \ell}^{m} (\beta, \bm{r}) \, .
\end{align}
If $n-\ell$ is \emph{even}, the Pochhammer symbol
$(-[n-\ell]/2)_{\sigma}$ causes a termination of the $\sigma$ summation
after a finite number of steps, and if $n-\ell$ is \emph{odd}, this is
accomplished by the Pochhammer symbol $(-[n-\ell-1]/2)_{\sigma}$. 
 
It is also possible to express Guseinov's complete and orthonormal
functions $\prescript{}{k}{\Psi}_{n, \ell}^{m} (\gamma, \bm{r})$ defined
by (\ref{Def_Psi_Guseinov}), which play a central role in Guseinov's work
on one-range addition theorems, as a finite sum of $B$ functions. We only
have to use \cite[Eq.\ (3.3-35) on p.\ 45]{Weniger/1982}
\begin{equation}
  \label{GLag->RBF2}
  \mathrm{e}^{-z}  \, L_{n}^{(\alpha)} (2z) \; = \; (\alpha+2n+1) \, 
  \sum_{\sigma=0}^{n} \, \frac{(-2)^{\sigma} \, \Gamma
    (\alpha+n+\sigma+1)}{(n-\sigma)! {\sigma}! \, \Gamma
    (\alpha+2\sigma+2)} \, \hat{k}_{\sigma+1/2} (z) 
\end{equation}
to obtain
\begin{align}
  \label{GusFun_Bfun}
  & \prescript{}{k}{\Psi}_{n, \ell}^{m} (\gamma, \bm{r}) \; = \; \left\{
    \frac{\gamma^{k+3} \, (n+\ell+k+1)!}{2^{k+1} \, (n-\ell-1)!}
  \right\}^{1/2} \frac{(2n+k+1) \, \Gamma (1/2) \, (\ell+1)!}  {\Gamma
    \bigl(\ell+2+k/2\bigr) \, \Gamma \bigl(\ell+[k+5]/2\bigr)}
  \notag \\
  & \qquad \times \sum_{\nu=0}^{n-\ell-1} \, \frac{(-n+\ell+1)_{\nu} \,
    (n+\ell+k+2)_{\nu} \, (\ell+2)_{\nu}}{\nu!  \,
    \bigl(\ell+2+k/2\bigr)_{\nu} \, \bigl(\ell+[k+5]/2\bigr)_{\nu}} \,
  B_{\nu+1,\ell}^{m} (\gamma, \hm{r}) \, .
\end{align}

Consequently, explicit expressions for multicenter integrals and addition
theorems of most exponentially decaying functions -- among them
Slater-type functions with integral principal numbers and Guseinov's
functions -- can be derived via the often simpler analogous results for
$B$ functions.

\typeout{==> Section: Guseinov's Rearrangements of One-Range Addition
  Theorems}
\section{Guseinov's Rearrangements of One-Range Addition Theorems}
\label{Sec:GuseinovsRearrangementsOfAdditionTheorems}

In \cite{Guseinov/1978,Guseinov/1980a,Guseinov/1985a,Guseinov/2001a,%
  Guseinov/2002c}, Guseinov derived one-range addition theorems for
Slater-type functions $\chi_{N, L}^{M} (\beta, \bm{r} \pm \bm{r}')$ with
integral or nonintegral principal quantum numbers $N$ by expanding them
in terms of his complete and orthonormal functions $\bigl\{
\prescript{}{k}{\Psi}_{n, \ell}^{m} (\gamma, \bm{r}) \bigr\}_{n, \ell,
  m}$ defined by (\ref{Def_Psi_Guseinov}) with an in general different
scaling parameter $\gamma \ne \beta > 0$:
\begin{subequations}
  \label{Gus_OneRangeAddTheorSTF_k}
  \begin{align}
    \label{Gus_OneRangeAddTheorSTF_k_a}
    \chi_{N, L}^{M} (\beta, \bm{r} \pm \bm{r}')
    & \; = \; \sum_{n \ell m} \,
     \prescript{}{k}{\mathbf{X}}_{n, \ell, m}^{N, L, M}
     (\gamma, \beta, \pm \bm{r}') \,
    \prescript{}{k}{\Psi}_{n, \ell}^{m} (\gamma, \bm{r}) \, ,
    \\
    \label{Gus_OneRangeAddTheorSTF_k_b}
    \prescript{}{k}{\mathbf{X}}_{n, \ell, m}^{N, L, M}
    (\gamma, \beta, \pm \bm{r}')
    & \; = \; \int \, \bigl[ \prescript{}{k}{\Psi}_{n, \ell}^{m}
    (\gamma, \bm{r}) \bigr]^{*} \, r^k \, \chi_{N, L}^{M} (\beta, \bm{r}
    \pm \bm{r}') \, \mathrm{d}^3 \bm{r} \, .
  \end{align}
\end{subequations}
As long as the principal quantum number $N$ is not too negative, which
will be tacitly assumed in the following text, the Slater-type function
$\chi_{N, L}^{M} (\beta, \bm{r} \pm \bm{r}')$ belongs to the weighted
Hilbert spaces $L_{r^k}^{2} (\mathbb{R}^3)$ defined by
(\ref{HilbertL_r^k^2}) with $k=-1, 0, 1, 2, \dots$. Accordingly, the
$k$-dependent one-range addition theorems
(\ref{Gus_OneRangeAddTheorSTF_k}), which are special cases of the general
addition theorem (\ref{OneRangeAddTheor_w}), converge in the mean with
respect to the norms of their corresponding weighted Hilbert spaces
$L_{r^k}^{2} (\mathbb{R}^3)$.

As already remarked at the end of Section
\ref{Sec:BasicFeaturesOfAdditionTheorems}, the central computational
problem occurring in the context of Guseinov's one-range addition
theorems (\ref{Gus_OneRangeAddTheorSTF_k}) is the efficient and reliable
evaluation of the overlap integrals (\ref{Gus_OneRangeAddTheorSTF_k_b}).

Normally, it is much easier to compute overlap integrals of exponentially
decaying functions with equal scaling parameters $\beta = \gamma$ than
with different parameters $\beta \ne \gamma$. However, Guseinov wanted to
have this additional degree of freedom for certain applications. For
example, in \cite{Guseinov/2005a} Guseinov represented the Coulomb
potential $1/\vert \bm{r} - \bm{r}' \vert$ as the limiting case $\beta
\to 0$ of the Yukawa potential $\exp (- \beta \vert \bm{r} - \bm{r}'
\vert)/\vert \bm{r} - \bm{r}' \vert$, which is proportional to the
Slater-type function $\chi_{0, 0}^{0} (\beta, \bm{r} - \bm{r}')$. In this
way, Guseinov \cite{Guseinov/2005a} formally obtained a one-range
addition theorem of the Coulomb potential in terms of his functions
$\prescript{}{k}{\Psi}_{n, \ell}^{m} (\gamma, \bm{r})$. But this approach
leads to other convergence and existence problems
\cite{Weniger/2007b,Weniger/2007c} which will be discussed in more
details in \cite{Weniger/2011a*}.

For the evaluation of the overlap integrals
(\ref{Gus_OneRangeAddTheorSTF_k_b}) occurring in his addition theorems
(\ref{Gus_OneRangeAddTheorSTF_k}a), Guseinov uses a simple approach which
was first used by Smeyers \cite{Smeyers/1966} in 1966. To the best of my
knowledge, this approach was adopted in 1978 by Guseinov \cite[Eqs.\ (6)
- (8)]{Guseinov/1978} and consistently used in his later publications.
Unfortunately, Smeyers' simple approach is not necessarily good, as
already emphasized in 1982 by Trivedi and Steinborn \cite[pp.\ 116 -
117]{Trivedi/Steinborn/1982}.

It follows at once from the explicit expression (\ref{GLag_1F1}) of the
generalized Laguerre polynomials that Guseinov's functions can be
expressed as finite sums of Slater-type functions with integral principal
quantum numbers:
\begin{subequations}
  \label{GusFun2STF}
  \begin{align}
    \label{GusFun2STF_a}
    \prescript{}{k}{\Psi}_{n, \ell}^{m} (\gamma, \bm{r}) & \; = \;
    \sum_{\nu=\ell+1}^{n} \,
    \prescript{}{k}{\mathbf{G}}_{\nu-\ell-1}^{(n, \ell)} (\gamma) \;
    \chi_{\nu, \ell}^{m} (\gamma, \bm{r}) \, ,
    \\
    \label{GusFun2STF_b}
    \prescript{}{k}{\mathbf{G}}_{j}^{(n, \ell)} (\gamma) & \; = \;
    2^{\ell} \, \left[ \frac{(2\gamma)^{k+3} \,
        (n+\ell+k+1)!}{(n-\ell-1)!} \right]^{1/2} \,
    \frac{(-n+\ell+1)_{j} \, 2^{j}}{(2\ell+k+j+2)! \, j!} \, .
  \end{align}
\end{subequations}
Accordingly, the overlap integrals (\ref{Gus_OneRangeAddTheorSTF_k_b})
can be expressed as finite sums of overlap integrals of Slater-type
functions: {\allowdisplaybreaks
\begin{subequations}
  \label{OverlapGusFun_STF}
  \begin{align}
    \prescript{}{k}{\mathbf{X}}_{n, \ell, m}^{N, L, M} (\gamma, \beta,
    \pm \bm{r}') & \; = \; \gamma^{-k} \, \sum_{\nu=\ell+1}^{n} \,
    \prescript{}{k}{\mathbf{G}}_{\nu-\ell-1}^{(n, \ell)} (\gamma) \;
    \mathbf{S}_{\nu+k, \ell, m}^{N, L, M} (\gamma, \beta, \pm \bm{r}')
    \, ,
    \\
    \mathbf{S}_{n, \ell, m}^{N, L, M} (\gamma, \beta, \pm \bm{r}') & \; =
    \; \int \, \bigl[ \chi_{n, \ell}^{m} (\gamma, \bm{r}) \bigr]^{*} \,
    \chi_{N, L}^{M} (\beta, \bm{r} \pm \bm{r}') \, \mathrm{d}^3 \bm{r}
    \, .
  \end{align}
\end{subequations}} 
Inserting this into the addition theorems
(\ref{Gus_OneRangeAddTheorSTF_k}) yields (compare \cite[Eqs.\ (15) and
(16)]{Guseinov/2002c}):
\begin{equation}
  \label{Gus_OneRangeAddTheorSTF_OvSTF_k}
  \chi_{N, L}^{M} (\beta, \bm{r} \pm \bm{r}') \; = \; \gamma^{-k} \,
  \sum_{n \ell m} \, \prescript{}{k}{\Psi}_{n, \ell}^{m} (\gamma, \bm{r})
  \, \sum_{\nu=\ell+1}^{n} \,
  \prescript{}{k}{\mathbf{G}}_{\nu-\ell-1}^{(n, \ell)} (\gamma) \;
  \mathbf{S}_{\nu+k, \ell, m}^{N, L, M} (\gamma, \beta, \pm
  \bm{r}') \, .  
\end{equation}

Guseinov's approach seems to have the advantage that existing programs
for overlap integrals of Slater-type functions can be used for the
evaluation of the overlap integrals (\ref{Gus_OneRangeAddTheorSTF_k_b})
involving Guseinov's functions. Unfortunately, this seemingly convenient
approach can easily lead to stability problems. Numerical instabilities
are extremely likely in expressions based on (\ref{GusFun2STF}) if the
indices $n$ of the Guseinov functions $\prescript{}{k}{\Psi}_{n,
  \ell}^{m} (\gamma, \bm{r})$ are large.

In his desire to reduce his whole formalism of one-range addition
theorems to Slater-type functions with integral principal quantum
numbers, Guseinov even expressed the functions $\prescript{}{k}{\Psi}_{n,
  \ell}^{m} (\gamma, \bm{r})$ on the right-hand side of
(\ref{Gus_OneRangeAddTheorSTF_OvSTF_k}) by Slater-type functions
according to (\ref{GusFun2STF}) (compare \cite[Eqs.\ (14) -
(16)]{Guseinov/2002c}):
\begin{align}
  \label{Gus_OneRangeAddTheorSTF_STF_k}
  \chi_{N, L}^{M} (\beta, \bm{r} \pm \bm{r}') \; = \; \gamma^{-k} \,
  \sum_{n \ell m} \, & \sum_{\nu'=\ell+1}^{n} \,
  \prescript{}{k}{\mathbf{G}}_{\nu'-\ell-1}^{(n, \ell)} (\gamma) \,
  \chi_{\nu', \ell}^{m} (\gamma, \bm{r})
  \notag \\
  \qquad \quad \; \times \, & \sum_{\nu=\ell+1}^{n} \,
  \prescript{}{k}{\mathbf{G}}_{\nu-\ell-1}^{(n, \ell)} (\gamma) \;
  \mathbf{S}_{\nu+k, \ell, m}^{N, L, M} (\gamma, \beta, \pm \bm{r}') \, .
\end{align}
In this version of Guseinov's $k$-dependent addition theorems, the
complete and orthogonal functions $\prescript{}{k}{\Psi}_{n, \ell}^{m}
(\gamma, \bm{r})$ do not occur any more and are replaced by Slater-type
functions $\chi_{n, \ell}^{m} (\gamma, \bm{r})$ with integral principal
quantum numbers $n \ge \ell+1$. Apparently, Guseinov considered this to
be a major achievement.

But there is a price to be paid. Firstly, introducing one finite inner
sum after the other does not look like a promising computational
strategy. Secondly, the numerical stability of Guseinov's transformations
is questionable, in particular in the case of (very) large quantum
numbers. 

Whenever we do calculations of nontrivial complexity with a fixed
precision, numerical stability is invariably a very important issue. Let
us for example assume that we work with a series expansion whose
coefficients are given by complicated expressions such as multiple nested
finite sums (this characterization certainly applies to Guseinov's
rearranged addition theorems). If alternating signs occur in these nested
sums, a loss of significant digits cannot be ruled out, and in the case
of (very) large indices, a catastrophic accumulation of rounding errors
may even produce completely nonsensical results.

Guseinov's rearranged addition theorems are based on the simple fact that
Guseinov's orthogonal functions $\prescript{}{k}{\Psi}_{n, \ell}^{m}
(\gamma, \bm{r})$ can according to (\ref{GusFun2STF}) be expressed as
finite sums of Slater-type functions with integral principal quantum
numbers. The linear combination (\ref{GusFun2STF}) is nothing but the
explicit expression (\ref{GLag_1F1}) of a generalized Laguerre polynomial
in disguise. The coefficients in (\ref{GLag_1F1}) and thus also the
coefficients in (\ref{GusFun2STF}) have strictly alternating signs.

Sign alternation of the coefficients of the classical orthogonal
polynomials is a necessary requirement for an orthogonality relationship
of the type of (\ref{GLag_Orthogonality}). But this sign alternation can
easily lead to a catastrophic accumulation of rounding errors, if we try
to compute an orthogonal polynomial with a large index from its explicit
expression. In practice, one usually avoids explicit expressions. It is
much better to evaluate orthogonal polynomials via their three-term
recursions (see for example \cite[Eq.\ (18.9.1) and Table
18.9.1]{Olver/Lozier/Boisvert/Clark/2010}). For example, Gautschi, who is
generally considered to be one of the leading experts on the stability of
linear recurrences, wrote in \cite[p.\ 277]{Gautschi/1993}:
\begin{quote}
  \emph{It is our experience, and the experience of many others, that the
  basic three-term recurrence relation for orthogonal polynomials is
  generally an excellent means of computing these polynomials, both
  within the interval of orthogonality and outside of it.}
\end{quote}

A FORTRAN SUBROUTINE OTHPL, which evaluates classical orthogonal
polynomials via their three-term recursions \cite[Eq.\ (18.9.1) and Table
18.9.1]{Olver/Lozier/Boisvert/Clark/2010}, is listed in \cite[pp.\ 23 -
24]{Zhang/Jin/1996}. Another important aspect is that a recursive
evaluation is much more efficient, in particular if whole strings of
orthogonal polynomials have to be evaluated simultaneously.

As long as the opposite is not explicitly proven, it makes sense to be
cautious and to assume that an expression like (\ref{GusFun2STF}), whose
coefficients $\prescript{}{k}{\mathbf{G}}_{j}^{(n, \ell)} (\gamma)$ have
strictly alternating signs, inherits the stability problems of the
explicit expression (\ref{GLag_1F1}) of the generalized Laguerre
polynomials from which it was derived. Similarly, the coefficients
$\prescript{}{k}{\mathbf{G}}_{\nu'-\ell-1}^{(n, \ell)} (\gamma)$ and
$\prescript{}{k}{\mathbf{G}}_{\nu-\ell-1}^{(n, \ell)} (\gamma)$ in the
inner $\nu'$ and $\nu$ sums in (\ref{Gus_OneRangeAddTheorSTF_STF_k}) have
strictly alternating signs.

I am not at all convinced that inner sums based on (\ref{GLag_1F1}),
which occur as coefficients in infinite series expansions like the one on
the right-hand side of (\ref{Gus_OneRangeAddTheorSTF_STF_k}) or later in
(\ref{Gus_OneRangeAddTheorSTF_STF_k_Rearr_1}) and
(\ref{Gus_OneRangeAddTheorSTF_STF_k_Rearr_2}), can be computed in a
numerically stable way for large values of the outer summation indices
$n$ and $\ell$. This has to be investigated (much) more thoroughly.

Another problem is that conventional programs for overlap integrals of
Slater-type functions, as they are for instance used in semiempirical
calculations, normally cannot be used in the case of (very) large
principal and angular momentum quantum numbers. Therefore, one should
look for alternative expressions for the overlap integrals
(\ref{Gus_OneRangeAddTheorSTF_k_b}), which are not based on
(\ref{GusFun2STF}) and which permit an efficient and reliable evaluation
even for (very) large values of the indices $n$ and $\ell$.

Slater-type functions are complete but not orthogonal. But this
nonorthogonality can also cause problems. There is a practically very
consequential aspect of orthogonal expansions which is often not
sufficiently appreciated. Let us assume that $f$ belongs to some Hilbert
space $\mathcal{H}$ with inner product $(\cdot \vert \cdot)$ and that the
functions $\{ \varphi_n \}_{n=0}^{\infty}$ are complete and orthonormal
in $\mathcal{H}$. Then, as discussed in Appendix
\ref{App:OrthogonalExpansions}, $f$ possesses the orthogonal expansion
(\ref{Expand_f_CONS}) which converges in the mean with respect to the
norm $\Vert \cdot \Vert$ of $\mathcal{H}$.  Moreover, the expansion
coefficients $(\varphi_n \vert f)$ satisfy Parseval's equality (see for
example \cite[Eq.\ (II.2) on p.\ 45]{Reed/Simon/1980})
\begin{equation}
  \label{ParsevalEquality}
\Vert f \Vert^2 \; = \;
\sum_{n=0}^{\infty} \, \vert (\varphi_n \vert f) \vert^2 \, .
\end{equation}
Parseval's equality implies that the inner products $(\varphi_n \vert f)$
are bounded in magnitude and that they vanish as $n \to \infty$. This may
well be the main reason why orthogonal expansions tend to be
computationally well behaved.

In the case of nonorthogonal expansions of the type of (\ref{f_InfExp}),
quite a few complications can happen. We cannot tacitly assume that the
expansion coefficients $C_n$ in (\ref{f_InfExp}) are necessarily bounded
in magnitude. These coefficients can have alternating signs and even
increase in magnitude with increasing index $n$ (examples can for
instance be found in \cite[Table I on p.\ 166]{Klahn/1981} or
\cite[Appendix E on pp.\ 162 - 164]{Filter/1978}). Such a behavior of the
expansion coefficients can easily lead to a cancellation of significant
digits or even to a catastrophic accumulation of rounding errors.

Guseinov approximated the Slater-type function $\chi_{N, L}^{M} (\beta,
\bm{r} \pm \bm{r}')$ by a truncation of the addition theorem
(\ref{Gus_OneRangeAddTheorSTF_STF_k}) including only the first
$\mathcal{N}$ terms of the outer $n$ summation and defined the complete
addition theorem as the limit $\mathcal{N} \to \infty$ of his
$\mathcal{N}$-dependent truncation (compare \cite[Eq.\
(13)]{Guseinov/2002c}):
\begin{align}
  \label{Gus_OneRangeAddTheorSTF_STF_k_Rearr_1}
  & \chi_{N, L}^{M} (\beta, \bm{r} \pm \bm{r}') \; = \; \gamma^{-k} \,
  \lim_{\mathcal{N} \to \infty} \, \sum_{n=1}^{\mathcal{N}} \, 
  \sum_{\ell=0}^{n-1} \, \sum_{m=-\ell}^{\ell} 
  \notag \\
  & \qquad \times \, \sum_{\nu'=\ell+1}^{n} \,
  \prescript{}{k}{\mathbf{G}}_{\nu'-\ell-1}^{(n, \ell)} (\gamma) \;
  \chi_{\nu', \ell}^{m} (\gamma, \bm{r}) \, \sum_{\nu=\ell+1}^{n} \,
  \prescript{}{k}{\mathbf{G}}_{\nu-\ell-1}^{(n, \ell)} (\gamma) \;
  \mathbf{S}_{\nu+k, \ell, m}^{N, L, M} (\gamma, \beta, \pm \bm{r}') \, .
\end{align}
In the $\mathcal{N}$-dependent part of this expression, Guseinov changed
the order of summations and formally expressed this finite sum as a
linear combination terms of Slater-type functions with integral principal
quantum numbers \cite[Eq.\ (15)]{Guseinov/2002c}. In this context, it is
in my opinion advantageous to change the order of the finite $n$ and
$\ell$ summations according to $\sum_{n=1}^{\mathcal{N}}
\sum_{\ell=0}^{n-1} \to \sum_{\ell=0}^{\mathcal{N}-1}
\sum_{n=\ell+1}^{\mathcal{N}}$. Then, we obtain:
\begin{align}
  \label{Gus_OneRangeAddTheorSTF_STF_k_Rearr_2}
  & \chi_{N, L}^{M} (\beta, \bm{r} \pm \bm{r}') \; = \; \gamma^{-k} \,
  \lim_{\mathcal{N} \to \infty} \, \sum_{\ell=0}^{\mathcal{N}-1} \,
  \sum_{m=-\ell}^{\ell} \, \sum_{t=\ell+1}^{\mathcal{N}} \, 
  \chi_{t, \ell+1}^{m} (\gamma, \bm{r})
  \notag \\
  & \qquad \times \, \sum_{p=t}^{\mathcal{N}} \,
  \prescript{}{k}{\mathbf{G}}_{t-\ell-1}^{(p, \ell)} (\gamma) \,
  \sum_{q=\ell+1}^{n} \,
  \prescript{}{k}{\mathbf{G}}_{q-\ell-1}^{(p, \ell)} (\gamma) \;
  \mathbf{S}_{q+k, \ell, m}^{N, L, M} (\gamma, \beta, \pm \bm{r}') \, .
\end{align}

As long as $\mathcal{N}$ is finite, the transformation of the
$\mathcal{N}$-dependent part of
(\ref{Gus_OneRangeAddTheorSTF_STF_k_Rearr_1}) to the
$\mathcal{N}$-dependent part of
(\ref{Gus_OneRangeAddTheorSTF_STF_k_Rearr_2}) is legitimate. But it is
not guaranteed that this transformation is still legitimate in the limit
$\mathcal{N} \to \infty$ and that one-range addition theorems can be
reformulated as expansions in terms of nonorthogonal Slater-type
functions with integral principal quantum numbers. Further details will
be given in Section \ref{Sec:OneRangeAdd->TwoRangeAdd}.

One-range addition theorems for Slater-type functions are fairly
complicated mathematical objects, whose series coefficients are
essentially overlap integrals. Thus, a detailed analysis of the existence
and convergence properties of such an addition theorem is certainly a
very demanding task. Consequently, it makes sense to look for
simplifications and to pursue an indirect approach. The situation becomes
much more transparent if we do not look at addition theorems for
Slater-type functions $\chi_{N, L}^{M} (\beta, \bm{r} \pm \bm{r}')$, but
rather at their angular projections
\begin{equation}
  \label{SFT_Mom}
  \Xi_{\ell, m}^{N, L, M} (\beta, r, \pm \bm{r}') \; = \; 
  \int_{\vert \bm{r} \vert = 1} \, 
  \left[ Y_{\ell}^{m} (\bm{r}/r) \right]^{*} \, 
  \chi_{N, L}^{M} (\beta, \bm{r} \pm \bm{r}') \, 
  \mathrm{d}^{3} \bm{r} \, .
\end{equation}
If we insert the addition theorem (\ref{Gus_OneRangeAddTheorSTF_k}) into
this integral and perform the integration over the surface of the unit
sphere in $\mathbb{R}^{3}$, we obtain with the help of the orthonormality
of the spherical harmonics an expansion in terms of the radial parts of
Guseinov's functions $\prescript{}{k}{\Psi}_{n, \ell}^{m} (\gamma,
\bm{r})$, which can be reformulated as an expansion in terms of
generalized Laguerre polynomials:
\begin{align}
  \label{SFT_Mom_GLagPol}
  \mathrm{e}^{\gamma r} \, 
  & \Xi_{\ell, m}^{N, L, M} (\beta, r, \pm \bm{r}') \; = \; 
  \sum_{n=\ell+1}^{\infty} \, 
   \prescript{}{k}{\mathbf{X}}_{n, \ell, m}^{N, L, M}
   (\gamma, \beta, \pm \bm{r}')
  \notag \\
  & \qquad \times \, \left[
  \frac{(2\gamma)^{k+3} (n-\ell-1)!}{(n+\ell+k+1)!} \right]^{1/2} \, 
  (\gamma r)^{\ell} \, L_{n-\ell-1}^{(2\ell+k+2)} (2 \gamma r) \, .
\end{align}
A further substantial simplifications takes place if we do not consider
the fairly complicated one-range addition theorems but rather their much
simpler one-center limits $\bm{r}' = \bm{0}$. Then, for fixed $\beta,
\gamma > 0$ the overlap integrals $\prescript{}{k}{\mathbf{X}}_{n, \ell,
  m}^{N, L, M} (\gamma, \beta, \pm \bm{r}')$ defined by
(\ref{Gus_OneRangeAddTheorSTF_k_b}) simplify to become numbers. Moreover,
because of the orthonormality of the spherical harmonics only the overlap
integrals with $\ell=L$ and $m=M$ can be nonzero.

\typeout{==> Section: The Transformation of Laguerre Series to Power
  Series}
\section{The Transformation of Laguerre Series to Power Series}
\label{Sec:LagSer->PowSer}

If Guseinov's rearrangements of the one-range addition theorems
(\ref{Gus_OneRangeAddTheorSTF_k}) for Slater-type functions $\chi_{N,
  L}^{M} (\beta, \bm{r} \pm \bm{r}')$ are legitimate, we ultimately
obtain an expansion of a Slater-type function with an in general
nonintegral principal quantum number $N \in \mathbb{R} \setminus
\mathbb{N}$ in terms of Slater-type functions $\bigl\{ \chi_{n, \ell}^{m}
(\gamma, \bm{r}) \bigr\}_{n \ell m}$ with integral principal quantum
numbers $n$ and a common scaling parameter $\gamma > 0$:
\begin{equation}
  \label{NISTO->ISTO_AddThm}
  \chi_{N, L}^{M} (\beta, \bm{r} \pm \bm{r}') \; = \; 
  \sum_{n \ell m} \, \mathbb{X}_{n, \ell, m}^{(N, L, M)} 
  (\beta, \gamma, \pm \bm{r}') \, \chi_{n, \ell}^{m} (\gamma, \bm{r}) \, .
\end{equation}
If we set $\bm{r}' = \bm{0}$, this expansion simplifies and we obtain
(\ref{GuExp_1}).

If the expansion (\ref{NISTO->ISTO_AddThm}) exists and converges for the
whole argument set $\mathbb{R}^{3} \times \mathbb{R}^{3}$, then it is
obviously a one-range addition theorem. Moreover,
(\ref{NISTO->ISTO_AddThm}) can be reformulated as a power series in
$\gamma r$ by shifting the exponential part $\exp (-\gamma r)$ of the
Slater-type functions $\chi_{n, \ell}^{m} (\gamma, \bm{r})$ to the
left-hand side:
\begin{equation}
  \label{NISTO->ISTO_PowSer}
  \mathrm{e}^{\gamma r} \, 
  \chi_{N, L}^{M} (\beta, \bm{r} \pm \bm{r}') \; = \; 
  \sum_{n \ell m} \, \mathbb{X}_{n, \ell, m}^{(N, L, M)} 
  (\beta, \gamma, \pm \bm{r}') \, (\gamma r)^{n-1} \, 
  Y_{\ell}^{m} (\bm{r}/r) \, .
\end{equation}
This expansion can be transformed to an infinite number of
$\ell$-dependent power series in $\gamma r$ for the angular projections
$\Xi_{\ell, m}^{N, L, M} (\beta, r, \pm \bm{r}')$ defined by
(\ref{SFT_Mom}):
\begin{equation}
  \label{SFT_Mom_PowSer}
  \mathrm{e}^{\gamma r} \, 
  \Xi_{\ell, m}^{N, L, M} (\beta, r, \pm \bm{r}') \; = \; 
  \sum_{n=\ell+1}^{\infty} \, \mathbb{X}_{n, \ell, m}^{(N, L, M)} 
  (\beta, \gamma, \pm \bm{r}') \, (\gamma r)^{n-1} \, . 
\end{equation}

A comparison of (\ref{SFT_Mom_GLagPol}) and (\ref{SFT_Mom_PowSer}) shows
that Guseinov's rearrangements ultimately corresponds to the
transformation of a Laguerre series
\begin{subequations}
  \label{f_Exp_GLag}
  \begin{align}
    \label{f_Exp_GLag_a}
    f (z) & \; = \; \sum_{n=0}^{\infty} \,
    \lambda_{n}^{(\alpha)} \, L_{n}^{(\alpha)} (z) \, ,
    \\
    \label{f_Exp_GLag_b}
    \lambda_{n}^{(\alpha)} & \; = \; \frac{n!}{\Gamma (\alpha+n+1)} \,
    \int_{0}^{\infty} \, z^{\alpha} \, \mathrm{e}^{-z} \,
    L_{n}^{(\alpha)} (z) \, f (z) \, \mathrm{d} z \, ,
  \end{align}
\end{subequations} 
which converges in the mean with respect to the norm of the weighted
Hilbert space $L^{2}_{\mathrm{e}^{-z} z^{\alpha}} \bigl([0, \infty)
\bigr)$ defined by (\ref{HilbertL^2_Lag}), to a power series
\begin{equation}
  \label{PowSer_f}
  f (z) \; = \; \sum_{n=0}^{\infty} \, \gamma_{n} \, z^{n} \, ,
\end{equation}
which -- if it exists -- converges pointwise in a suitable subset of the
complex plane. 

There is, however, the problem that analyticity in the sense of complex
analysis and the existence of a Laguerre expansion are completely
unrelated concepts. Therefore, we cannot expect that the transformation
from (\ref{f_Exp_GLag}) to (\ref{PowSer_f}) is necessarily possible in
the case of an essentially arbitrary function $f \colon \mathbb{C} \to
\mathbb{C}$.

The convergence and existence problems, which occur in this context, were
recently studied in depth in \cite{Weniger/2008}. There, some simple
sufficient conditions based on the decay rate and the sign pattern of the
Laguerre series coefficients $\lambda_{n}^{(\alpha)}$ as $n \to \infty$
were formulated. These conditions, which extend previous results by
Gottlieb and Orszag \cite[p.\ 42]{Gottlieb/Orszag/1977} and by Doha
\cite[p.\ 5452]{Doha/2003}, respectively, make it possible to decide
whether the transformation of a Laguerre series of the type of
(\ref{f_Exp_GLag}) to a power series produces a mathematically meaningful
result or not.

In order to understand better the possible pitfalls and dangers of such a
transformation, let us first consider a partial sum of the infinite
Laguerre series (\ref{f_Exp_GLag}):
\begin{align}
  \label{FinSum_GLag}
  f_{M} (z) & \; = \; \sum_{n=0}^{M} \,
  \lambda_{n}^{(\alpha)} \, L_{n}^{(\alpha)} (z)
  \\
  \label{FinSum_GLag_Tr_1}
  & \; = \; \sum_{n=0}^{M} \, \lambda_{n}^{(\alpha)} \,
  \frac{(\alpha+1)_{n}}{n!} \, \sum_{\nu=0}^{n} \,
  \frac{(-n)_{\nu}}{(\alpha+1)_{\nu}} \, \frac{z^{\nu}}{\nu!} \, ,
  \qquad M \in \mathbb{N}_{0} \, .  
\end{align}
In this partial sum, the power $z^{p}$ with $0 \le p \le M$ occurs in the
Laguerre polynomials $L_{p}^{(\alpha)} (z)$, $L_{p+1}^{(\alpha)} (z)$,
\dots, $L_{M}^{(\alpha)} (z)$. Thus, we have add up all contributions
with $\nu=p$ on the right-hand side of (\ref{FinSum_GLag_Tr_1}) to obtain
the coefficients of $z^{p}$. A short calculation yields \cite[Eq,\
(3.13)]{Weniger/2008}:
\begin{equation}
  \label{RearrFinSum_GLag}
  f_{M} (z) \; = \; \sum_{\nu=0}^{M} \, 
  \frac{(-z)^{\nu}}{\nu!} \,
  \sum_{\mu=0}^{M-\nu} \, \frac{(\alpha+\nu+1)_{\mu}}{\mu!} \,
  \lambda_{\mu+\nu}^{(\alpha)} \, .
\end{equation}

Since the truncated Laguerre series $f_{M} (z)$ is simply a polynomial of
degree $M$ in $z$, it is always possible to reformulate it by
interchanging the order of the nested finite sums. No convergence and/or
existence problems can occur.

Unfortunately, this does not mean that the transformation from
(\ref{FinSum_GLag}) to (\ref{RearrFinSum_GLag}) is also possible in the
limit $M \to \infty$. In this case, the finite inner sum on the
right-hand side of (\ref{RearrFinSum_GLag}) becomes an infinite series
which can a diverge if the coefficients $\lambda_{n}^{(\alpha)}$ do not
decay sufficiently rapidly as $n \to \infty$. Moreover, the fact that
$f_{M} (z)$ is a mathematically meaningful object does not imply that $f
(z)$ possesses a power series expansion of the type of (\ref{PowSer_f})
which converges in a suitable subset of the complex plane. The expression
(\ref{RearrFinSum_GLag}) for $f_{M} (z)$ possesses the following general
structure:
\begin{equation}
  \label{TruncPowSer_f}
  f_{M} (z) \; = \; \sum_{\mu=0}^{M} \, 
  \gamma_{\mu}^{(M)} \, z^{\mu} \, .
\end{equation}
The coefficients $\gamma_{n}^{(M)}$ depend explicit on the
summation limit $M$, i.e., we have in general
$\gamma_{n}^{(M)} \ne \gamma_{n}^{(M+1)} \ne
\gamma_{n}^{(M+2)}$ $\ne \dots$ for all $n \in \mathbb{N}_{0}$.
Thus, we have to show explicitly that the $\gamma_{n}^{(M)}$
possess for all $n \in \mathbb{N}_{0}$ well defined limits $\gamma_{n} =
\gamma_{n}^{(\infty)} = \lim_{M \to \infty}
\gamma_{n}^{(M)}$. Only if all these limits exist, $f (z)$ can
possess a convergent power series expansion about $z=0$. This, however,
requires that the Laguerre series coefficients $\lambda_{n}^{(\alpha)}$
decay sufficiently rapidly as $n \to \infty$.

Let us now consider the transformation of the infinite Laguerre expansion
(\ref{f_Exp_GLag}). The power $z^{p}$ with $p \ge 0$ occurs in all
Laguerre polynomials $L_{n}^{(\alpha)} (z)$ with $n \ge p$. We obtain the
corresponding power series coefficient $\gamma_{p}$, if we add up all
contributions with $\nu=p$ in the infinite series
\begin{equation}
  \label{InfSer_GLag_Tr_1}
  f (z) \; = \; \sum_{n=0}^{\infty} \, \lambda_{n}^{(\alpha)} \,
  \frac{(\alpha+1)_{n}}{n!} \, \sum_{\nu=0}^{n} \,
  \frac{(-n)_{\nu}}{(\alpha+1)_{\nu}} \, \frac{z^{\nu}}{\nu!} \, .
\end{equation}
A short calculation yields \cite[Eq,\ (3.14)]{Weniger/2008}:
\begin{equation}
  \label{Rearr_f_Exp_GLag}
  f (z) \; = \;
  \sum_{\nu=0}^{\infty} \, \frac{(-z)^{\nu}}{\nu!} \,
  \sum_{\mu=0}^{\infty} \, \frac{(\alpha+\nu+1)_{\mu}}{\mu!} \,
  \lambda_{\mu+\nu}^{(\alpha)} \, .
\end{equation}

If we compare (\ref{RearrFinSum_GLag}) and (\ref{Rearr_f_Exp_GLag}),
which corresponds to the limit $M \to \infty$ in
(\ref{RearrFinSum_GLag}), we immediately see that the rearrangement of an
infinite Laguerre expansion is not necessarily possible since the power
series coefficients $\gamma_{n}$ are now given by infinite series. If the
inner $\mu$ series in (\ref{Rearr_f_Exp_GLag}) do not produce finite
results, we end up with a formal power series with expansion coefficients
that are infinite in magnitude.

Thus, the existence of the power series (\ref{PowSer_f}) for $f (z)$
depends crucially on the decay rate and the sign pattern of the Laguerre
series coefficients $\lambda_{n}^{(\alpha)}$ as $n \to \infty$. In
\cite{Weniger/2008}, three different prototypes of large index behavior
of the coefficients $\lambda_{n}^{(\alpha)}$ were discussed:
\begin{itemize}
\item algebraic decay with ultimately monotone signs,
\item exponential or factorial decay,
\item algebraic decay with ultimately strictly alternating signs.
\end{itemize}

If the coefficients $\lambda_{n}^{(\alpha)}$ decay algebraically as $n
\to \infty$ and ultimately have monotone signs, the inner $\mu$ series in
(\ref{Rearr_f_Exp_GLag}) diverge for sufficiently large outer indices
$\nu$ \cite[Section 4]{Weniger/2008}. Thus, a function $f (z)$
represented by a Laguerre expansion with ultimately monotone and
algebraically decaying series coefficients $\lambda_{n}^{(\alpha)}$
cannot be analytic in a neighborhood of the origin $z=0$.

If the coefficients $\lambda_{n}^{(\alpha)}$ decay exponentially or even
factorially, the inner $\mu$ series in (\ref{Rearr_f_Exp_GLag}) converge
(often quite rapidly), and the function $f (z)$ under consideration is
analytic in a suitable neighborhood of the origin $z=0$ \cite[Section
5]{Weniger/2008}. This is pretty much the most advantageous situation we
can hope for.

These conclusions are in agreement with short remarks by Gottlieb and
Orszag \cite[p.\ 42]{Gottlieb/Orszag/1977} and by Doha \cite[p.\
5452]{Doha/2003}, respectively, who had stated without detailed proof
that such a Laguerre series converges faster than algebraically if the
function under consideration is analytic at the origin. However, both
Gottlieb and Orszag \cite[p.\ 42]{Gottlieb/Orszag/1977} and Doha
\cite[p.\ 5452]{Doha/2003}, respectively, had failed to see that there is
a different scenario, which also produces convergent power series
expansions about $z=0$.

Let us assume that the coefficients $\lambda_{n}^{(\alpha)}$ decay
algebraically in magnitude and ultimately have strictly alternating
signs. Then, the inner $\mu$ series in (\ref{Rearr_f_Exp_GLag}) still
diverge at least for sufficiently large values of the outer index $\nu$.
However, this divergence does not imply that a convergent power series
about $z=0$ does not exist. The strictly alternating signs make it
possible to associate something finite to the divergent $\mu$ series by
employing suitable summation techniques for divergent series. A highly
condensed review of summation techniques with a special emphasis on
sequence transformations can be found in \cite[Appendices A and
B]{Weniger/2008}, and in \cite{Weniger/2010a} there is an admittedly
incomplete discussion of the usefulness of sequence transformations
beyond the summation of divergent series.

In \cite[Section 4]{Weniger/2008}, several known generating functions of
the generalized Laguerre polynomials were recovered by summing divergent
inner $\mu$ series representing power series coefficients. In these
comparatively simple cases, the divergent inner $\mu$ series could always
be expressed by generalized hypergeometric series with argument
$-1$. However, these generalized hypergeometric series only converge in
the interior of the unit circle. Thus, analytic continuation from the
interior of the circle of convergence of such a hypergeometric series to
its boundary was sufficient to accomplish a summation. In the case of
simpler hypergeometric series, this is not too difficult.

A skeptical reader might therefore conclude that a summation is only
feasible in the case of comparatively simple problems. This is not
true. In \cite[Section 6]{Weniger/2008} it was shown that it is also
possible to sum divergent inner $\mu$ series by purely numerical
techniques. Particularly good results were obtained by the nonlinear
$\mathcal{S}$ transformation which I had introduced in \cite[Eq.\
(8.4-4)]{Weniger/1989} (a highly condensed review of the historical
development was given in \cite[Section 2]{Weniger/2010b}). Some authors
call this $\mathcal{S}$ transformation the Weniger transformation (see
for example \cite{Borghi/2007,Borghi/2008c,Borghi/2009,Cvetic/Yu/2000,%
  Li/Zang/Tian/2009,Temme/2007} or \cite[Eq.\ (9.53) on p.\
287]{Gil/Segura/Temme/2007}). This terminology was also used in the
recently published NIST Handbook of Mathematical Functions \cite[Chapter
3.9(v) Levin's and Weniger's
Transformations]{Olver/Lozier/Boisvert/Clark/2010} (see also the
companion NIST Digital Library of Mathematical Functions under
{\verb+http://dlmf.nist.gov/3.9#v+}).

The example of divergent but summable inner $\mu$ series shows that
fairly sophisticated mathematical techniques may be needed to accomplish
the transformation of a Laguerre expansion to a power series.

It should be noted that the structure of the formulas, which effect the
transformation of a Laguerre series to power series, is of a more general
nature, and that very similar transformation formulas occur also in
completely different contexts. In \cite[Appendix B]{Weniger/2010b} it was
shown that formulas having analogous structures occur as long as the
transformation matrices are triangular and satisfy certain orthogonality
conditions. In \cite{Weniger/2010b}, transformations between factorial
series and inverse power series were considered, and the transformation
matrices involved Stirling numbers of the first and second kind,
respectively.

\typeout{==> Section: One-Center Expansions for Slater-type Functions}
\section{One-Center Expansions for Slater-type Functions}
\label{Sec:OneCenterExpansionSTF}

As already mentioned above, the one-center expansion for Slater-type
functions with in general nonintegral principal quantum numbers used by
Guseinov and Mamedov \cite[Eq.\ (4)]{Guseinov/Mamedov/2008a} was
originally derived by Guseinov \cite[Eq.\ (21)]{Guseinov/2002c} as the
one-center limit of a rearranged addition theorem for Slater-type
functions of the type of (\ref{Gus_OneRangeAddTheorSTF_STF_k_Rearr_2})
\cite[Eq.\ (15)]{Guseinov/2002c}. It is, however, both simpler and for
our purposes much more instructive to derive first one-center expansions
of Slater-type function with both integral and nonintegral principal
quantum numbers in terms of Guseinov's complete and orthonormal
Laguerre-type functions $\bigl\{ \prescript{}{k}{\Psi}_{n, \ell}^{m}
(\gamma, \bm{r}) \bigr\}_{n \ell m}$ defined by
(\ref{Def_Psi_Guseinov}). In the second step, it is investigated whether
and under which conditions these orthogonal expansions can be transformed
to expansions in terms of Slater-type functions $\bigl\{ \chi_{n,
  \ell}^{m} (\gamma, \bm{r}) \bigr\}_{n \ell m}$ with integral principal
quantum numbers $n$.

A convenient starting point for the construction of the one-center
expansion of a Slater-type function with an essentially arbitrary
principal quantum number in terms of Guseinov's functions is the
following Laguerre series \cite[Eq.\ (6.12)]{Weniger/2007b}:
\begin{align}
  \label{ExpoPow2GLag}
  z^{\rho} \, \mathrm{e}^{u z} & \; = \; (1-u)^{-\alpha-\rho-1} \,
  \frac{\Gamma (\alpha+\rho+1)}{\Gamma (\alpha+1)}
  \notag \\
  & \qquad \times \, \sum_{n=0}^{\infty} \, {}_2 F_1 \left(-n,
    \alpha+\rho+1; \alpha+1; \frac{1}{1-u} \right) \, L_{n}^{(\alpha)}
  (z) \, ,
  \notag \\
  & \qquad \qquad \rho \in \mathbb{R} \setminus \mathbb{N}_{0} \, ,
  \qquad \Re (\alpha+2\rho) > - 1 \, , \! \qquad u \in (-\infty, 1/2) \, .
\end{align}
The restriction $u \in (-\infty, 1/2)$ is necessary to guarantee the
existence of some integrals occurring in the derivation of this
expansion. If we assume $\rho = m$ with $m \in \mathbb{N}_{0}$, no
immediately obvious simplification occurs in (\ref{ExpoPow2GLag}). But
for $u=0$, we obtain a terminating Gaussian hypergeometric series ${}_{2}
F_{1}$ with unit argument that can be expressed in closed form with the
help of Gauss' summation theorem \cite[p.\
40]{Magnus/Oberhettinger/Soni/1966} and we obtain a much simpler
expansion \cite[Eq.\ (16) on p.\
214]{Erdelyi/Magnus/Oberhettinger/Tricomi/1953b}:
\begin{equation}
  \label{GenPow2GLag}
  z^{\rho} \; = \; \frac{\Gamma (\rho+\alpha+1)}{\Gamma (\alpha+1)} \,
  \sum_{n=0}^{\infty} \, \frac{(-\rho)_n}{(\alpha+1)_n} \,
  L_{n}^{(\alpha)} (z) \, ,
  \quad \rho \in \mathbb{R} \setminus \mathbb{N}_0 \, ,
  \quad \Re (\alpha+2\rho) > - 1 \, .
\end{equation}
If we now set $\rho=m$ with $m \in \mathbb{N}_0$, the infinite series
terminates because of the Pochhammer symbol $(-m)_n$ and we obtain
instead a finite sum:
\begin{equation}
  \label{IntPow_GlagPol}
  z^{m} \; = \;
  (\alpha+1)_m \, \sum_{n=0}^{m} \, \frac{(-m)_n}{(\alpha+1)_n}
  \, L_{n}^{(\alpha)} (z) \, ,
  \qquad m \in \mathbb{N}_0 \, , \quad \Re (\alpha)+2m > - 1 \, .
\end{equation}

The Laguerre expansions listed above can all be transformed to expansions
of Slater-type functions in terms of Guseinov's functions. The expansion
(\ref{ExpoPow2GLag}) yields the most general case, i.e., we obtain a
$k$-dependent one-center expansion of a Slater-type function $\chi_{N,
  L}^{M} (\beta, \bm{r})$ with a nonintegral principal quantum number $N
\in \mathbb{R} \setminus \mathbb{N}$ in terms of Guseinov's functions
with different scaling parameters $\beta \ne \gamma > 0$ \cite[Eq.\
(5.9)]{Weniger/2007c}:
\begin{align}
  \label{Expand_NISTF2Gusfun_DiffScaPar_1}
  \chi_{N, L}^{M} (\beta, \bm{r}) & \; = \; \frac
  {(2\gamma)^{L+(k+3)/2} \, \beta^{N-1}}{[\beta+\gamma]^{N+L+k+2}}
  \, \frac{\Gamma (N+L+k+2)}{(2L+k+2)!}
  \notag \\
  & \qquad \times \sum_{\nu=0}^{\infty} \, \left[
    \frac{(\nu+2L+k+2)!}{\nu!} \right]^{1/2} \,
  \prescript{}{k}{\Psi}_{\nu+L+1, L}^{M} (\gamma, \bm{r})
  \notag \\
  & \qquad \qquad \times {}_2 F_1 \left(-\nu, N+L+k+2; 2L+k+3;
    \frac{2\gamma}{\beta+\gamma} \right) \, .
\end{align}  
This expansion can also be derived by performing the one-center limit
$\bm{r}' = \bm{0}$ in the addition theorem
(\ref{Gus_OneRangeAddTheorSTF_k}).

If we assume in (\ref{Expand_NISTF2Gusfun_DiffScaPar_1}) that the
principal quantum number $N$ of the Slater-type function is a positive
integer satisfying $N \ge L+1$, no immediately obvious simplification
occurs. But in the case of equal scaling parameters $\beta = \gamma > 0$,
which corresponds to $u=0$ in (\ref{ExpoPow2GLag}), the expansion
(\ref{Expand_NISTF2Gusfun_DiffScaPar_1}) simplifies considerably,
yielding \cite[Eq.\ (5.7)]{Weniger/2007c}:
\begin{align}
  \label{NISTF2Gusfun_EqScaPar}
  \chi_{N, L}^{M} (\beta, \bm{r}) & \; = \; \frac
  {(2\gamma)^{-(k+3)/2}}{2^{N-1}} \, \Gamma (N+L+k+2)
  \notag \\
  & \qquad \times \, \sum_{\nu=0}^{\infty} \,
  \frac{(-N+L+1)_{\nu}}{\bigl[ (\nu+2L+k+2)! \, \nu! \bigr]^{1/2}}
  \, \prescript{}{k}{\Psi}_{\nu+L+1, L}^{M} (\beta, \bm{r}) \, ,
  \notag \\
  & \qquad \qquad N \in \mathbb{R} \setminus \mathbb{N} \, ,
    \qquad \beta > 0 \, , \qquad k = -1, 0, 1, 2, \dots \, .
\end{align}
If $N \in \mathbb{N}$ and $N \ge L+1$, the infinite series on the
right-hand side terminates because of the Pochhammer symbol
$(-N+L+1)_{\nu}$. The resulting finite sum can also be derived directly
via (\ref{IntPow_GlagPol}).

With the help of the sufficient conditions formulated in
\cite{Weniger/2008} and discussed in Section \ref{Sec:LagSer->PowSer}, we
can analyze whether the Laguerre-type functions
$\prescript{}{k}{\Psi}_{n, \ell}^{m} (\gamma, \bm{r})$ in the one-center
expansions given above can be replaced by Slater-type functions $\chi_{n,
  \ell}^{m} (\gamma, \bm{r})$ with integral principal quantum numbers.
But it is simpler and also more transparent to investigate instead
whether the equivalent Laguerre expansions (\ref{ExpoPow2GLag}),
(\ref{GenPow2GLag}), and (\ref{IntPow_GlagPol}) can be transformed to
power series.

The most simple situation occurs if we replace in the finite sum
(\ref{IntPow_GlagPol}) for the integral power $z^{m}$ with $m \in
\mathbb{N}_{0}$ the generalized Laguerre polynomials by powers via
(\ref{GLag_1F1}) and rearrange the order of the summations \cite[Eqs.\
(3.5) and (3.6)]{Weniger/2008}. Then we arrive at the trivial identity
$z^{m}=z^{m}$, which proves the correctness of (\ref{IntPow_GlagPol}) but
provides no new insight.

In the case of the Laguerre expansion (\ref{GenPow2GLag}) for a
nonintegral power $z^{\rho}$ with $\rho \in \mathbb{R} \setminus
\mathbb{N}_{0}$, we face a completely different situation. Firstly, we
have no \emph{a priori} reason to assume that (\ref{GenPow2GLag})
converges pointwise. Hilbert space theory only guarantees that this
expansion converges in the mean with respect to the norm of the weighted
Hilbert space $L^{2}_{\mathrm{e}^{-z} z^{\alpha}} \bigl([0, \infty)
\bigr)$, Moreover, the power function $z^{\rho}$ is not analytic at $z=0$
in the case of nonintegral $\rho \in \mathbb{R} \setminus
\mathbb{N}_{0}$. Thus, a mathematically meaningful power series about
$z=0$ cannot exist, and the transformation of the Laguerre expansion
(\ref{GenPow2GLag}) via (\ref{Rearr_f_Exp_GLag}) also must fail.

This is indeed the case.  We only need the asymptotic approximation
\cite[Eq.\ (6.1.47) on p.\ 257]{Abramowitz/Stegun/1972}
\begin{equation}
  \label{AsyGammaRatio}
  \Gamma(z+a)/\Gamma(z+b) \; = \; z^{a-b} \,
  \bigl[ 1 + \mathrm{O} (1/z) \bigr] \, , \qquad z \to \infty \, ,
\end{equation}
to obtain the following asymptotic estimate for the coefficients in
(\ref{GenPow2GLag}):
\begin{equation}
  \label{AsyCF_GenPow2GLag}
  \frac{\Gamma (\rho+\alpha+1)}{\Gamma (\alpha+1)} \, 
  \frac{(-\rho)_{n}}{(\alpha+1)_{n}} \; = \; 
  \frac{\Gamma (\rho+\alpha+1)}{\Gamma (-\rho)} \, n^{-\alpha-\rho-1} 
  \, \left[ 1 + \mathrm{O} \bigl( n^{-1} \bigr) \right] \, ,
  \quad n \to \infty \, .
\end{equation}
This asymptotic estimate shows that the coefficients in
(\ref{GenPow2GLag}) decay algebraically as $n \to \infty$. Moreover,
these coefficients ultimately have the same sign. Thus, on the basis of
the sufficient conditions formulated in \cite{Weniger/2008} and reviewed
in Section \ref{Sec:LagSer->PowSer} we can conclude that the inner $\mu$
series in (\ref{Rearr_f_Exp_GLag}) do not converge for larger values of
the outer index $\nu$.

We can also insert the explicit expression (\ref{GLag_1F1}) for the
generalized Laguerre polynomials into infinite series (\ref{GenPow2GLag})
and rearrange the order of summations, which yields after some algebra
\cite[Eq.\ (3.7)]{Weniger/2008}:
\begin{equation}
  \label{Chk_x^m_GlagPol_6}
  z^{\rho} \; = \; \frac{\Gamma (\rho+\alpha+1)}{\Gamma (\alpha+1)}
  \, \sum_{k=0}^{\infty} \, (-1)^k \, \frac{(-\rho)_k}{(\alpha+1)_k} \,
  \frac{z^k}{k!} \, {}_1 F_0 (k-\rho; 1) \, .
\end{equation}
Superficially, it looks as if we succeeded in constructing a power series
expansion for the nonintegral power $z^{\rho}$. However, the generalized
hypergeometric series ${}_1 F_0$ with unit argument is the limiting case
$z \to 1$ of the so-called binomial series \cite[p.\
38]{Magnus/Oberhettinger/Soni/1966}:
\begin{equation}
  \label{BinomSer}
  {}_1 F_0 (a; z) \; = \; 
  \sum_{m=0}^{\infty} \, \frac{(a)_{m}}{m!} \, z^{m} \; = \;
  \sum_{m=0}^{\infty} \, \binom{-a}{m} (-z)^m \; = \; (1-z)^{-a} \, ,
  \qquad \vert z \vert < 1 \, .
\end{equation}
If we set $a=k-\rho$ with $k \in \mathbb{N}_{0}$ and $\rho \in \mathbb{R}
\setminus \mathbb{N}_{0}$, we obtain for the ${}_1 F_0$ in
(\ref{Chk_x^m_GlagPol_6}):
\begin{equation}
  \label{Lim_Ser_Chk_x^m_GlagPol_6}
  {}_1 F_0 (k-\rho; 1) \; = \; \lim_{z \to 1} \, (1-z)^{\rho-k} \; = \;
  \begin{cases}
    \infty \, , \qquad \rho < 0 \, , \\
    0 \, , \qquad \; \, k < \rho \ge 0 \, , \\
    \infty \, , \qquad k > \rho \ge 0 \, .
  \end{cases}
\end{equation}
Thus, the power series (\ref{Chk_x^m_GlagPol_6}) for $z^{\rho}$ is purely
formal since it contains infinitely many series coefficients that are
infinite in magnitude.

The most general case is the Laguerre expansion (\ref{ExpoPow2GLag}) for
$z^{\rho} \mathrm{e}^{u z}$ with $\rho \in \mathbb{R} \setminus
\mathbb{N}_{0}$ and $u \in (-\infty, 1/2)$. It is immediately obvious
that $z^{\rho} \mathrm{e}^{u z}$ is analytic at $z=0$ if $\rho$ is
integral, i.e., if $\rho = m$ with $m \in \mathbb{N}_{0}$, and it is
nonanalytic if $\rho$ is nonintegral. This implies that the inner $\mu$
series in (\ref{Rearr_f_Exp_GLag}) diverge if $\rho$ is nonintegral, and
it converges if $\rho$ is a nonnegative integer. However, $\rho$ occurs
on the right-hand side of (\ref{ExpoPow2GLag}) apart from the prefactor
$(1-u)^{-\alpha-\rho-1} \Gamma (\alpha+\rho+1)/\Gamma (\alpha+1)$ only in
the terminating hypergeometric series ${}_2 F_1 (-n, \alpha+\rho+1;
\alpha+1; 1/(1-u))$. Since the prefactor does not affect the convergence
and existence of the resulting expansions, we have to analyze the
asymptotics of the terminating hypergeometric series as $n \to \infty$.

Thus, with the Laguerre series coefficients
\begin{align}
  \label{LagCF_rho}
  \lambda_{n}^{(\alpha)} & \; = \; (1-u)^{-\alpha-\rho-1} \, 
  \frac{\Gamma (\alpha+\rho+1)}{\Gamma (\alpha+1)} 
  \notag \\ 
  & \qquad \times \, 
  {}_2 F_1 \left(-n, \alpha+\rho+1; \alpha+1; \frac{1}{1-u} \right) \, ,
  \qquad \rho \in \mathbb{R} \setminus \mathbb{N}_{0} \, ,  
\end{align}
the inner $\mu$ series in (\ref{Rearr_f_Exp_GLag}) diverge for
sufficiently large values of the outer index $\nu$, and with the
coefficients
\begin{align}
  \label{LagCF_m}
  \lambda_{n}^{(\alpha)} & \; = \; (1-u)^{-\alpha-m-1} \, (\alpha+1)_{m}
  \notag \\ 
  & \qquad \times \, 
   {}_2 F_1 \left(-n, \alpha+m+1; \alpha+1; \frac{1}{1-u} \right) \, ,
  \qquad m \in \mathbb{N}_{0}  
\end{align}
the inner $\mu$ series in (\ref{Rearr_f_Exp_GLag}) converge. This is
certainly a surprising observation which indicates that the difference
between the almost identical terminating hypergeometric series ${}_2 F_1$
in (\ref{LagCF_rho}) and (\ref{LagCF_m}), respectively, is greater than
it appears at first sight.

Nevertheless, this puzzle can be resolved by analyzing the large $n$
asymptotics of the terminating hypergeometric series in (\ref{LagCF_rho})
and (\ref{LagCF_m}), respectively. The study of large parameters of a
Gaussian hypergeometric series is an old problem of special function
theory with an extensive literature (see for example
\cite{OldeDaalhuis/2003a,OldeDaalhuis/2003b,OldeDaalhuis/2010,Temme/2003}
and references therein). 

Large parameter estimates of that kind turned out be useful in the
context of multicenter integrals. In
\cite[Appendix]{Grotendorst/Weniger/Steinborn/1986}, we derived and used
large parameter estimates for some special Gaussian hypergeometric series
to analyze the rate of convergence of certain series expansions for
multicenter integrals.

If we apply the linear transformation \cite[p.\
47]{Magnus/Oberhettinger/Soni/1966}
\begin{equation}
  \label{LTr_1}
  {}_2 F_1 (a, b; c; z) \; = \; (1-z)^{-a} \, 
  {}_2 F_1 \bigl( a, c-b; c; z/(z-1) \bigr)  
\end{equation}
to the hypergeometric series in (\ref{LagCF_m}), we obtain:
\begin{align}
  \label{LT_2F1_A}
  & {}_2 F_1 \left(-n, \alpha+m+1; \alpha+1; \frac{1}{1-u} \right)
  \notag \\
  & \qquad 
  \; = \; \left( \frac{u}{u-1} \right)^{n} \, 
  {}_2 F_1 \left(-n, -m; \alpha+1; \frac{1}{u} \right) \, ,
  \qquad m, n \in \mathbb{N}_{0} \, .
\end{align}
If $n$ becomes large and $m$ is fixed, the terminating hypergeometric
series on the right-hand side can be expressed as follows:
\begin{equation}
  {}_2 F_1 \left(-n, -m, \alpha+1; \frac{1}{u} \right) \; = \;
  \sum_{\mu=0}^{m} \, \frac{(-n)_{\mu} (-m)_{\mu}}{(\alpha+1)_{\mu}} \,
  \frac{u^{-\mu}}{\mu!} \, .
\end{equation}
The asymptotically dominant contribution on the right-hand side is the
last term with $\mu=m$. With the help of (\ref{AsyGammaRatio}) we find
$(-n)_{m} = \mathrm{O} (n^{m})$ as $n \to \infty$. Since $u \in (-\infty,
1/2)$ implies $u/(u-1) \in (-1, 1)$, we can conclude that the right-hand
side of (\ref{LT_2F1_A}) decays exponentially as $n \to \infty$ because
of the prefactor $[u/(u-1)]^{n}$. Thus, the inner $\mu$ series in
(\ref{Rearr_f_Exp_GLag}) converge if the coefficients
$\lambda_{n}^{(\alpha)}$ are chosen according to (\ref{LagCF_m}).

If we apply the linear transformation (\ref{LTr_1}) to the hypergeometric
series in (\ref{LagCF_rho}), we do not obtain an expression that would be
useful for our purposes. Therefore, we apply instead the analytic
continuation formula \cite[pp.\ 47 - 48]{Magnus/Oberhettinger/Soni/1966}
\begin{align}
  \label{AC_1}
  & {}_2 F_1 (a, b; c; z) \; = \; \frac {\Gamma (c) \Gamma (c-a-b)}
  {\Gamma (c-a) \Gamma (c-b)} \,
  {}_2 F_1 (a, b; a+b-c+1; 1-z) \notag \\
  & \qquad + \, \frac {\Gamma (c) \Gamma (a+b-c)} {\Gamma (a) \Gamma
    (b)} \,
  (1-z)^{c-a-b} \notag \\
  & \qquad \qquad \times \,
  {}_2 F_1 (c-a, c-b; c-a-b+1; 1-z) \, , \notag \\
  & \qquad \qquad \qquad
  \vert \arg (1-z) \vert < \pi \, , \qquad c - a - b \ne
  \pm m \, , \quad m \in \mathbb{N}_0 \, ,
\end{align}
to the hypergeometric series in (\ref{LagCF_rho}) and obtain:
\begin{align}
  \label{AC_2F1_1}
  & {}_2 F_1 \left(-n, \alpha+\rho+1; \alpha+1; \frac{1}{1-u} \right)  
  \notag \\
  & \qquad \; = \; \frac{(-\rho)_{n}}{(\alpha+1)_{n}} \,
  {}_2 F_1 \left( \alpha+\rho+1, -n; \rho-n+1; \frac{u}{u-1} \right) \, .
\end{align}
Application of the linear transformation (\ref{LTr_1}) then yields:
\begin{align}
  \label{LT_2F1_B}
  & {}_2 F_1 \left(-n, \alpha+\rho+1; \alpha+1; \frac{1}{1-u} \right)  
  \notag \\
  & \qquad \; = \; (1-u)^{\alpha+\rho+1} \, 
  \frac{(-\rho)_{n}}{(\alpha+1)_{n}} \,
  {}_2 F_1 \left(\rho+1, \alpha+\rho+1; \rho-n+1; u \right) \, .
\end{align}
If we set $u=0$ in (\ref{LT_2F1_B}), the hypergeometric series on the
right-hand side terminates after the first term. Thus, we see once more
that (\ref{ExpoPow2GLag}) simplifies for $u=0$ to yield
(\ref{GenPow2GLag}).

With the help of (\ref{AsyGammaRatio}) we find $(\rho-n+1)_{m} =
\mathrm{O} (n^{m})$ as $n \to \infty$. Thus, the hypergeometric series on
the right-hand side of (\ref{LT_2F1_B}) can for arbitrary $k \in
\mathbb{N}_{0}$ be expressed as follows:
\begin{align}
  \label{Asy_2F1_u}
  & {}_2 F_1 \left(\rho+1, \alpha+\rho+1; \rho-n+1; u \right)
  \notag \\
  & \qquad \; = \; 
  \sum_{\kappa=0}^{k} \, \frac{(\rho+1)_{\kappa} (\alpha+\rho+1)_{\kappa}}
   {(\rho-n+1)_{\kappa} \kappa!} \, u^{\kappa} \, + \, 
  \mathrm{O} \bigl( n^{-k-1} \bigr) \, , 
  \qquad n \to \infty \, .  
\end{align}
For $u \in (-1, 1/2)$, this hypergeometric series converges and we have a
convergent asymptotic expansion as $n \to \infty$. For $u \in (-\infty,
1]$, the hypergeometric series diverges, but (\ref{Asy_2F1_u})
nevertheless holds in the sense of an asymptotic expansion as $n \to
\infty$. If we set $k=0$ in (\ref{Asy_2F1_u}) and use
(\ref{AsyCF_GenPow2GLag}), we find that the Laguerre series coefficients
in (\ref{LagCF_rho}) satisfy the leading order asymptotics
\begin{equation}
  \label{Asy_lanmda_GenCase}
  \lambda_{n}^{(\alpha)} \; \sim \; 
  \frac{\Gamma (\rho+\alpha+1)}{\Gamma (\alpha+1)} \, 
  \frac{(-\rho)_{n}}{(\alpha+1)_{n}} \; \sim \;
  \frac{\Gamma (\rho+\alpha+1)}{\Gamma (-\rho)} \, n^{-\alpha-\rho-1} 
  \, , \qquad n \to \infty \, .
\end{equation}
This asymptotic estimate, which does not depend on $u$, shows once more
that $z^{\rho} \mathrm{e}^{u z}$ is not analytic at $z=0$ if $\rho$ is
nonintegral. The inner $\mu$ series in (\ref{Rearr_f_Exp_GLag}) diverge
for sufficiently large values of the outer index $\nu$.

Comparison of (\ref{Asy_lanmda_GenCase}) with (\ref{AsyCF_GenPow2GLag})
shows that the Laguerre series coefficients $\lambda_{n}^{(\alpha)}$ in
(\ref{GenPow2GLag}), which correspond to $z^{\rho} \mathrm{e}^{u z}$ with
$u \in (-\infty, 1/2)$, and the coefficients in (\ref{GenPow2GLag}),
which correspond to $z^{\rho}$ or to $u=0$ in $z^{\rho} \mathrm{e}^{u
  z}$, possess the same leading order asymptotics as $n \to \infty$ that
does not depend of $u$.

\typeout{==> Section: The Transformation of One-Range to Two-Range
  Addition Theorems}
\section{The Transformation of One-Range to Two-Range Addition Theorems}
\label{Sec:OneRangeAdd->TwoRangeAdd}

As discussed in the previous Sections, the legitimacy of Guseinov's
rearrangement of a $k$-dependent one-range addition theorem
(\ref{Gus_OneRangeAddTheorSTF_k}), whose series expansions
(\ref{SFT_Mom_GLagPol}) for the angular projections (\ref{SFT_Mom}) are
essentially Laguerre series of the type of (\ref{f_Exp_GLag}), can be
checked by analyzing the convergence of the transformation formula
(\ref{Rearr_f_Exp_GLag}). One only has to determine the asymptotic sign
pattern and the asymptotic decay rate of the expansion coefficients
$\lambda_{n}^{(\alpha)}$ of the corresponding Laguerre series and employ
the sufficient convergence criteria formulated in \cite{Weniger/2008}.

As discussed in Section \ref{Sec:OneCenterExpansionSTF}, this approach
works in a very satisfactory way in the case of the Laguerre series
(\ref{ExpoPow2GLag}) for $z^{\rho} \mathrm{e}^{u z}$ or the equivalent
one-center expansion (\ref{Expand_NISTF2Gusfun_DiffScaPar_1}) for the
Slater-type function $\chi_{N, L}^{M} (\beta, \bm{r})$ with an in general
nonintegral principal quantum number $N \in \mathbb{R} \setminus
\mathbb{N}$.

The one-center expansions considered in Section
\ref{Sec:OneCenterExpansionSTF} have the highly advantageous feature that
they are comparatively simple. Therefore, we can understand the large
index asymptotics of the expansion coefficients via the large index
asymptotics of some Gaussian hypergeometric series, whose derivation is
not straightforward but nevertheless not too difficult. In this way, we
can explain the rate of convergence or divergence of the rearrangements
of the Laguerre series (\ref{ExpoPow2GLag}).

Guseinov's $k$-dependent one-range addition theorems are genuine
two-center problems. Therefore, the situation is much more difficult and
we are confronted with nontrivial technical problems. The expansion
coefficients of such an addition theorem are according to
(\ref{OneRangeAddTheor}), (\ref{OneRangeAddTheor_w}), and
(\ref{Gus_OneRangeAddTheorSTF_k}) overlap integrals, which are fairly
complicated functions $\mathbb{R}^{3} \to \mathbb{C}$ whose asymptotic
sign patterns and asymptotic decay rates cannot be determined easily. But
even in this troublesome two-center case, we can arrive at some definite
conclusions by pursuing an indirect approach based on the analysis of the
singularities of the function which is to be expanded.

Our starting point is a $k$-dependent one-range addition theorem
(\ref{Gus_OneRangeAddTheorSTF_k}) for a Slater-type function $\chi_{N,
  L}^{M} (\beta, \bm{r} \pm \bm{r}')$ with an in general nonintegral
principal quantum number $N \in \mathbb{R} \setminus \mathbb{N}$. We
assume that we succeeded in constructing the series expansions
(\ref{SFT_Mom_GLagPol}) for the angular projections (\ref{SFT_Mom}) of
the one-range addition theorem (\ref{Gus_OneRangeAddTheorSTF_k}).

For the moment, let us also assume that the expansion coefficients of the
series expansions (\ref{SFT_Mom_GLagPol}), which are Laguerre series of
the type of (\ref{f_Exp_GLag}), have asymptotic sign patterns and
asymptotic decay rates which according to the criteria formulated in
\cite{Weniger/2008} guarantee that the transformation formula
(\ref{Rearr_f_Exp_GLag}) produces functions $\mathbb{C} \to \mathbb{C}$
that are analytic in a neighborhood the origin $z=0$.

Superficially, it appears that under these circumstances the legitimacy
of Guseinov's rearrangement is guaranteed: A $k$-dependent one-range
addition theorem (\ref{Gus_OneRangeAddTheorSTF_k}), which is an expansion
in terms of Guseinov's complete and orthonormal functions $\bigl\{
\prescript{}{k}{\Psi}_{n, \ell}^{m} (\gamma, \bm{r}) \bigr\}_{n \ell m}$,
is transformed to a one-range addition theorem of the type of
(\ref{NISTO->ISTO_AddThm}), which is an expansions in terms of
Slater-type functions $\bigl\{ \chi_{n,L}^{M} (\gamma, \bm{r}) \bigr\}_{n
  \ell m}$ with integral principal quantum numbers $n$ and a common
scaling parameter $\gamma > 0$.

But such a conclusion is premature. The transformed expansion
(\ref{NISTO->ISTO_AddThm}) in terms of Slater-type functions can only be
one-range addition theorem, i.e., a map $\mathbb{R}^{3} \times
\mathbb{R}^{3} \to \mathbb{C}$, if it converges for \emph{all} $\bm{r},
\bm{r}' \in \mathbb{R}^{3}$. This requires that the power series
(\ref{SFT_Mom_PowSer}) for the angular projections of the expansion
(\ref{NISTO->ISTO_PowSer}) for $\exp (\gamma r) \chi_{N, L}^{M} (\beta,
\bm{r} \pm \bm{r}')$ converge for \emph{all} $r, r' \in [0, \infty)$.
This is a very demanding requirement, which cannot be satisfied in the
case of exponentially decaying functions such as Slater-type
functions. In the two-center case, it is irrelevant whether the principal
quantum number of the Slater-type function is integral or nonintegral.

A function $f \colon \mathbb{C} \to \mathbb{C}$ is analytic at the origin
$z=0$ in the sense of complex analysis if it has a power series in $z$
that converges in some neighborhood of $z=0$. Or to put it differently:
Such an $f$ is analytic at $z=0$ if its power series in $z$ has a circle
of convergence with a \emph{nonzero} radius. As is well known, we cannot
tacitly assume that the radius of convergence of a power series is
necessarily infinite, or equivalently, we cannot assume that $f$ is
necessarily an entire function that is analytic for all $z \in
\mathbb{C}$.

Therefore, we should look for features of functions analytic at $z=0$,
which rule out an infinite radius of convergence but which do not
interfere with the existence of a Laguerre series of the type of
(\ref{f_Exp_GLag}). My subsequent arguments are based on the simple, but
nevertheless very consequential fact that power series and Laguerre
series differ substantially in the way how they are affected by
singularities of the function which is to be expanded.

Power series converge pointwise in their circles of convergence, and in
the interior of these circles they not only converge uniformly but they
can also be used for the computation of higher derivatives. Since,
however, the higher derivatives of a function ultimately become infinite
in magnitude at a singularity, the radius of the circle of convergence is
determined by the location of that singularity which is closest to the
expansion point.

If the function, which is to be expanded, possesses a singularity
somewhere in the complex plane $\mathbb{C}$, the radius of convergence of
its power series cannot be infinite. This has immediate and undesirable
consequence for integrals over the positive real semi-axis $[0, \infty)$
as they typically occur in the theory of Laguerre polynomials or in the
radial parts of the three- and six-dimensional integrals in electronic
structure theory. If the semi-infinite integration interval is not
contained completely in the circle of convergence, the term-wise
integration of such a power series either leads to convergence to a wrong
result or to divergence.

In that respect, Laguerre series have much more convenient properties
precisely because they in general do not converge pointwise. As discussed
in more detail in Appendix \ref{App:GeneralizedLaguerrePolynomials}, the
existence of a Laguerre series of the type of (\ref{f_Exp_GLag}) for some
function $f \colon \mathbb{C} \to \mathbb{C}$ is guaranteed as long as
$f$ belongs to the weighted Hilbert space $L^{2}_{\mathrm{e}^{-z}
  z^{\alpha}} \bigl([0, \infty) \bigr)$ defined by
(\ref{HilbertL^2_Lag}). Of course, $f$ must not have a non-integrable
singularity on the integration interval $[0, \infty)$, but singularities
away from the positive real semi-axis cause no problems. Loosely
speaking, we can say that a Laguerre series simply ignores all
singularities which are not located on the integration contour. Power
series cannot do that. For them, \emph{all} singularities matter even if
they are far away from the integration contour.

These differences between power series and Laguerre series, respectively,
can be illustrated by considering the following class of functions:
\begin{equation}
  \label{Def:F_eta}
  F_{\eta} (z; u, \theta) \; = \; 
 \bigl[ z^{2} + u^{2} - 2 z u \cos \theta \bigr]^{\eta} \, ,
 \qquad z, u \in \mathbb{C} \, , \quad \eta, \theta \in \mathbb{R} \, .
\end{equation}
For $\eta = 1/2$, the function $F_{\eta} (z; u, \theta)$ is nothing but
the explicit expression for the difference $\vert \bm{x} - \bm{y} \vert =
[x^{2} + y^{2} - 2xy \cos \theta]^{1/2}$ of two vectors $\bm{x}, \bm{y}
\in \mathbb{R}^{3}$ in spherical polar coordinates in disguise, and for
$\eta = -1/2$, it corresponds to the Coulomb or Newton potential $1/\vert
\bm{x} - \bm{y} \vert$.

As long as $F_{\eta} (z; u, \theta)$ does not have a non-integrable
singularity on the positive real semi-axis, it belongs to the weighted
Hilbert space $L^{2}_{\mathrm{e}^{-z} z^{\alpha}} \bigl([0, \infty)
\bigr)$ defined in (\ref{HilbertL^2_Lag}). Therefore, $F_{\eta} (z; u,
\theta)$ possesses a Laguerre series of the type of (\ref{f_Exp_GLag}),
although I was not able to find a closed form expression for the
coefficients of this expansion.

If $\eta$ is a nonnegative integer, $\eta = n$ with $n \in
\mathbb{N}_{0}$, $F_{\eta} (z; u, \theta)$ is a polynomial in $z$ and
therefore an analytic function for all $z \in \mathbb{C}$. But if $\eta
\in \mathbb{R} \setminus \mathbb{N}_{0}$, $F_{\eta} (z; u, \theta)$ has
singularities at
\begin{equation}
  \label{z_1,2}
  z_{1,2} \; = \; 
  \Bigl\{ \cos \theta \pm \sqrt{[\cos \theta]^{2}-1} \Bigr\} 
  \, u \, .
\end{equation}
Accordingly, the radius of convergence of its power series about $z=0$ is
equal to $\vert u \vert$.

The function $F_{\eta} (z; u, \theta)$ essentially corresponds to the
well known generating function
\begin{equation}
  \label{GenFun_GegPol}
  \bigl[ 1 - 2 x t + t^2 \bigr]^{-\lambda} \; = \;
  \sum_{n=0}^{\infty} \, C_{n}^{\lambda} (x) \, t^{n} \, ,
  \qquad \vert t \vert < 1 \, , \quad \lambda \ne 0 \, ,
\end{equation}
of the Gegenbauer polynomials \cite[p.\
222]{Magnus/Oberhettinger/Soni/1966}. Thus, $F_{\eta} (z; u, \theta)$
possesses -- depending on the relative magnitudes of $\vert z \vert$ and
$\vert u \vert$ -- two complementary power series expansions. For $\vert
z/u \vert < 1$, it possesses a convergent power series in $z/u$,
\begin{equation}
  \label{F_small_z}
  F_{\eta} (z; u, \theta) \; = \; u^{2 \eta} \, \sum_{n=0}^{\infty} \, 
  C_{n}^{-\eta} \bigl( \cos \theta \bigr) \, (z/u)^{n} \, ,
\end{equation}
and for $\vert z/u \vert > 1$, it possesses a convergent \emph{inverse}
power series in $z/u$,
\begin{equation}
  \label{F_large_z}
  F_{\eta} (z; u, \theta) \; = \; z^{2 \eta} \, \sum_{n=0}^{\infty} \, 
  C_{n}^{-\eta} \bigl( \cos \theta \bigr) \, (u/z)^{n} \, ,
\end{equation}
which can also be interpreted as a convergent power series in $u/z$.

Let us now assume that the singularities (\ref{z_1,2}) of $F_{\eta} (z;
u, \theta)$ with $\eta \in \mathbb{R} \setminus \mathbb{N}$ do not lie on
the integration interval $[0, \infty)$ and that we succeeded in
constructing its Laguerre series of the type of (\ref{f_Exp_GLag}),
possibly by purely numerical means. Then, the application of the
transformation formula (\ref{Rearr_f_Exp_GLag}) to this Laguerre series
produces -- if necessary with the help of nonlinear sequence
transformations as described in \cite[Section 6]{Weniger/2008} -- the
power series (\ref{F_small_z}), which converges for $\vert z/u \vert < 1$
and which diverges for $\vert u/z \vert < 1$.

The small $z$ series (\ref{F_small_z}) accomplishes at least for $\vert
z/u \vert < 1$ a separation of the variables $z$ and $u$ and therefore
resembles a two-range addition theorem. For $\vert z/u \vert > 1$, the
large $z$ series (\ref{F_large_z}) also accomplishes this separation, but
I see no obvious way of computing the large $z$ series (\ref{F_large_z})
from the Laguerre series for $F_{\eta} (z; u, \theta)$. Instead, we would
have to construct a Laguerre series in $u$ -- this is possible since $z$
and $u$ play a symmetrical role in $F_{\eta} (z; u, \theta)$ -- from
which we can compute the large $z$ series (\ref{F_large_z}) which is also
a small $u$ series.

These considerations show that if we want to represent $F_{\eta} (z; u,
\theta)$ by power series in $z$, a two-range scenario cannot be
avoided. It does not matter if we start from a Laguerre series which
provides a unique representation of $F_{\eta} (z; u, \theta)$ that is
computationally useful in integrals over the whole real semi-axis. The
singularities (\ref{z_1,2}) of $F_{\eta} (z; u, \theta)$ with $\eta \in
\mathbb{R} \setminus \mathbb{N}$ imply that there can be no power series
in $z$ which converges for all $z \in \mathbb{C}$.

In the context of addition theorems, it may be of interest that the
generating function (\ref{GenFun_GegPol}) of the Gegenbauer polynomials
can be used for the construction of two-range addition theorems in a
relatively straightforward way. If we set $\lambda = -\nu/2$, $t =
r_{<}/r_{>}$, and $x = \cos \theta$ in the generating function
(\ref{GenFun_GegPol}), we obtain the following Gegenbauer expansion for
the general power function
\begin{equation}
  \label{GenPowFun_GegExp}
  \vert \bm{r}_{<} \pm \bm{r}_{>} \vert^{\nu} \; = \; r_{>}^{\nu} \,
  \sum_{n=0}^{\infty} \, (\mp 1)^{n} \, C_{n}^{-\nu/2} (\cos \theta) \,
  ( r_{<}/r_{>} )^{n} \, , \qquad \nu \in \mathbb{R} \, .
\end{equation}
The two-range form of this Gegenbauer expansion is a direct consequence
of the convergence condition $\vert t \vert < 1$ in (\ref{GenFun_GegPol})
which translates to the convergence condition $r_{<}/r_{>} < 1$.

We can easily construct an addition theorem from the Gegenbauer expansion
(\ref{GenPowFun_GegExp}): We only have to replace the Gegenbauer
polynomials by Legendre polynomials. However, the practical realization
of this obvious idea had apparently not been so easy. As discussed by
Steinborn and Filter \cite[pp.\ 269 - 270]{Steinborn/Filter/1975c}, many
authors had quite a few problems with the determination of explicit
expressions for the coefficients of the expansion of Gegenbauer
polynomials in terms of Legendre polynomials. Also Steinborn and Filter
constructed very messy expressions for these coefficients which are
restricted to certain superscripts of the Gegenbauer polynomial
\cite[Section 3]{Steinborn/Filter/1975b}.

This is somewhat strange because already at that time a much more
convenient expression for these expansion coefficients had been available
in the mathematical literature. In Exercise 4 on p.\ 284 of Rainville's
book \cite{Rainville/1971}, one finds the following relationship, where
$\Ent{m/2}$ denotes the integral part of $m/2$ (compare \cite[Eq.\
(5.2)]{Weniger/Steinborn/1989b}):
\begin{equation}
  \label{Gegenbauer_2_Legendre}
  C_{m}^{\mu} (x) \; = \;
  \sum_{s=0}^{\Ent{m/2}} \, \frac{(\mu)_{m-s} \,
  (\mu-1/2)_{s}}{(3/2)_{m-s} \, s!} \, (2m - 4 s + 1) \, P_{m-2s} (x) \, .
\end{equation}
This result can be proved via the explicit expression \cite[Eq.\ 7.313.7
on p.\ 836]{Gradshteyn/Rhyzhik/1994} for the integral $\int_{-1}^{1}
(1-x)^{\alpha} (1-x)^{\nu-1/2} C_{m}^{\mu} (x) C_{n}^{\nu} (x) \mathrm{d}
x$. One only has to set $\nu=1/2$ and perform the limit $\alpha \to 0$,
which requires, however, some algebraic trickery.

If we now insert (\ref{Gegenbauer_2_Legendre}) into
(\ref{GenPowFun_GegExp}) and rearrange the order of summations, we obtain
after some algebra the following expansion in terms of Legendre
polynomials:
\begin{align}
  \label{GenPowFun_LegExp}
  & \vert  \bm{r}_{<} \pm \bm{r}_{>} \vert^{\nu} \; = \; r_{>}^{\nu} \,
  \sum_{\ell=0}^{\infty} \, (\mp 1)^{\ell} \, P_{\ell} (\cos \theta) \,
  ( r_{<}/r_{>} )^{\ell}
  \notag
  \\
  & \qquad \times \, \frac{(-\nu/2)_{\ell}}{(3/2)_{\ell}} \,
  {}_{2} F_{1} \bigl( \ell-\nu/2, - [\nu+1]/2; \ell+3/2;
  [r_{<}/r_{>}]^{2} \bigr) \, .
\end{align}
If $\nu$ is an even integer, $\nu = 2n$ with $n \in \mathbb{N}_{0}$,
$\vert \bm{r}_{<} \pm \bm{r}_{>} \vert^{\nu}$ is a polynomial in both
$r_{<}$ and $r_{<}$ and therefore analytic. The infinite $\ell$ series in
(\ref{GenPowFun_LegExp}) terminates because of the Pochhammer symbol
$(-\nu/2)_{\ell} = (-n)_{\ell}$, implying $\ell \le n$. Similarly, the
Gaussian hypergeometric series ${}_{2} F_{1}$ in (\ref{GenPowFun_LegExp})
terminates since $\ell-\nu/2 = \ell-n$ is either a negative integer or
zero.

In (\ref{GenPowFun_LegExp}), we only have to replace the Legendre
polynomials by spherical harmonics via the so-called spherical harmonic
addition theorem (\ref{Ylm_AddThm}) to obtain an expansion in terms of
spherical harmonics, which had originally been derived by Sack \cite[Eq.\
(19)]{Sack/1964a} by solving a partial differential equation and which
converges as long as $r_{<}/r_{>} < 1$ holds:
\begin{align}
  \label{GenPowFun_AddThm}
  \vert \bm{r}_{<} \pm \bm{r}_{>} \vert^{\nu} & \; = \;
  4\pi r_{>}^{\nu+1} \, \sum_{\ell=0}^{\infty} (\mp 1)^{\ell} \,
  \sum_{m=-\ell}^{\ell} \,
  \bigl[ \mathcal{Y}_{\ell}^{m} (\bm{r}_{<}) \bigr]^{*} \,
  \mathcal{Z}_{\ell}^{m} (\bm{r}_{>})
  \notag \\
  & \qquad \times \, \frac{(-\nu/2)_{\ell}}{(3/2)_{\ell}} \,
  {}_{2} F_{1} \bigl( \ell-\nu/2, - [\nu+1]/2; \ell+3/2;
  [r_{<}/r_{>}]^{2} \bigr) \, .
\end{align}
This two-range addition theorem simplifies considerably and also assumes
a one-range form if $\nu$ is a positive even integer, $\nu = 2n$ with $n
\in \mathbb{N}$ (see above or also \cite[pp.\ 1258 -
1259]{Weniger/2005}). But for arbitrary $\nu \in \mathbb{R}$,
(\ref{GenPowFun_AddThm}) is a two-range addition theorem. This is a
direct consequence of its derivation via the generating function
(\ref{GenFun_GegPol}) of the Gegenbauer polynomials. which is a power
series in $t$ with a nonzero, but finite radius of convergence.

This observation suggests the following interpretation: A two-range
addition theorem for a function $f (\bm{r}_{<} \pm \bm{r}_{>})$ having
its only singularity at $\bm{r}_{<} \pm \bm{r}_{>} = \bm{0}$ corresponds
to a possibly rearranged power series in $r_{<}$ with a finite radius of
convergence, and this radius is determined by the condition $r_{<} <
r_{>}$. This interpretation is also confirmed by the differential
operator (\ref{ST_TransOp}) which had been the central tool for the
derivation of two-range addition theorems in
\cite{Weniger/2000a,Weniger/2002} or in \cite[Section 7]{Weniger/2005}.

This interpretation also applies to the addition theorem
(\ref{GenPowFun_AddThm}). For arbitrary $\nu \ne 0, 2, 4, \dots$, the
general power function $\vert \bm{r}_{<} \pm \bm{r}_{>} \vert^{\nu}$ has
a singularity for $\bm{r}_{<} \pm \bm{r}_{>} = \bm{0}$, and this
singularity determines the radius of convergence of its power series in
$r_{<}$ and enforces a two-range form
\cite{Weniger/2000a,Weniger/2002,Weniger/2005}.

These conclusions about the nature of the addition theorem for the
general power function $\vert \bm{r}_{<} \pm \bm{r}_{>} \vert^{\nu}$ help
us to understand some essential features of addition theorems for
Slater-type functions that converge pointwise. For the sake of
simplicity, let us first consider the so-called $1 s$ function
\begin{equation}
  \label{Def_STF_1s}
  \exp \bigl(-\beta \vert \bm{r} - \bm{r}' \vert\bigr) \; = \; 
  \exp \bigl( -\beta [r^{2} + r^{\prime 2} - 2 r r' \cos \theta]^{1/2} 
  \bigr) 
\end{equation}
and let us also assume $r' > 0$. If Guseinov's rearrangements are
legitimate, then $\exp \bigl(-\beta \vert \bm{r} - \bm{r}' \vert\bigr)$
must possess an expansion of the type of (\ref{NISTO->ISTO_AddThm}) in
terms of Slater-type functions $\bigl\{ \chi_{n, \ell}^{m} (\gamma,
\bm{r}) \bigr\}_{n \ell m}$, which converges for the whole argument set
$\mathbb{R}^{3} \times \mathbb{R}^{3}$.  Obviously, this is equivalent to
requiring that $\exp\bigl(\gamma r \bigr) \exp \bigl( -\beta \vert \bm{r}
- \bm{r}' \vert\bigr)$ possesses a power series about $r = 0$, which
converges for all $r \in [0, \infty)$. Since, however, $\exp\bigl(\gamma
r \bigr)$ is an entire function, whose power series in $r$ converges in
the whole complex plane $\mathbb{C}$, we can ignore it for the moment. It
is sufficient to analyze whether the $1 s$ function (\ref{Def_STF_1s})
possesses a convergent power series in $r$ and for which values of $r$
and $r'$ this series converges.

If we expand the exponential on the right-hand side of
(\ref{Def_STF_1s}), we obtain:
\begin{equation}
  \label{STF_1s_expand}
  \exp \bigl(-\beta \vert \bm{r} - \bm{r}' \vert\bigr) \; = \;
  \sum_{\kappa=0}^{\infty} \, \frac{(-\beta)^{\kappa}}{\kappa!} \,
  [r^{2} + r^{\prime 2} - 2 r r' \cos \theta]^{\kappa/2} \, .
\end{equation}
If the index $\kappa$ is even, $\kappa = 2k$ with $k \in \mathbb{N}_{0}$,
then $[r^{2} + r^{\prime 2} - 2 r r' \cos \theta]^{k}$ is a polynomial in
$r$ which is obviously analytic for all $r \in [0, \infty)$. But if
$\kappa$ is odd, $\kappa = 2k+1$ with $k \in \mathbb{N}_{0}$, we are in
trouble. We can use the generating function (\ref{GenFun_GegPol}) of the
Gegenbauer polynomials to obtain a power series expansion in $r/r'$
\begin{align}
  \label{GegenbauerExpandTerms}
  [r^{2} + {r'}^{2} - 2 r r' \cos \theta ]^{k+1/2} & \; = \; 
  {r'}^{2k+1} \, 
  \bigl[1 + (r/r')^{2} - 2(r/r') \cos \theta \bigr]^{k+1/2}
  \notag \\
  & \; = \; {r'}^{2k+1} \, \sum_{j=0}^{\infty} \, 
   C_{j}^{-k-1/2} \bigl( \cos \theta \bigr) \, (r/r')^{j} \, . 
\end{align}
Unfortunately, the Gegenbauer expansion (\ref{GegenbauerExpandTerms})
converges only if $r/r' < 1$ holds. Therefore, the expansion obtained in
this way corresponds to the case $\vert \bm{r} \vert < \vert \bm{r}'
\vert$ of the two-range addition theorem for $\vert \bm{r} \pm \bm{r}'
\vert^{k+1/2}$ which is a special case of (\ref{GenPowFun_AddThm}).

This has some far-reaching consequences for the analyticity of the $1 s$
function $\exp \bigl(-\beta \vert \bm{r} - \bm{r}' \vert\bigr)$ at
$r=0$. The odd powers in the power series expansion (\ref{STF_1s_expand})
are singular for $\bm{r} - \bm{r}' = \bm{0}$. This fact makes it
impossible to construct for either $\exp \bigl(-\beta \vert \bm{r} -
\bm{r}' \vert\bigr)$ or $\exp (\gamma r) \exp \bigl(-\beta \vert \bm{r} -
\bm{r}' \vert\bigr)$ with $r' > 0$ a power series in $r$ that converges
for all $r \in [0, \infty)$. This is only possible in the one-center case
$\bm{r}' = \bm{0}$. Consequently, a one-range addition theorem of the
type of (\ref{NISTO->ISTO_AddThm}), which converges pointwise for all
$\bm{r} \in \mathbb{R}^{3}$ or equivalently for all $r \in [0, \infty)$,
cannot exist for the $1 s$ function if $\bm{r}' \ne \bm{0}$.

An explicit expression for two-range addition theorem for the $1 s$
function (\ref{Def_STF_1s}) can be derived easily via the following
Gegenbauer-type addition theorem for the modified Bessel function
$w^{-\nu} K_{\nu} (\gamma w)$ with $ w = \bigl[ \rho^{2} + r^{2} - 2 r
\rho \cos \theta \bigr]^{1/2}$, $0 < \rho < r$, and $\nu \in \mathbb{C}
\setminus \mathbb{N}_{0}$ \cite[pp.\ 106 -
107]{Magnus/Oberhettinger/Soni/1966}:
\begin{equation}
  \label{BesK_GegenbauerAddThm}
  w^{-\nu} K_{\nu} (\gamma w) \; = \; 2^{\nu} \, \gamma^{-\nu} \,
  \Gamma (\nu) \, (r \rho)^{-\nu} \,
   \sum_{n=0}^{\infty} \, C_{n}^{\nu} (\cos \theta ) \,
   I_{\nu+n} (\gamma \rho) \, K_{\nu+n} (\gamma r) \, .  
\end{equation}
Here, $I_{\nu+n} (\gamma \rho)$ and $K_{\nu+n} (\gamma r)$ are modified
Bessel function of the first and second kind, respectively \cite[p.\
66]{Magnus/Oberhettinger/Soni/1966}.

The modified Bessel function $w^{-\nu} K_{\nu} (\gamma w)$ in
(\ref{BesK_GegenbauerAddThm}) is essentially a reduced Bessel function
$\hat{k}_{\nu} (\gamma w)$ defined by (\ref{Def:RBF}). On the basis of
(\ref{Ylm_AddThm}), (\ref{Gegenbauer_2_Legendre}), and
(\ref{BesK_GegenbauerAddThm}), the following two-range addition theorem
for reduced Bessel functions with half-integral orders can be derived in
a fairly straightforward way (\cite[Eq.\ (3.4)]{Steinborn/Filter/1975c}
or, as an improved version \cite[Eq.\ (5.5)]{Weniger/Steinborn/1989b}):
\begin{align}
  \label{RBF_AddThm_HalfInt}
  & \hat{k}_{n-1/2} \bigl(\beta \vert \bm{r}_{<} \pm \bm{r}_{>} \vert
  \bigr) \; = \; \frac{(-1)^{n} 8\pi}{(2n-1)!!} \, (\beta r_<)^{n-1/2}
  (\beta r_>)^{n-1/2} 
  \notag
  \\
  & \qquad \times \, \sum_{\ell=0}^{\infty} \, \sum^{\ell}_{m=-\ell} \,
  (\mp 1)^{\ell} \, \bigl[ Y^{m^{\star}}_{\ell}(\bm{r}_{<}/r_{<})
  \bigr]^{*} \, Y^m_{\ell} (\bm{r}_{>}/r_{>}) 
  \notag 
  \\
  & \quad \qquad \times \, \sum^{n}_{\nu=0} \, \frac{(-n)_{\nu}
    (1/2-n)_{\ell+\nu}} {{\nu}! (3/2)_{\ell+\nu}} \, (\ell+2\nu-n+1/2)
  \notag
  \\
  & \qquad \qquad \times \, I_{\ell+2\nu-n+1/2}(\beta r_{<})
  K_{\ell+2\nu-n+1/2}(\beta r_{>}) \, .
\end{align}
This addition theorem was quite consequential for my later scientific
interests. In my diploma thesis \cite{Weniger/1977}, which was published
in condensed form in \cite{Steinborn/Weniger/1977}, I used this addition
theorem for the evaluation of simple multicenter integrals of reduced
Bessel functions.

If we set $m=0$ in (\ref{RBF_HalfInt}), we obtain $\hat{k}_{1/2} (z) =
\mathrm{e}^{-z}$. Thus, we only have to set $n=1$ in
(\ref{RBF_AddThm_HalfInt}) to obtain a two-range addition theorem for the
$1 s$ function. At first sight, neither the addition theorem
(\ref{RBF_AddThm_HalfInt}) nor its special case with $n=1$ looks like a
power series expansion in $r_{<}$. However, the modified Bessel functions
$I_{\ell+2\nu-n+1/2}(\beta r_{<})$ is defined by a power series in $\beta
r_{<}$ \cite[p.\ 66]{Magnus/Oberhettinger/Soni/1966}, which shows that
the addition theorem (\ref{RBF_AddThm_HalfInt}) is nothing but an
infinite multitude of $\ell$-dependent rearranged power series expansions
in $r_{<}$.

The same conclusions hold for the following two-range addition theorem of
a $B$ function \cite[Eq.\ (5.11)]{Weniger/2002} which can be viewed to be
an anisotropic generalization of the addition theorem
(\ref{RBF_AddThm_HalfInt}) for a reduced Bessel function:
\begin{align}
  \label{B_Fun_AddThm}
  & B_{n, \ell}^{m} (\beta, \mathbf{r}_{<} \pm \mathbf{r}_{>}) \; = \;
  \frac{(2\pi)^{3/2}}{(-2)^{n+\ell}} \, 
  \sum_{\ell_1=0}^{\infty} \, \sum_{m=- \ell_1}^{\ell_1} \,
  (\mp 1)^{\ell_1} \, \, \left[ \mathcal{Y}_{\ell_1}^{m_1} (\mathbf{r}_{<})
  \right]^{*}
  \notag 
  \\
  & \qquad \times \, 
  \sum_{q=0}^{n+\ell} \, \frac {(-2)^q} {(n+\ell-q)!} \, 
  (\beta r_{<})^{n+\ell-\ell_1-q-1/2} \,
  I_{n+\ell+\ell_1-q+1/2} (\beta r_{<}) 
  \notag 
  \\
  & \qquad \quad \times \,
  \sum_{\ell_2=\ell_2^{\mathrm{min}}}^{\ell_2=\ell_2^{\mathrm{max}}} \!
  {}^{(2)} \, \langle \ell_2 m+m_1 \vert \ell_1 m_1 \vert \ell m \rangle 
  \notag 
  \\
  & \qquad \qquad \times
  \, \sum_{s=0}^{\min (q, \Delta \ell_2)} \, (-1)^s \,
  {\binom{\Delta \ell_2} {s}} \, B_{q-\ell_2-s,\ell_2}^{m+m_1} (\beta,
  \mathbf{r}_{>}) \, .
\end{align}
Other two-range addition theorems of $B$ functions are discussed in
\cite[Sections IV and V]{Weniger/Steinborn/1989b}.

In (\ref{B_Fun_AddThm}), $\langle \ell_2 m+m_1 \vert \ell_1 m_1 \vert
\ell m \rangle$ is a so-called Gaunt coefficient \cite{Gaunt/1929} which
corresponds to the integral of the product of three spherical harmonics
over the surface of the unit sphere in $\mathbb{R}^{3}$. The selection
rules of this Gaunt coefficient (see for example \cite[Section
3]{Weniger/Steinborn/1982} or \cite[Appendix C]{Weniger/2005}) imply that
$\Delta \ell_{2} = (\ell+\ell_{1}-\ell_{2})/2$ in (\ref{B_Fun_AddThm}) is
always either zero or a positive integer. The symbol $\sum \! {}^{(2)}$
in (\ref{B_Fun_AddThm}) indicates that the summation proceeds in steps of
two.

Since a Slater-type function with an integral principal quantum number
can according to (\ref{STF->Bfun}) be expressed as a finite sum of $B$
functions, we can conclude that the two-range addition theorem of a
Slater-type function obtained by forming linear combinations of
(\ref{B_Fun_AddThm}) is nothing but an infinite multitude of rearranged
power series expansions in $r_{<}$.

These considerations apply also to Slater-type functions $\chi_{N, L}^{M}
(\beta, \bm{r}_{<} \pm \bm{r}_{>})$ with nonintegral principal quantum
numbers $N$ or to $\exp (\gamma r) \chi_{N, L}^{M} (\beta, \bm{r}_{<} \pm
\bm{r}_{>})$. These functions are obviously singular for $\bm{r}_{<} \pm
\bm{r}_{>} = \bm{0}$, which implies that their power series expansions
about $r_{<}=0$ can only converge for $r_{<} < r_{>}$. Consequently, an
expansion of the type of (\ref{NISTO->ISTO_AddThm}) in terms of
Slater-type functions $\{ \chi_{n,\ell}^{m} (\gamma, \bm{r}) \}_{n, \ell,
  m}$ with integral principal quantum numbers $n$ and a common scaling
parameter $\gamma > 0$, that converges \emph{pointwise} for \emph{all}
$\bm{r}, \bm{r}' \in \mathbb{R}^{3}$, cannot exist.

These considerations can be generalized further: Let us assume that $f
(\bm{r} \pm \bm{r}')$ with $r' > 0$ is singular for $\bm{r} \pm \bm{r}' =
\bm{0}$ but analytic elsewhere. Accordingly, a power series expansion for
either $f (\bm{r} \pm \bm{r}')$ or for $\mathrm{e}^{\gamma r} f (\bm{r}
\pm \bm{r}')$ about $r=0$ can only converge for $r < r'$. This rules out
the existence of an expansion of $f (\bm{r} \pm \bm{r}')$ in terms of
Slater-type functions $\chi_{n,\ell}^{m} (\gamma, \bm{r}) \}$ with
integral principal quantum numbers $n$ and a common scaling parameter
$\gamma > 0$, that converges pointwise for all $\bm{r}, \bm{r}' \in
\mathbb{R}^{3}$.

As shown in \cite{Weniger/2000a,Weniger/2002} or in \cite[Section
7]{Weniger/2005}, pointwise convergent addition theorems are nothing but
rearranged Taylor expansions. Thus, the assumed singularity of such an $f
(\bm{r} \pm \bm{r}')$ for $\bm{r} \pm \bm{r}' = \bm{0}$ implies that a
pointwise convergent addition theorem must have a two-range form. 

Slater-type functions as well as all the other commonly used
exponentially decaying basis functions have a singularity at the
origin. Consequently, their pointwise convergent addition theorems must
have a two-range form. In contrast, the Gaussian function $\exp (- \beta
\vert \bm{r} \pm \bm{r}' \vert^{2})$ is analytic for all $\bm{r}, \bm{r}'
\in \mathbb{R}^{3}$. Consequently, it possesses a one-range addition
theorem that converges pointwise \cite[Eq.\
(8)]{Kaufmann/Baumeister/1989}.

It makes no difference if we start from a Laguerre series for a given
function. If we apply the transformation formula (\ref{Rearr_f_Exp_GLag})
to the series expansions (\ref{SFT_Mom_GLagPol}) for the angular
projections (\ref{SFT_Mom}) of Guseinov's one-range addition theorem
(\ref{Gus_OneRangeAddTheorSTF_k}) for $\chi_{N, L}^{M} (\beta, \bm{r} \pm
\bm{r}')$, we obtain power series expansion in $r$ for the angular
projections which converge or diverge, depending on the relative
magnitude of $\bm{r}$ and $\bm{r}'$.

Let us now assume that $f (\bm{r} \pm \bm{r}')$ is a function which is
singular for $\bm{r} \pm \bm{r}' = \bm{0}$ and that we know its two-range
addition theorem. If we accept the premise that its two-range addition
theorem is nothing but a multitude of rearranged power series expansions
in $r$ that converge for $r/r' < 1$, then the uniqueness of a power
series in the interior of its circle of convergence implies that
Guseinov's rearrangements of one-range addition theorems cannot produce
anything new. We obtain a two-range addition theorem for $\exp (\gamma r)
f (\bm{r} \pm \bm{r}')$ by forming the Cauchy product of the power series
for $\exp (\gamma r)$ with the possibly rearranged $\ell$-dependent power
series in $r$ that occur in the two-range addition theorem for $f (\bm{r}
\pm \bm{r}')$. In the final step, we only have to multiply the resulting
addition theorem for $\exp (\gamma r) f (\bm{r} \pm \bm{r}')$ by $\exp (-
\gamma r)$ or its power series to arrive at the possibly rearranged
addition theorem for $f (\bm{r} \pm \bm{r}')$ from which we started.

The discussion of this Section may create the false impression that
singularities of basis functions for electronic structure calculations
are something very negative. This is certainly true in the case of
analytical manipulations since virtually all manipulations becomes more
difficult in the presence of singularities. However, Kato
\cite{Kato/1957} had shown that the singularities of an atomic or
molecular Hamiltonian translate to corresponding singularities of the
eigenfunctions, commonly called cusps. But a cusp is just another word
for a singularity. Thus, the ability of basis functions to reproduce the
singularities of exact wave functions is of considerable importance for
the rate of convergence of an electronic structure calculation.

Gaussian functions do not have singularities like exponentially decaying
functions. In my opinion, this is the main reason why their multicenter
integrals can be evaluated much more easily than the corresponding
integrals of exponentially decaying functions. At the same time, the
absence of singularities is also a major drawback of Gaussian
functions. Many Gaussian functions are needed to approximate functions
possessing singularities with sufficient accuracy.

\typeout{==> Section: Numerical Implications of Truncated Expansions}
\section{Numerical Implications of Truncated Expansions}
\label{Sec:NumericalImplications}

The analysis of Sections \ref{Sec:Introduction} and
\ref{Sec:OneCenterExpansionSTF} shows that the one-center expansion
(\ref{GuExp_1}) of a Slater-type function $\chi_{N, L}^{M} (\beta,
\bm{r})$ in terms of Slater-type functions $\bigl\{ \chi_{n,L}^{M}
(\gamma, \bm{r}) \bigr\}_{n=L+1}^{\infty}$ with integral principal
quantum numbers $n \in \mathbb{N}$ and a common scaling parameter $\gamma
> 0$ does not exist if the principal quantum number $N$ is
nonintegral. The leading coefficients of (\ref{GuExp_1}) are zero, and
the higher coefficients are infinite in magnitude.

The nonexistance of (\ref{GuExp_1}) is a direct consequence of the fact
that the radial part of $\chi_{N, L}^{M} (\beta, \bm{r})$ with $N \in{
  \mathbb{R}} \setminus \mathbb{N}$ is not analytic at $r=0$. As shown in
Section \ref{Sec:OneCenterExpansionSTF}, this nonexistance can also be
shown by applying the transformation formula (\ref{Rearr_f_Exp_GLag}) to
the radial part of a $k$-dependent one-center expansion
(\ref{Expand_NISTF2Gusfun_DiffScaPar_1}) for $\chi_{N, L}^{M} (\beta,
\bm{r})$. In the case of nonintegral quantum numbers, the transformation
formula (\ref{Rearr_f_Exp_GLag}) produces power series coefficients that
are either zero or infinite in magnitude.

In Section \ref{Sec:OneRangeAdd->TwoRangeAdd} it was shown that the
rearrangement of a $k$-dependent one-range addition theorem
(\ref{Gus_OneRangeAddTheorSTF_k}), which is a much more complicated
expansion than its one-center limit
(\ref{Expand_NISTF2Gusfun_DiffScaPar_1}), also does not produce the
desired result. Both $\chi_{N, L}^{M} (\beta, \bm{r} \pm \bm{r}')$ and
$\exp (\gamma r) \chi_{N, L}^{M} (\beta, \bm{r} \pm \bm{r}')$ with $r' >
0$ are analytic at $r=0$. Consequently, the application of the
transformation formula (\ref{Rearr_f_Exp_GLag}) to the series expansions
of the angular projections of $\exp (\gamma r) \chi_{N, L}^{M} (\beta,
\bm{r} \pm \bm{r}')$ produces mathematically meaningful power series
expansions in $r$. Unfortunately, these power series expansions have a
finite radius of convergence and they converge only for $r < r'$. This
follows at once from the fact that both $\chi_{N, L}^{M} (\beta, \bm{r}
\pm \bm{r}')$ and $\exp (\gamma r) \chi_{N, L}^{M} (\beta, \bm{r} \pm
\bm{r}')$ are singular for $\bm{r} \pm \bm{r}' = \bm{0}$. Accordingly,
Guseinov's rearrangements transform a \emph{one-range addition theorem}
for $\chi_{N, L}^{M} (\beta, \bm{r} \pm \bm{r}')$ to a \emph{two-range
  addition theorem}. This is certainly not what Guseinov had tried to
achieve.

But these observations do not tell the whole truth, and in particular
they do not imply that Guseinov's approach is necessarily doomed. There
is overwhelming evidence that Guseinov never transformed a
\emph{complete} one-range addition theorem of the type of
(\ref{Gus_OneRangeAddTheorSTF_k}) containing an infinite number of
terms. Instead, Guseinov only transformed truncations of either addition
theorems or their one-center limits, which all contain a finite number of
terms only. Thus, Guseinov exclusively did his rearrangements with an
analog of (\ref{RearrFinSum_GLag}), which transforms a truncated Laguerre
series of the type of (\ref{FinSum_GLag}) to a polynomial of the type of
(\ref{TruncPowSer_f}). To the best of my knowledge, Guseinov never used
the complete transformation formula (\ref{Rearr_f_Exp_GLag}) and only
claimed -- probably on the basis of an insufficient amount of numerical
evidence -- that his transformed truncations remain meaningful in the
limit of infinite expansion lengths, but he never provided convincing
evidence supporting his claim.

The radial parts of the angular projections of Guseinov's truncations are
apart from a common exponential $\exp (-\gamma r)$ finite linear
combinations of generalized Laguerre polynomials in $2 \gamma r$. Thus,
for finite truncation orders $\mathcal{N}$, Guseinov's rearrangements are
legitimate and produce polynomials in $r$.

In actual calculations, we always have to truncate infinite series
expansions after a finite number of terms unless we are fortunate enough
to find a way of expressing a series in closed form. Therefore, a
skeptical reader might argue that Guseinov's approach is completely
satisfactory from a practical point of view, and that my insistence on
the convergence and existence of infinite series expansions, which in
actual calculations have to be truncated and thus do not really occur in
practice, is nothing but a mathematical over-sophistication.

However, Guseinov's rearranged addition theorems do not only suffer from
the fact that in limit of infinite expansions lengths they either do not
exist or that they lose their one-range nature. These rearrangements
cause also other problems which become particularly evident if one tries
to achieve a (very) high accuracy by including a large number of terms in
Guseinov's truncations.

One serious problem was already discussed in Section
\ref{Sec:GuseinovsRearrangementsOfAdditionTheorems}. As explained there,
I have grave doubts that a Guseinov function $\prescript{}{k}{\Psi}_{n,
  \ell}^{m} (\gamma, \bm{r})$ can be expressed in a numerically stable
way as a finite sum of Slater-type functions $\bigr\{ \chi_{n', \ell}^{m}
(\gamma, \bm{r}) \bigl\}_{n'=\ell+1}^{n}$ via (\ref{GusFun2STF}) if the
index $n$ is large. This applies also to the substitution of overlap
integrals involving Guseinov functions by finite sums of overlap
integrals involving Slater-type functions via
(\ref{OverlapGusFun_STF}). Both (\ref{GusFun2STF}) and
(\ref{OverlapGusFun_STF}) are based on the explicit expression
(\ref{GLag_1F1}) of the generalized Laguerre polynomial $L_{n}^{(\alpha)}
(z)$, which tends to become numerically unstable if its index $n$ becomes
large. The reason is that the coefficients of orthogonal polynomials have
strictly alternating signs.

The substitution of a function set, which is complete and orthonormal in
a given Hilbert space, by a function set, which is only complete, but not
orthogonal, also causes some nontrivial problems. In Section
\ref{Sec:GuseinovsRearrangementsOfAdditionTheorems} it was emphasized
that orthogonal expansions tend to be computationally well behaved
because Parseval's equality (\ref{ParsevalEquality}) guarantees that
their series coefficients are bounded in magnitude and vanish for large
indices. In contrast, the series coefficients of nonorthogonal expansions
are not necessarily bounded in magnitude and do not necessarily vanish
with increasing index.

These complications with unbounded coefficients can be demonstrated
convincingly by considering a truncation of the comparatively simple
one-center expansion (\ref{Expand_NISTF2Gusfun_DiffScaPar_1}). This
ansatz corresponds to the following approximation of a Slater-type
function $\chi_{N, L}^{M} (\beta, \bm{r})$ with an in general nonintegral
principal quantum number $N \in \mathbb{R} \setminus \mathbb{N}$ by a
finite sum of Guseinov functions:
\begin{align}
  \label{Expand_NISTF2Gusfun_DiffScaPar_1_Trunc}
  \chi_{N, L}^{M} (\beta, \bm{r}) & \; \approx \; \frac
  {(2\gamma)^{L+(k+3)/2} \, \beta^{N-1}}{[\beta+\gamma]^{N+L+k+2}}
  \, \frac{\Gamma (N+L+k+2)}{(2L+k+2)!}
  \notag \\
  & \qquad \times \sum_{\nu=0}^{\mathcal{N}} \, \left[
    \frac{(\nu+2L+k+2)!}{\nu!} \right]^{1/2} \,
  \prescript{}{k}{\Psi}_{\nu+L+1, L}^{M} (\gamma, \bm{r})
  \notag \\
  & \qquad \qquad \times {}_2 F_1 \left(-\nu, N+L+k+2; 2L+k+3;
    \frac{2\gamma}{\beta+\gamma} \right) \, .
\end{align}  
If $\chi_{N, L}^{M} (\beta, \bm{r})$ belongs to the weighted Hilbert
space $L_{r^k}^{2} (\mathbb{R}^3)$ defined by (\ref{HilbertL_r^k^2}),
this $\mathcal{N}$-dependent approximation is mathematically meaningful,
and we can be sure that it converges in the mean with respect to the norm
of $L_{r^k}^{2} (\mathbb{R}^3)$ as $\mathcal{N} \to \infty$.

We only have to cancel the spherical harmonics and remove the exponential
on the right-hand side of (\ref{Expand_NISTF2Gusfun_DiffScaPar_1_Trunc})
to see that the finite sum on the right-hand side corresponds to a
truncated Laguerre series of the type of (\ref{FinSum_GLag}). If we now
apply the transformation formula (\ref{RearrFinSum_GLag}) to this
truncated Laguerre series, we ultimately obtain an approximation of the
Slater-type function $\chi_{N, L}^{M} (\beta, \bm{r})$ with $N \in
\mathbb{R} \setminus \mathbb{N}$ as a finite linear combination of
Slater-type functions $\chi_{n, L}^{M} (\gamma, \bm{r})$ with integral
principal quantum numbers $n$.

For finite values of the truncation order $\mathcal{N}$, all coefficients
of this approximation to $\chi_{N, L}^{M} (\beta, \bm{r})$ are well
defined and finite. Unfortunately, this does not imply that this
expression remains well behaved in the limit $\mathcal{N} \to \infty$.
For large values of $\mathcal{N}$, the leading coefficients of the
resulting expression approach zero and the higher coefficients diverge in
magnitude.

This can also be demonstrated by inserting the expansion coefficients of
the Laguerre series (\ref{GenPow2GLag}) for $z^{\rho} \mathrm{e}^{u z}$
with nonintegral $\rho \in \mathbb{R} \setminus \mathbb{N}_{0}$ into the
$\mu$ series in (\ref{Rearr_f_Exp_GLag}). If we replace the exact
expressions for the Laguerre series coefficients of the later terms of
the $\mu$ series with indices $\mu \ge M$ by their leading order
asymptotic approximations (\ref{Asy_lanmda_GenCase}), we obtain the
following leading order asymptotic approximation:
\begin{equation}
  \label{ZetaTypeTail}
  \sum_{\mu=M}^{\infty} \, \frac{(\alpha+\nu+1)_{\mu}}{\mu!} \, 
  \lambda_{\mu+\nu}^{(\alpha)} \; \sim \; 
  \frac{\Gamma (\rho+\alpha+1)}{\Gamma (-\rho) \Gamma (\alpha+\nu+1)} \, 
  \sum_{\mu=M}^{\infty} \, \mu^{\nu-\rho-1}  \, ,
  \quad M \to \infty \, .
\end{equation}
The infinite series on the right-hand side is nothing but the tail of the
Dirichlet series $\zeta (s) = \sum_{n=0}^{\infty} (n+1)^{-s}$ for the
Riemann zeta function with $s = \rho-\nu+1$ (see for example \cite[p.\
21]{Magnus/Oberhettinger/Soni/1966}). As is well known, the Dirichlet
series for $\zeta (s)$ converges for $\Re (s) > 1$ and diverges for $\Re
(s) \le 1$. Thus, the series (\ref{ZetaTypeTail}) converges for $\nu <
\rho$, and it diverges for $\nu > \rho$.

If $\Re (s)$ is only slightly larger than 1, the convergence of the
Dirichlet series $\sum_{n=0}^{\infty} (n+1)^{-s}$ can become
prohibitively slow (the horrifying example of the Dirichlet series with
$s = 1.01$ is discussed in \cite[p.\ 194]{Weniger/Kirtman/2003}). The
slowest convergence on the right-hand side of (\ref{ZetaTypeTail}) occurs
for the largest value of $\nu$ satisfying $\nu < \rho$, i.e., for $\nu =
\Ent{\rho}$, where $\Ent{\rho}$ is the integral part of $\rho$. In this
case, convergence can become so slow that it is practically impossible to
evaluate the series on the right-hand of (\ref{ZetaTypeTail}) with
sufficient accuracy by adding up its terms successively. Instead, one has
to employ suitable convergence acceleration techniques as for example the
Euler-Maclaurin formula (see for example \cite[Section
2]{Weniger/Kirtman/2003} and references therein). In this context, it may
be of interest that the Euler-Maclaurin formula for the Riemann zeta
function and other asymptotic approximations to truncation errors of
series representations for special functions can also be derived by
solving systems of linear equations \cite{Weniger/2007a}.

A skeptical reader might argue that my digression on the convergence
properties of the Dirichlet series for the Riemann zeta is of no interest
in the context of molecular electronic structure. However, I personally
became interested in the Riemann zeta function because of certain
infinite series expansions occurring in expressions for molecular
integrals of exponentially decaying functions involving the Coulomb
potential. The convergence properties of these expansions closely
resemble that of the Dirichlet series for $\zeta (s)$ with $\Re (s)$ not
much larger than 1 (see for example \cite[Tables I, II, V, VI, and
VII]{Grotendorst/Weniger/Steinborn/1986} or \cite[Table
1]{Steinborn/Weniger/1990}).

The slow convergence of the series on the right-side of
(\ref{ZetaTypeTail}) for $\nu$ only slightly smaller than $\rho$ makes it
(very) hard or even practically impossible to observe by purely numerical
means that the leading terms of the formal power series for $z^{\rho}
\mathrm{e}^{u z}$ with nonintegral $\rho \in \mathbb{R} \setminus
\mathbb{N}_{0}$ vanish. 

Moreover, the series on the right-hand side of (\ref{ZetaTypeTail})
diverges for $\nu > \rho$, but it diverges quite slowly if $\nu$ is only
slightly larger than $\rho$. Quite a few terms are needed to observe the
divergence of such an infinite series in the case of small or moderately
large values of $\nu$. There is also the additional complication that it
is by no means easy to establish unambiguously the divergence of a series
of the type of (\ref{ZetaTypeTail}) if only a finite number of terms are
available \cite{Nozaki/1988}.

A further complication occurs if we do not rearrange an infinite Laguerre
series of the type of (\ref{f_Exp_GLag}) via the complete transformation
formula (\ref{Rearr_f_Exp_GLag}), which uses \emph{all} Laguerre series
coefficients $\lambda_{n}^{(\alpha)}$ with $n \in \mathbb{N}_{0}$, but
via (\ref{RearrFinSum_GLag}) which transforms the partial sum $f_{M} (z)$
of the Laguerre series (\ref{f_Exp_GLag}) to a polynomial of degree $M$
in $z$. Since the transformation formula (\ref{RearrFinSum_GLag}) only
uses the coefficients $\lambda_{0}^{(\alpha)}$, $\lambda_{1}^{(\alpha)}$,
\dots, $\lambda_{M}^{(\alpha)}$ and since $M$ is in practice only
moderately large, neither the vanishing of power series coefficients with
$\nu < \rho$ nor the divergence of the coefficients with $\nu > \rho$ can
be observed easily (everything remains finite). Consequently, I would not
be surprised if Guseinov and Mamedov, who certainly had done test
calculations, simply overlooked the nonexistence of their one-center
expansion \cite[Eq.\ (4)]{Guseinov/Mamedov/2008a} in the limit
$\mathcal{N} \to \infty$.

It would have been interesting if Guseinov and Mamedov had applied
sequence transformations, because this could have helped them to see the
vanishing or the divergence of their coefficients more clearly. My
suggestion may sound paradoxical because sequence transformations are
normally used to accelerate convergence or to associate a finite value to
a divergent sequence or series. It was, however, shown in recent articles
by Beckermann, Kalyagin, Matos, and Wielonsky
\cite{Beckermann/Kalyagin/Matos/Wielonsky/2011}, Beckermann, Matos, and
Wielonsky \cite{Beckermann/Matos/Wielonsky/2008}, Brezinski
\cite{Brezinski/2004}, Brezinski and Redivo Zaglia
\cite{Brezinski/RedivoZaglia/2009}, and Guilpin, Gacougnolle, and Simon
\cite{Guilpin/Gacougnolle/Simon/2004} that sequence transformations can
also be used to determine the location of discontinuities of functions
more precisely or to show them more clearly. By a slight abuse of
language, such an application of sequence transformations could be called
\emph{acceleration of divergence}.

Guseinov's and Mamedov's inability of observing any problems with their
rearranged one-center expansions highlights once more the dangers of
relying entirely on numerical test calculations without trying to
understand the subtleties of the underlying mathematics.

One should also take into account that the apparent convergence of the
sum of the leading terms of an infinite series to the correct limit does
not prove the existence of this series, let alone its converges to the
correct limit. As discussed in Appendix \ref{App:Semiconvergence}, the
phenomenon of \emph{semiconvergence} -- initial apparent convergence of
the leading terms of an infinite series followed by divergence if more
terms are included -- is well established in the literature.

The examples in Appendix \ref{App:Semiconvergence} on semiconvergence and
related phenomena should suffice to convince even a skeptic that
misinterpretations of seemingly convincing numerical evidence can happen
easily. It also happened to me. In \cite{Weniger/1990} I misinterpreted
my summation results obtained by applying Wynn's epsilon algorithm
\cite{Wynn/1956a}, the $d$ variant \cite[Eq.\ (7.3-9)]{Weniger/1989} of
Levin's transformation \cite{Levin/1973}, and the $d$ variant of the
so-called $\mathcal{S}$ transformation \cite[Eq.\ (8.4-4)]{Weniger/1989}
to the partial sum of the first 22 perturbation series coefficients of
the factorially divergent Rayleigh-Schr\"{o}dinger perturbation series
for the ground state energy of the quartic anharmonic oscillator. All
calculations were done in FORTRAN 77 on a Cyber 180-995 E with a
precision of approximately 29 decimal digits (for more details, see
\cite[pp.\ 7 - 8]{Weniger/2007d}).

Unfortunately, my conclusion in \cite{Weniger/1990} that Levin's
transformation produces a convergent result -- although perfectly
plausible at that time -- was premature and based on the incomplete
evidence provided by the first 22 perturbation series coefficients. In
\cite{Weniger/Cizek/Vinette/1993}, we repeated my previous calculations
using now 200 perturbation series coefficients calculated exactly with
the help of Maple's rational arithmetic, and we did the summation
calculations in Maple with a precision of up to 1000 decimal digits and
transformation orders as high as $k=199$. These calculations showed
unambiguously that Levin's transformation failed to produce convergent
results in the case of higher transformation orders, and that this
failure could not be attributed to numerical instabilities. The
divergence of Levin's transformation was also confirmed in \cite[Table
2]{Weniger/1992}. A similar divergence of Levin's transformation was
later observed by \v{C}\'{\i}\v{z}ek, Zamastil, and Sk\'{a}la \cite[p.\
965]{Cizek/Zamastil/Skala/2003} in the case of the hydrogen atom in an
external magnetic field.

Of course, my misinterpretation in \cite{Weniger/1990} or similar
problems of other authors do not rule out the possibility that carefully
conducted purely numerical investigations can provide valuable
theoretical insight. In \cite{Bender/Weniger/2001} we formulated with the
help of some numerical techniques developed in \cite{Weniger/2000b} the
conjecture that the factorially divergent perturbation expansion of a
certain non-Hermitian $\mathcal{PT}$-symmetric anharmonic oscillator is a
Stieltjes series. Recently, our conjecture, whose correctness implies the
Pad\'{e} summability of this perturbation expansion, was proved
rigorously by Grecchi, Maioli, and Martinez
\cite{Grecchi/Maioli/Martinez/2009}.

Let us now assume that we rearrange a truncation
(\ref{Expand_NISTF2Gusfun_DiffScaPar_1_Trunc}) of the one-center
expansion (\ref{Expand_NISTF2Gusfun_DiffScaPar_1}) for $\chi_{N, L}^{M}
(\beta, \bm{r})$ with nonintegral principal quantum number $N \in
\mathbb{R} \setminus \mathbb{N}$. If the truncation order $\mathcal{N}$
in (\ref{Expand_NISTF2Gusfun_DiffScaPar_1_Trunc}) is small, its increase
will certainly improve the accuracy of the rearranged
truncation. However, for sufficiently large values of $\mathcal{N}$, the
behavior of the rearranged truncations changes. The leading order
asymptotic approximation (\ref{ZetaTypeTail}) implies that the accuracy
ultimately deteriorates with increasing $\mathcal{N}$, and for
$\mathcal{N} \to \infty$, the rearranged truncations ultimately become
mathematically meaningless. Thus, rearranged truncations of the
one-center expansion (\ref{Expand_NISTF2Gusfun_DiffScaPar_1}) are
semiconvergent with respect to a variation of $\mathcal{N}$ for
nonintegral principal quantum numbers $N \in \mathbb{R} \setminus
\mathbb{N}$.

The semiconvergence of the rearranged truncations implies that they can
be used for computational purposes at least for sufficiently small
truncation orders $\mathcal{N}$. But obviously, one should be careful. It
is necessary to investigate for which values of $\mathcal{N}$ the
intrinsic pathologies of the rearranged expansions become
intolerable. The leading order asymptotic approximation
(\ref{ZetaTypeTail}) indicates that even fairly large values of
$\mathcal{N}$ should produce acceptable results, but additional and in
particular more detailed numerical investigations, which also try to
estimate the detrimental effect of possible numerical instabilities,
would certainly be desirable.

One can look at the rearrangements of truncations
(\ref{Expand_NISTF2Gusfun_DiffScaPar_1_Trunc}) also from a different
perspective. As discussed in Appendix \ref{App:OrthogonalExpansions},
\emph{finite} approximations to a function $f \in \mathcal{H}$ of the
type of (\ref{f_FinAppr}) in terms of a function set, that is complete
but nonorthogonal in some Hilbert space $\mathcal{H}$, can be constructed
by minimizing the mean square deviation (\ref{MeanSquareDev}), although
we cannot tacitly assume that these finite approximations can be extended
to infinite expansions of the type of (\ref{f_InfExp}).

Thus, is certainly legitimate to \emph{approximate} a Slater-type
function $\chi_{N, L}^{M} (\beta, \bm{r})$ with a nonintegral principal
quantum number $N \in \mathbb{R} \setminus \mathbb{N}$ by a finite linear
combination of Slater-type functions $\chi_{n, L}^{M} (\gamma, \bm{r})$
with integral principal quantum numbers $n$. However, a determination of
the coefficients of these approximations via a purely numerical
minimization of the mean square deviation does not look like a promising
computational strategy in the context of addition theorems or multicenter
integrals. In such a case, Guseinov's approach is most likely the better
alternative.

The truncations (\ref{Expand_NISTF2Gusfun_DiffScaPar_1_Trunc}) of the
one-center expansion (\ref{Expand_NISTF2Gusfun_DiffScaPar_1}) are simple
enough to permit a detailed mathematical analysis of Guseinov's
rearrangements. But we would of course be much more interested in
understanding the subtleties of the rearrangements of the truncations
\begin{equation}
  \label{Gus_OneRangeAddTheorSTF_k_Trunc}
  \chi_{N, L}^{M} (\beta, \bm{r} \pm \bm{r}')
  \; \approx \; \sum_{n=1}^{\mathcal{N}} \, 
  \sum_{\ell=0}^{n-1} \, \sum_{m=-\ell}^{\ell} \,
  \prescript{}{k}{\mathbf{X}}_{n, \ell, m}^{N, L, M}
  (\gamma, \beta, \pm \bm{r}') \,
  \prescript{}{k}{\Psi}_{n, \ell}^{m} (\gamma, \bm{r})
\end{equation}
of a $k$-dependent addition theorem
(\ref{Gus_OneRangeAddTheorSTF_k}). The overlap integral
$\prescript{}{k}{\mathbf{X}}_{n, \ell, m}^{N, L, M} (\gamma, \beta, \pm
\bm{r}')$ in (\ref{Gus_OneRangeAddTheorSTF_k_Trunc}) is defined by by
(\ref{Gus_OneRangeAddTheorSTF_k_b}).
 
Since the $k$-dependent addition theorems
(\ref{Gus_OneRangeAddTheorSTF_k}) are genuine two-center problems, it is
not at all easy to do a rigorous analysis of the mathematical properties
of rearrangements of the truncations
(\ref{Gus_OneRangeAddTheorSTF_k_Trunc}). Let me emphasize once more that
there is a fundamental difference between the one-center case analyzed in
Section \ref{Sec:OneCenterExpansionSTF} and the two-center case analyzed
in Section \ref{Sec:OneRangeAdd->TwoRangeAdd}. In the limit of infinite
truncation orders $\mathcal{N}$, Guseinov's rearrangements of one-center
expansions for Slater-type functions with nonintegral principal quantum
numbers produce mathematically meaningless expansions, and for finite
values of $\mathcal{N}$ these rearrangements are semiconvergent.

In the two-center case, Guseinov's rearrangements of one-range addition
theorems produce mathematically meaningful expansions, but they only
converge for $r < r'$. Consequently, these expansions correspond to the
small $r$ parts of two-range addition theorems. This is certainly not
what Guseinov had tried to achieve.

Apart from being an undesirable result, the two-center nature of
rearrangements of the truncations (\ref{Gus_OneRangeAddTheorSTF_k_Trunc})
in the limit of infinite truncation orders $\mathcal{N}$ is also a
possible source of problems. For finite values of $\mathcal{N}$, the
truncations (\ref{Gus_OneRangeAddTheorSTF_k_Trunc}) are obviously
one-range addition theorems: The vector $\bm{r}$ occurs exclusively in
the Guseinov functions $\prescript{}{k}{\Psi}_{n, \ell}^{m} (\gamma,
\bm{r})$, and $\bm{r}'$ occurs exclusively in the overlap integral
$\prescript{}{k}{\mathbf{X}}_{n, \ell, m}^{N, L, M} (\gamma, \beta, \pm
\bm{r}')$. This applies also to their rearrangements. The vectors
$\bm{r}$ and $\bm{r}'$ are still separated even if the Guseinov functions
are replaced by Slater-type functions with integral principle quantum
numbers via (\ref{GusFun2STF}) and the overlap integrals involving
Guseinov functions by overlap integrals of Slater-type functions via
(\ref{OverlapGusFun_STF}).

But in the limit of infinite truncation orders $\mathcal{N}$, a change
resembling a phase transition takes place: The resulting rearranged
expansions lose their advantageous one-range nature and converge to
two-range addition theorems. As is well known, the decay rate of the
coefficients $\gamma_{n}$ of a power series $f (z) = \sum_{n=0}^{\infty}
\gamma_{n} z^{n}$ as $n \to \infty$ determines the convergence type of
this series: If $\gamma_{n}$ decays factorially as $n \to \infty$, $f
(z)$ is entire, and if $\gamma_{n}$ decays only exponentially, the radius
of convergence of its power series is finite. Thus, the expansion
coefficients of the Slater-type functions $\chi_{n, \ell}^{m} (\gamma,
\bm{r})$ in Guseinov's rearranged addition theorems apparently decay at
most exponentially since these expansions only converge for $r < r'$.

Otherwise, very little can be said about the $\mathcal{N}$-dependence of
Guseinov's rearrangements of the truncations
(\ref{Gus_OneRangeAddTheorSTF_k_Trunc}). The problem is that in Section
\ref{Sec:OneRangeAdd->TwoRangeAdd} the mathematical properties of
Guseinov's rearranged addition theorems were analyzed via the singularity
of a Slater-type function $\chi_{N, L}^{M} (\beta, \bm{r} \pm \bm{r}')$
for $\bm{r} \pm \bm{r}' = \bm{0}$, but not via the transformation formula
(\ref{Rearr_f_Exp_GLag}). We would need the large index asymptotics of
the overlap integrals $\prescript{}{k}{\mathbf{X}}_{n, \ell, m}^{N, L, M}
(\gamma, \beta, \pm \bm{r}')$ in (\ref{Gus_OneRangeAddTheorSTF_k}).  On
the basis of our current level of understanding, such an asymptotic
analysis seems to be out of reach. We do not have anything resembling a
two-center analog of the leading order asymptotic approximation
(\ref{ZetaTypeTail}) which turned out to be so very useful in the
one-center case.

Guseinov's rearrangements of one-range addition theorems can be used to
construct \emph{approximations} to a Slater-type function $\chi_{N,
  L}^{M} (\beta, \bm{r} \pm \bm{r}')$ with both integral and nonintegral
principal quantum numbers in terms of a \emph{finite} number of
Slater-type functions $\chi_{n,\ell}^{m} (\gamma, \bm{r})$ with integral
principal quantum numbers $n$. As in the one-center case, such a finite
approximation of the type of (\ref{f_FinAppr}) can -- at least in
principle -- be constructed by minimizing the mean square deviation
(\ref{MeanSquareDev}), although such a numerical determination of the
expansion coefficients is most likely not a very good idea in the context
of multicenter integrals.

But there remains a principal problem. If we use a two-range addition
theorem in a multicenter integral and do not take into account its
two-range nature by splitting up the integration interval of the
resulting radial integration, we may well end up either with convergence
to the wrong limit or even with a divergent series expansion for the
multicenter integral.

For finite truncation orders $\mathcal{N}$, Guseinov's rearrangements of
the truncations (\ref{Gus_OneRangeAddTheorSTF_k_Trunc}) are one-range
addition theorems since the vectors $\bm{r}$ and $\bm{r}'$ are completely
separated. Thus, these rearranged truncations can safely be used in
multicenter integrals, and it is not necessary to split up the
integration contour. But in the limit $\mathcal{N} \to \infty$, these
rearrangements lose their convenient one-range property. Consequently,
their careless use in a multicenter integral without splitting up the
integration contour may produce either a wrong or a divergent result.

It makes sense to assume that the ultimate two-range nature of the
rearrangements of the truncations (\ref{Gus_OneRangeAddTheorSTF_k_Trunc})
becomes noticeable already for sufficiently large, but finite values of
$\mathcal{N}$. It is therefore conceivable that multicenter integrals,
whose integrands contain a rearrangement of a truncations
(\ref{Gus_OneRangeAddTheorSTF_k_Trunc}), are semiconvergent with respect
to a variation of $\mathcal{N}$. Obviously, the exact behavior of these
integrals as $\mathcal{N} \to \infty$ certainly does not only depend on
the approximation to the addition theorem, but also on the remaining
integrand. This certainly makes a detailed analysis even more difficult.

These considerations are for the moment essentially speculation. A
sufficiently rigorous analysis of this possible semiconvergence cannot be
done yet. We lack some necessary mathematical tools such as a two-center
analog of the leading order asymptotic expansion (\ref{ZetaTypeTail})
which was so very useful in the one-center case.

\typeout{==> Section: Summary and Outlook}
\section{Summary and Outlook}
\label{Sec:SummaryOutlook}

Infinite dimensional function spaces and in particular Hilbert spaces are
of considerable importance not only in quantum theory, but also in
approximation theory and in functional analysis. Accordingly, there is an
extensive literature discussing these function spaces, both from a purely
mathematical point of view and also from the perspective of quantum
physics.

It is an obvious idea to try to understand the properties of these
admittedly complicated function spaces by emphasizing the analogies to
the simpler $n$-dimensional real and complex vector spaces
$\mathbb{R}^{n}$ and $\mathbb{C}^{n}$, respectively. Often, this approach
provides valuable insight. Nevertheless, there are dangers. Infinite
dimensional function spaces possess certain peculiarities which do not
exist in the case of $n$-dimensional vector spaces. Naive generalizations
and overly optimistic analogies can therefore be badly misleading.

For example, in the $n$-dimensional vector spaces $\mathbb{R}^{n}$ or
$\mathbb{C}^{n}$ any set of $n$ linearly independent vectors can be used
as a basis, which means that every vector belonging to either
$\mathbb{R}^{n}$ or $\mathbb{C}^{n}$ can be expressed as a linear
combination of $n$ linearly independent vectors. There is also no
principal problem if we want to replace one basis by another: All
linearly independent sets of $n$ vectors are in that respect
equivalent. Orthogonality of the basis vectors in $n$-dimensional vector
spaces is convenient since it greatly simplifies certain operations, but
it is not really essential.

In infinite dimensional function spaces, the situation is much more
complicated. Firstly, a basis now consists of an infinite number of
elements which always raises the question of convergence. Moreover, there
are different convergence types, which are in general incompatible.
Secondly, a basis in a function space has to be complete, i.e., the span
of this basis has to be dense in the function space.

As discussed in Appendix \ref{App:OrthogonalExpansions}, the completeness
of a basis suffices to guarantee that finite expansion of the type of
(\ref{f_FinAppr}) exist. In addition, completeness implies that the
corresponding mean square deviation (\ref{MeanSquareDev}) can be made as
small as we like by increasing the length of the finite
expansion. Therefore, it is a seemingly obvious conclusion that the
completeness of a basis implies the existence of infinite expansions of
the type of (\ref{f_InfExp}) in terms of this basis. Unfortunately, this
conclusion is wrong. The existence of an infinite expansion is only
guaranteed if the basis is not only complete, but also orthogonal. These
facts are well known, but nevertheless often ignored. They are also the
basis of this article.

In \cite{Guseinov/1978,Guseinov/1980a,Guseinov/1985a,Guseinov/2001a,%
  Guseinov/2002c}, Guseinov derived one-range addition theorems for
Slater-type functions $\chi_{N, L}^{M} (\beta, \bm{r} \pm \bm{r}')$ with
integral and nonintegral principal quantum numbers $N$ by expanding them
in terms of his Laguerre-type functions $\bigl\{
\prescript{}{k}{\Psi}_{n, \ell}^{m} (\gamma, \bm{r}) \bigr\}_{n, \ell,
  m}$. For a given $k = -1, 0, 1, 2, \dots$, these functions defined by
(\ref{Def_Psi_Guseinov}) are complete and orthonormal in the
corresponding weighted Hilbert space $L_{r^k}^{2} (\mathbb{R}^3)$ defined
by (\ref{HilbertL_r^k^2}). Guseinov's approach is mathematically sound as
long as the Slater-type function $\chi_{N, L}^{M} (\beta, \bm{r} \pm
\bm{r}')$ belongs to $L_{r^k}^{2} (\mathbb{R}^3)$. Such a one-range
addition theorem converges in the mean with respect to the norm of the
Hilbert space $L_{r^k}^{2} (\mathbb{R}^3)$, but not necessarily
pointwise.

However, Guseinov considered it to be advantageous to replace in his
$k$-dependent addition theorems the complete and orthonormal functions
$\bigl\{ \prescript{}{k}{\Psi}_{n, \ell}^{m} (\gamma, \bm{r}) \bigr\}_{n,
  \ell, m}$, whose radial parts are according to (\ref{Def_Psi_Guseinov})
essentially generalized Laguerre polynomials, by complete, but
nonorthogonal Slater-type functions $\bigl\{ \chi_{n, \ell}^{m} (\gamma,
\bm{r}) \bigr\}_{n, \ell, m}$ with integral principal quantum numbers via
(\ref{GusFun2STF}). Ultimately, Guseinov's approach corresponds to the
replacement of a Laguerre series of the type of (\ref{f_Exp_GLag}) by a
power series of the type of (\ref{PowSer_f}). Unfortunately, this
approach is not necessarily legitimate and can easily lead to nontrivial
convergence and existence problems, which had already been discussed in
\cite{Weniger/2008}, albeit in a less detailed way.

Because of these principal problems, Guseinov cannot tacitly assume that
the transformation of his one-range addition theorems, which are
expansions in terms of his complete and orthonormal Laguerre-type
functions, to expansions of the type of (\ref{NISTO->ISTO_AddThm}) in
terms of the complete, but nonorthogonal Slater-type functions are
necessarily legitimate. This has to be demonstrated explicitly, but
Guseinov had only done this in a very superficial way. There is
considerable evidence that Guseinov and co-workers had only done some
purely numerical tests.

The deficiencies of Guseinov's approach become particularly evident in
the case of the one-center expansion (\ref{GuExp_1}) for a Slater-type
function $\chi_{N, L}^{M} (\beta, \bm{r})$ with an in general nonintegral
principal quantum number $N$ in terms of Slater-type functions $\bigl\{
\chi_{n, L}^{M} (\gamma, \bm{r}) \bigr\}_{n=L+1}^{\infty}$ with integral
principal quantum numbers $n$ and an in general different common scaling
parameter $\gamma \ne \beta > 0$. Guseinov \cite[Eq.\
(21)]{Guseinov/2002c} had constructed this expansion by performing the
one-center limit of a rearranged truncated addition theorem for
Slater-type functions \cite[Eq.\ (15)]{Guseinov/2002c}. This one-center
limit was later used by Guseinov and Mamedov
\cite{Guseinov/Mamedov/2008a} for the construction of series expansions
for overlap integrals of Slater-type functions with nonintegral principal
quantum numbers.

It is, however, trivially simple to show that Guseinov's one-center
expansion (\ref{GuExp_1}) does not exist if the principal quantum number
$N$ is nonintegral. As discussed in Section \ref{Sec:Introduction}, this
follows at once from the simple and yet consequential fact that
expansions in terms of Slater-type functions with integral principal
quantum numbers and a common scaling parameter are nothing but power
series expansions about $r=0$ in disguise. Moreover, every power series
is also a Taylor series for some function (see for example
\cite{Meyerson/1981}).

This fact is extremely helpful because the factors, which govern the
analyticity of a function, are fairly well understood. It is trivially
simple to show that the radial part of $\exp (\gamma r) \chi_{N, L}^{M}
(\beta, \bm{r})$ is not analytic in the sense of complex analysis at
$r=0$ if the principal quantum number $N$ is not a positive integer
satisfying $N \ge L+1$. Thus, a power series in $r$ for $\exp (\gamma r)
\chi_{N, L}^{M} (\beta, \bm{r})$ with $N \in \mathbb{R} \setminus
\mathbb{N}$ cannot exist, which implies that the one-center expansion
(\ref{GuExp_1}) for $\chi_{N, L}^{M} (\beta, \bm{r})$ does not exist if
$N \in \mathbb{R} \setminus \mathbb{N}$.

We arrive at the same conclusion if we apply the transformation formula
(\ref{Rearr_f_Exp_GLag}) to the $k$-dependent expansions
(\ref{Expand_NISTF2Gusfun_DiffScaPar_1}) for $\chi_{N, L}^{M} (\beta,
\bm{r})$ in terms of Guseinov's complete and orthonormal functions or to
the equivalent Laguerre series (\ref{ExpoPow2GLag}) for $z^{\rho}
\mathrm{e}^{u z}$, from which (\ref{Expand_NISTF2Gusfun_DiffScaPar_1})
was derived. As discussed in Section \ref{Sec:OneCenterExpansionSTF}, the
large index asymptotics of the coefficients in the Laguerre series
(\ref{ExpoPow2GLag}) clearly shows that the expansion (\ref{GuExp_1}) for
$\chi_{N, L}^{M} (\beta, \bm{r})$ in terms of Slater-type functions does
not exist if $N$ is nonintegral.

In principle, the same strategy could also be pursued if one-range
addition theorems for $\chi_{N, L}^{M} (\beta, \bm{r} \pm \bm{r}')$,
which are $k$-dependent two-center expansions in terms of Guseinov's
complete and orthonormal function $\bigl\{ \prescript{}{k}{\Psi}_{n,
  \ell}^{m} (\gamma, \bm{r}) \bigr\}_{n, \ell, m}$, are rearranged. The
expansion coefficients of these addition theorems are overlap
integrals. On the basis of our current level of mathematical
understanding, it is, however, very difficult or even practically
impossible to determine the large index asymptotics of these complicated
integrals. Fortunately, by means of an indirect approach it is
nevertheless possible to arrive at some useful conclusions about the
validity of Guseinov's rearrangements of his one-range addition theorems.

My key argument in Section \ref{Sec:OneRangeAdd->TwoRangeAdd} is again
analyticity in the sense of complex analysis. If a function $f \colon
\mathbb{C} \to \mathbb{C}$ has a singularity somewhere in the complex
plane, its power series about the origin cannot have an infinite radius
of convergence. In contrast, singularities affect Laguerre expansions
only if they are nonintegrable and located on the integration contour
$[0, \infty)$.

Both $\exp (-\beta \vert \bm{r} \pm \bm{r}' \vert )$ as well as $\exp (
\gamma r ) \exp (-\beta \vert \bm{r} \pm \bm{r}' \vert )$ are for $r' >
0$ analytic at $r=0$. Accordingly, these functions possess power series
expansions in $r$ which converge in a vicinity of the expansion point
$r=0$. However, $\exp (-\beta \vert \bm{r} \pm \bm{r}' \vert )$ is
singular for $\bm{r} \pm \bm{r}' = \bm{0}$. Thus, the power series
expansions for both $\exp (-\beta \vert \bm{r} \pm \bm{r}' \vert )$ and
$\exp ( \gamma r ) \exp (-\beta \vert \bm{r} \pm \bm{r}' \vert )$
converge only for $r < r'$. This implies that a one-range addition
theorem for the $1 s$ function $\exp (-\beta \vert \bm{r} \pm \bm{r}'
\vert )$, which converges pointwise for all $\bm{r}, \bm{r}' \in
\mathbb{R}^{3}$ or equivalently for all $r, r' \in [0, \infty)$, cannot
exist.

These considerations apply also to other Slater-type functions $\chi_{N,
  L}^{M} (\beta, \bm{r} \pm \bm{r}')$ with integral or nonintegral
principal quantum numbers $N$. If $r' > 0$, these functions as well as
the related functions $\exp (\gamma r) \chi_{N, L}^{M} (\beta, \bm{r} \pm
\bm{r}')$ are analytic at $r=0$, but they also have a singularity at
$\bm{r} \pm \bm{r}' = \bm{0}$. Therefore, an expansion of the type of
(\ref{NISTO->ISTO_AddThm}) in terms of Slater-type functions with
integral principal quantum numbers and a common scaling parameter, that
converges pointwise for all $\bm{r}, \bm{r}' \in \mathbb{R}^{3}$, cannot
exist. Instead, we obtain the small $r$ part of a two-range addition
theorem, either by doing a Taylor expansion of $\chi_{N, L}^{M} (\beta,
\bm{r} \pm \bm{r}')$ about $r=0$ or by applying the transformation
formula (\ref{Rearr_f_Exp_GLag}) to the angular projections
(\ref{SFT_Mom_GLagPol}).

The results of Sections \ref{Sec:OneCenterExpansionSTF} and
\ref{Sec:OneRangeAdd->TwoRangeAdd} can be summarized as follows: The
one-center expansion (\ref{GuExp_1}) for $\chi_{N, L}^{M} (\beta,
\bm{r})$ does not exist if $N$ is nonintegral, and the expansion
(\ref{NISTO->ISTO_AddThm}) for $\chi_{N, L}^{M} (\beta, \bm{r} \pm
\bm{r}')$, which looks like a one-range addition theorem, has a two-range
form since it converges only for $\vert \bm{r} \vert < \vert \bm{r}'
\vert$. Therefore, one might be tempted to dismiss Guseinov's
rearrangements of expansions in terms of generalized Laguerre polynomials
as being both useless and dangerous.

However, the situation is more complicated than it may look at first
sight. As discussed in Section \ref{Sec:NumericalImplications}, Guseinov
apparently never rearranged infinite one-center expansions of the type of
(\ref{Expand_NISTF2Gusfun_DiffScaPar_1}) or infinite one-range addition
theorems of the type of (\ref{Gus_OneRangeAddTheorSTF_OvSTF_k}), although
this can be done in a systematic way with the help of the transformation
formula (\ref{Rearr_f_Exp_GLag}). 

Instead, Guseinov exclusively rearranged \emph{finite} truncations of his
infinite expansions such as the truncation
(\ref{Gus_OneRangeAddTheorSTF_k_Trunc}) of the addition theorem
(\ref{Gus_OneRangeAddTheorSTF_STF_k_Rearr_1}) or the truncation
(\ref{Expand_NISTF2Gusfun_DiffScaPar_1_Trunc}) of its one-center limit
(\ref{Expand_NISTF2Gusfun_DiffScaPar_1}), all with a finite truncation
order $\mathcal{N}$. This is highly consequential. If we replace in a
truncated Laguerre series of the type of (\ref{FinSum_GLag}) the
generalized Laguerre polynomials by powers, for example via
(\ref{RearrFinSum_GLag}), there can be no convergence or existence
problems because the resulting expression is simply a polynomial.

Accordingly, Guseinov's rearrangements of truncated one-center and
two-center expansions produce approximations to $\chi_{N, L}^{M} (\beta,
\bm{r})$ and $\chi_{N, L}^{M} (\beta, \bm{r} \pm \bm{r}')$, respectively,
that consist of a \emph{finite} number of Slater-type functions $\chi_{n,
  \ell}^{m} (\gamma, \bm{r})$ with integral principal quantum numbers and
a common scaling parameter. Since only a finite number of terms is
transformed in these rearrangements, there can be no existence
problems. Nevertheless, Guseinov's approach is affected bu some serious
problems beyond the possible numerical stability problems discussed in
Section \ref{Sec:GuseinovsRearrangementsOfAdditionTheorems}.

In Section \ref{Sec:NumericalImplications} it is shown that
rearrangements of the truncations
(\ref{Expand_NISTF2Gusfun_DiffScaPar_1_Trunc}) of the one-center
expansion (\ref{Expand_NISTF2Gusfun_DiffScaPar_1}) are
\emph{semiconvergent} with respect to the truncation order $\mathcal{N}$
if the principal quantum number $N$ of the Slater-type function is
nonintegral. It follows from the leading order asymptotic approximation
(\ref{ZetaTypeTail}) that the accuracy of these approximations first
increases with increasing truncation order $\mathcal{N}$, but for larger
values of $\mathcal{N}$ the accuracy decreases again, and in the limit of
infinite truncation orders, everything goes to pieces.

The situation is not nearly so well understood in the case of
rearrangements of truncations (\ref{Gus_OneRangeAddTheorSTF_k_Trunc}) of
the $k$-dependent addition theorem
(\ref{Gus_OneRangeAddTheorSTF_STF_k_Rearr_1}). For finite truncation
orders $\mathcal{N}$, these rearrangements are one-range addition
theorems, but as $\mathcal{N} \to \infty$ they converge to two-range
addition theorems.

It makes sense to assume that the ultimate two-range nature of these
rearrangements becomes noticeable in integrals already for sufficiently
large, but finite values of $\mathcal{N}$. Thus, it is conceivable that
at least certain multicenter integrals containing such rearranged
truncations might turn out to be semiconvergent with respect to
$\mathcal{N}$. For the moment, these considerations are essentially
speculations, since substantial mathematical knowledge is still
lacking. 

As is well known from the literature, the use of semiconvergent
expansions can offer computational benefits under favorable
circumstances. In the case of Guseinov's rearranged truncations of
one-center and two-center expansions this may also be the case. However,
the use of semiconvergent expansions or of other approximations, whose
limit of infinite truncation order either does not exist or has
undesirable features, clearly involves some risks. So, before using such
an approximation in an actual calculation, we firstly must try to
understand the inherent risks, and secondly, we must convince ourselves
that we will be able to handle these risks. It would be extremely
negligent to ignore these risks and treat a semiconvergent expansion like
a convergent expansion. 

The examples mentioned above or the ones given in Appendix
\ref{App:Semiconvergence} should suffice to convince even a skeptic
reader that the apparent convergence of the leading terms of an infinite
series \emph{alone} does not prove anything. We also need some additional
mathematical insight indicating convergence.

Numerical demonstrations have obvious limitations. It is always desirable
to augment them by sufficiently rigorous mathematical investigations.
Unfortunately, rigorous proofs are extremely difficult in a research
topic as complex and computer oriented as electronic structure theory. In
the vast majority of all problems in electronic structure theory, a
rigorous mathematical analysis is out of reach and we have to be content
with numerical demonstrations, in spite of their obvious limitations and
their capacity of misleading us. However, the construction of addition
theorems and their application in multicenter integrals is a mathematical
problem, and -- as shown in this article -- quite a few things can be
understood on the basis of mathematical considerations. Guseinov's
strategy of relying entirely on numerical demonstrations without trying
to understand the underlying mathematics is not acceptable from a
methodological point of view.


\begin{appendix}
\typeout{==> Appendix: General Aspects of Series Expansions}
\section{General Aspects of Series Expansions}
\label{App:GeneralAspectsOfSeriesExpansions}

For the sake of simplicity, let us consider functions $F \colon
\mathbb{C} \to \mathbb{C}$. A \emph{series expansion}
$\sum_{n=0}^{\infty} u_{n} \mathcal{U}_{n} (z)$ for such a function $F
(z)$ requires a sequence $\{ u_n \}_{n=0}^{\infty}$ of fixed coefficient
and a sequence $\{ \mathcal{U}_n (z) \}_{n=0}^{\infty}$ of known
functions $\mathcal{U}_{n} \colon \mathbb{C} \to \mathbb{C}$.

To make the expansion $\sum_{n=0}^{\infty} u_{n} \mathcal{U}_{n} (z)$
useful, it has to represent $F (z)$ in some sense. Thus, we assume that
$\sum_{n=0}^{\infty} u_{n} \mathcal{U}_{n} (z)$ converges to $F (z)$
according to some specified convergence type and we write:
\begin{equation}
  \label{Def_SerExp_F}
  F (z) \; = \; \sum_{n=0}^{\infty} \, u_{n} \,\mathcal{U}_{n} (z) \, .
\end{equation}
Numerous different convergence types occur in practice. Important
examples are pointwise convergence, which is typical of classical complex
analysis or also of two-range addition theorems, convergence in the mean
with respect to the norm of some Hilbert space, which is typical of most
one-range addition theorems, or even distributional or weak convergence
in the sense of Schwartz \cite{Schwartz/1966a}. 

Therefore, the indiscriminate use of the ``$=$'' sign for all convergence
types is potentially misleading since it suggests a uniqueness which does
not exist. Depending on the convergence type, an ``$=$'' sign can have a
completely different meaning, or to put it differently, series expansions
of the type of (\ref{Def_SerExp_F}) can have very different mathematical
properties.  Moreover, different convergence types are in general
incompatible, i.e., the convergence of a series expansion for a given
convergence type does not imply that this series converges also with
respect to another convergence type.

It is, however, overly restrictive to insist that a series expansion of
the type of (\ref{Def_SerExp_F}) must converge in some sense to be
practically useful. Divergent, but summable series expansions are simply
to useful to be discarded, not only in mathematics, but in particular in
quantum physics (see for example \cite{LeGuillou/Zinn-Justin/1990} and
references therein). In numerous scientific applications, there is no
alternative to the summation of divergent series. A condensed review of
divergent series and their summation can be found in \cite[Appendices A
and B]{Weniger/2008}.

Let us now assume that we have a sequence of approximate expressions of
the following kind:
\begin{equation}
  \label{Def_F_N}
  F_{N} (z) \; = \;  \sum_{n=0}^{N} \, u_{n}^{(N)} \,
  \mathcal{U}_{n} (z) \, , \qquad N \in \mathbb{N}_{0} \, .
\end{equation}
The superscript $N$ in $u_{n}^{(N)}$ indicates that the coefficients may
depend explicitly on the summation limit $N$, i.e., we in general have
$u_{n}^{(N)} \ne u_{n}^{(N+1)} \ne u_{n}^{(N+2)} \ne \dots$ for fixed $n,
N \in \mathbb{N}_{0}$.

Let us now also assume that the approximants $F_{N} (z)$ converge to $F
(z)$ as $N \to \infty$ in some sense. This raises the question whether
the resulting expression $F (z) = \lim_{N \to \infty} F_{N} (z)$
constitutes an expansion of $F (z)$ in terms of the functions $\{
\mathcal{U}_n (z) \}_{n=0}^{\infty}$.

The answer is that this is in general not true. The convergence of the
approximants $F_{N} (z)$ to $F (z)$ as $N \to \infty$ only means that we
can make the difference between $F_{N} (z)$ and $F (z)$, whose exact
meaning depends on the convergence type, as small as we like by
increasing $N$ as much as necessary. This does not guarantee that the
coefficients $u_{n}^{(N)}$ in (\ref{Def_F_N}) have for all $n \in
\mathbb{N}_{0}$ unique limits $u_{n} = u_{n}^{(\infty)} = \lim_{N \to
  \infty} u_{n}^{(N)}$. Thus, the existence of a convergent sequence of
approximants of the type of (\ref{Def_F_N}) does not imply the existence
of a series expansion of the type of (\ref{Def_SerExp_F}).

\typeout{==> Appendix: Glory and Misery of Power Series}
\section{Glory and Misery of Power Series}
\label{App:GloryMiseryPowSer}

It is probably justified to claim that power series
\begin{equation}
  \label{Def:PowSer_z_0}
  f (z) \; = \; \sum_{\nu=0}^{\infty} \, \frac{f^{(\nu)} (z_{0})}{\nu!} 
   \, (z-z_{0})^{\nu} \; = \; 
   \sum_{\nu=0}^{\infty} \, c_{\nu} \, (z-z_{0})^{\nu}
\end{equation}
are the most important \emph{analytical} tools not only in mathematical
analysis, but also in the mathematical treatment of scientific and
engineering problems.

As is well known, power series can be differentiated \emph{term-by-term}
under relatively mild conditions. Accordingly, it is an obvious idea to
try to solve differential equations in terms of power series. A large
part of special function theory consists of the construction and analysis
of power series solutions to the ordinary differential equations that are
of relevance in mathematical physics.  Since power series can also be
integrated \emph{term-by-term} under very mild conditions, they are also
indispensable for the construction of explicit expressions for integrals
involving functions that possess power series representations.

While the usefulness of power series in analytical manipulations cannot
be overemphasized, it is nevertheless also true that a power series
representation for a given function is at best a mixed blessing from a
purely numerical point of view. The problem is that power series
expansions converge in circles about the expansion point. As is well
known, the radius of such a circle of convergence is determined by the
location the closest singularity of the function under consideration.
Thus, the radius of convergence of a power series expansion can be zero,
finite and infinite.

In general, a power series in $z-z_{0}$ is numerically useful only if $z$
and $z_{0}$ differ slightly, i.e., in the immediate vicinity of the
expansion point $z_{0}$. Then, a few terms of the series normally suffice
to produce excellent approximations. But close to the boundary of its
circle of convergence, the rate of convergence of such a power series
expansion can become prohibitively slow. In my opinion, the current
popularity of Pad\'{e} approximants, which normally converge much more
rapidly than the partials sums from which they are constructed (see for
example the monograph of Baker and Graves-Morris
\cite{Baker/Graves-Morris/1996} and references therein), is largely due
to the combined effect of the undeniable analytical usefulness of power
series and their (very) limited usefulness as computational tools.

A nonzero, but finite radius of convergence of a power series can also
cause serious problems in integrals. As is well known, a series expansion
for the integrand can be integrated termwise if it converges uniformly
for the whole integration interval. Otherwise, we have to be prepared
that termwise integration either generates a wrong result or even a
divergent series expansion for the integral. Thus, if we want to replace
a part of an integrand by its power series expansion, we are on the safe
side only if the integration interval is completely contained in the
circle of convergence. In particular in the case of infinite or
semi-infinite integration intervals, this is normally not the case.

\typeout{==> Appendix: Orthogonal Expansions}
\section{Orthogonal Expansions}
\label{App:OrthogonalExpansions}

The analyticity of a function $f \colon \mathbb{C} \to \mathbb{C}$ in the
sense of complex analysis is undeniably a highly desirable feature. In
the interior of suitable subsets of $\mathbb{C}$, $f$ can be represented
by power series expansions which converge pointwise and uniformly, and it
is also comparatively easy to compute derivatives of $f$ in this way.
Nevertheless, it is often advantageous to use instead of power series
alternative expansions that converge in a weaker sense.

Let $\mathcal{V}$ be an infinite dimensional vector space with inner
product $( \cdot \vert \cdot ) \colon \mathcal{V} \times \mathcal{V} \to
\mathbb{C}$ and its corresponding norm $\Vert \cdot \Vert \colon
\mathcal{V} \to \mathbb{R}_{+}$ defined by $\Vert f \Vert = (f \vert
f)^{1/2}$ with $f \in \mathcal{V}$. If every Cauchy sequence in
$\mathcal{V}$ converges with respect to the norm $\Vert \cdot \Vert$ to
an element of $\mathcal{V}$, then $\mathcal{V}$ is called a \emph{Hilbert
  space}.

Let us now assume that $f$ is an element of some Hilbert space
$\mathcal{H}$ and that the functions $\{ \varphi_m \}_{m=0}^{\infty}$ are
linearly independent and complete in $\mathcal{H}$. Then, we can
construct approximations
\begin{equation}
  \label{f_FinAppr}
f_M \; = \; \sum_{m=0}^{M} C_{m}^{(M)} \varphi_m
\end{equation}
to $f$, where $M$ is a \emph{finite} integer. The expansion coefficients
$C_{m}^{(M)}$, which in general depend on the summation limit $M$, are
chosen in such a way that the mean square deviation 
\begin{equation}
  \label{MeanSquareDev}
  \Vert f - f_M \Vert^2 \; = \; (f - f_M \vert f - f_M)
\end{equation}
becomes minimal.

The finite approximation (\ref{f_FinAppr}) converges to $f$ as $M \to
\infty$ if (\ref{MeanSquareDev}) can be made as small as we like by
increasing $M$. It therefore looks natural to assume that $f$ possesses
an \emph{infinite expansion}
\begin{equation}
  \label{f_InfExp}
  f \; = \; \sum_{m=0}^{\infty} C_{m} \varphi_m
\end{equation}
in terms of the linearly independent and complete functions $\{ \varphi_m
\}_{m=0}^{\infty}$ with well defined expansion coefficients $C_{m} =
\lim_{M \to \infty} C_{m}^{(M)}$.

As discussed in Appendix \ref{App:GeneralAspectsOfSeriesExpansions}, this
naturally looking assumption is not necessarily true.  In general, the
coefficients $C_{m}^{(M)}$ in (\ref{f_FinAppr}) do not only depend on
$m$, $f$, and $\{ \varphi_m \}_{m=0}^{\infty}$, but also on the summation
limit $M$. It is not \emph{a priori} clear whether the coefficients
$C_{m}^{(M)}$ in (\ref{f_FinAppr}) possess well defined limits $C_{m} =
\lim_{M \to \infty} C_{m}^{(M)}$, or whether an infinite expansion of the
type of (\ref{f_InfExp}) exists. In fact, expansions of that kind may or
may not exist.

It is one of the central results of approximation theory that for
arbitrary $f \in \mathcal{H}$ the mean square deviation $\Vert f - f_M
\Vert^2$ becomes minimal if the functions $\{ \varphi_m
\}_{m=0}^{\infty}$ are not only linearly independent and complete, but
also orthonormal satisfying $(\varphi_m \vert \varphi_{m'}) = \delta_{m
  {m'}}$ for all $m, m' \in \mathbb{M}_{0}$, and if the coefficients are
chosen according to $C_{m}^{(M)} = (\varphi_m \vert f)$ (see for example
\cite[Theorem 9 on p.\ 51]{Davis/1989}).

If the functions $\{ \varphi_m \}_{m=0}^{\infty}$ are complete and
orthonormal in the Hilbert space $\mathcal{H}$ and if the expansion
coefficients in (\ref{f_FinAppr}) are chosen according to $C_{m}^{(M)} =
(\varphi_m \vert f)$, then the expansion coefficients do not depend on
$M$. Thus, the limit $M \to \infty$ is possible, and $f \in \mathcal{H}$
possesses an \emph{infinite} series expansion
\begin{equation}
  \label{Expand_f_CONS}
f \; = \; \sum_{m=0}^{\infty} \, (\varphi_m \vert f) \, \varphi_m
\end{equation}
and this expansion converges in the mean with respect to the norm $\Vert
\cdot \Vert$ of $\mathcal{H}$.

The fact, that the completeness of a function set $\{ \varphi_m
\}_{m=0}^{\infty}$ in an infinite dimensional Hilbert space $\mathcal{H}$
alone does not suffice to guarantee the existence of expansions of the
type of (\ref{f_InfExp}), is highly consequential. Nevertheless, it is
occasionally overlooked, although this insufficiency is well documented
both in the mathematical literature (see for example \cite[Theorem 10 on
p.\ 54]{Davis/1989} or \cite[Section 1.4]{Higgins/1977}) as well as in
the literature on electronic structure calculations
\cite{Klahn/1975,Klahn/1981,Klahn/Bingel/1977a,Klahn/Bingel/1977b,%
  Klahn/Bingel/1977c,Klahn/Morgan/1984}). Horrifying examples of
nonorthogonal expansions with pathological properties can be found in
\cite[Section III.I]{Klahn/1981}.

If the Hilbert space $\mathcal{H}$ is an infinite dimensional vector
space consisting of function $f, g \colon \mathbb{C} \to \mathbb{C}$, the
inner product $( f \vert g )$ of $\mathcal{H}$ is usually identified with
an integral $\int_{a}^{b} w (z) [f (z)]^{*} g (z) \mathrm{d} z$, where $w
(z)$ is an appropriate positive weight function.  Moreover, the complete
orthonormal functions $\{ \varphi_m \}_{m=0}^{\infty}$ in $\mathcal{H}$
are normally related to a suitable subclass $\{ \mathcal{P}_{m} (z)
\}_{m=0}^{\infty}$ of the classical orthogonal polynomials of
mathematical physics, as specified by the integration limits $a$ and $b$
and the weight function $w (z)$. In this case, the general orthogonal
expansion (\ref{Expand_f_CONS}) boils down to the expansion of a function
$f (z)$ in terms of orthogonal polynomials:
\begin{equation}
  \label{F_OrthoPol}
  f (z) \; = \; \sum_{m=0}^{\infty} \, 
  \mathcal{C}_{m} \, \mathcal{P}_{m} (z)
\end{equation}

It is generally accepted that orthogonal expansions are extremely useful
mathematical tools and that they have many highly advantageous features.
This is, however, not the whole truth, in particular if we want to
approximate functions. Hilbert space theory only guarantees that an
orthogonal expansion converges in the mean with respect to the
corresponding norm $\Vert \cdot \Vert$, but not necessarily pointwise or
even uniformly. Thus, convergence in the mean is a comparatively weak
form of convergence, and orthogonal expansions are not necessarily a good
choice if we are predominantly interested in the local properties of a
function. However, convergence in the mean is usually completely
satisfactory for the evaluation of integrals.

\typeout{==> Appendix: Generalized Laguerre Polynomials}
\section{Generalized Laguerre Polynomials}
\label{App:GeneralizedLaguerrePolynomials}

The surface spherical harmonics $Y_{\ell}^{m} (\theta, \phi)$ are
complete and orthonormal with respect to an integration over the surface
of the unit sphere in $\mathbb{R}^3$ (an explicit proof can for instance
be found in \cite[Section III.7.6]{Prugovecki/1981}). Since more complex
Hilbert spaces can be constructed by forming tensor products of simpler
Hilbert spaces (see for example \cite[Section II.6.5]{Prugovecki/1981}),
we only have to find suitable radial functions that are complete and
orthogonal with respect to an integration from $0$ to $\infty$ (see also
\cite[Lemma 6 on p.\ 31]{Klahn/Bingel/1977b}).  Thus, we more or less
automatically arrive at function sets based on the generalized Laguerre
polynomials.

The \emph{generalized Laguerre polynomials} $L_{n}^{(\alpha)} (z)$ with
$\Re (\alpha) > - 1$ and $n \in \mathbb{N}_{0}$ are orthogonal with
respect to an integration over the positive real semiaxis $[0, \infty)$
with weight function $w (z) = z^{\alpha} \exp (-z)$. In the mathematical
literature, they are defined either via the \emph{Rodrigues relationship}
\begin{equation}
  \label{GLag_Rodrigues}
L_{n}^{(\alpha)} (z) \; = \; z^{-\alpha} \, \frac{\mathrm{e}^{z}}{n!} \,
\frac{\mathrm{d}^n}{\mathrm{d} z^n} \,
\bigl[ \mathrm{e}^{-z} z^{n+\alpha} \bigr]
\end{equation}
or as a terminating confluent hypergeometric series ${}_{1} F_{1}$:
\begin{equation}
  \label{GLag_1F1}
  L_{n}^{(\alpha)} (z) \; = \;
  \frac{(\alpha+1)_n}{n!} \, {}_1 F_1 (-n; \alpha+1; z) \; = \;
  \frac{(\alpha+1)_n}{n!} \, \sum_{\nu=0}^{n} \, 
  \frac{(-n)_{\nu}}{(\alpha+1)_{\nu}} \, \frac{z^{\nu}}{\nu!} \, .
\end{equation}
The generalized Laguerre polynomials satisfy for $\Re (\alpha) > - 1$
and $m, n \in \mathbb{N}_0$ the orthogonality relationship
\begin{equation}
  \label{GLag_Orthogonality}
\int_{0}^{\infty} \, z^{\alpha} \, \mathrm{e}^{-z} \,
L_{m}^{(\alpha)} (z) \, L_{n}^{(\alpha)} (z) \, \mathrm{d} z \; = \;
\frac{\Gamma (\alpha+n+1)}{n!} \, \delta_{m n} \, .
\end{equation}
Accordingly, the polynomials
\begin{equation}
  \label{Def:GenLag_Normalized}
  \mathcal{L}_{n}^{(\alpha)} (z) \; = \;
  \left[ \frac{n!}{\Gamma (\alpha+n+1)} \right]^{1/2} \,
  L_{n}^{(\alpha)} (z) \, ,
  \qquad n \in \mathbb{N}_0 \, , \quad \alpha > - 1 \, ,
\end{equation}
are for $\Re (\alpha) > - 1$ \emph{orthonormal} with respect to an
integration over the interval $[0, \infty)$ involving the weight function
$z^{\alpha} \exp (-z)$:
\begin{equation}
  \label{Orthonorm_CalL}
  \int_{0}^{\infty} \, \mathrm{e}^{-z} \, z^{\alpha} \,
  \mathcal{L}_{m}^{(\alpha)} (z) \,  \mathcal{L}_{n}^{(\alpha)} (z) \,
  \mathrm{d} z \; = \; \delta_{m n} \, .
\end{equation}
Alternatively, we can also use the functions
\begin{equation}
  \label{Def:GenLag_CONS}
  \Phi_{n}^{(\alpha)} (z) \; = \;
  \left[ \frac{n!}{\Gamma (\alpha+n+1)} \right]^{1/2} \,
  \mathrm{e}^{-z/2} \, z^{\alpha/2} \, L_{n}^{(\alpha)} (z) \, ,
  \qquad n \in \mathbb{N}_0 \, ,
\end{equation}
which are for $\Re (\alpha) > - 1$ \emph{orthonormal} with respect to an
integration over the interval $[0, \infty)$:
\begin{equation}
  \label{Orthonorm_Phi}
  \int_{0}^{\infty} \, \Phi_{m}^{(\alpha)} (z) \,
  \Phi_{n}^{(\alpha)} (z) \, \mathrm{d} z \; = \; \delta_{m n} \, .
\end{equation}

The completeness of the generalized Laguerre polynomials in the weighted
Hilbert space
\begin{equation}
  \label{HilbertL^2_Lag}
  L^{2}_{\mathrm{e}^{-z} z^{\alpha}} \bigl([0, \infty) \bigr) \; = \;
  \Bigl\{ f \colon [0, \infty) \to \mathbb{C} \Bigm\vert \,
  \int_{0}^{\infty} \,\mathrm{e}^{-z} \, z^{\alpha} \, \vert f (z)
  \vert^2 \, \mathrm{d} z < \infty \Bigr\}
\end{equation}
is a classic result of mathematical analysis (see for example
\cite[p.\ 33]{Higgins/1977}, \cite[pp.\ 349 - 351]{Sansone/1977}, or
\cite[pp.\ 235 - 238]{Tricomi/1970}).

In general, Laguerre expansions converge only in the mean, but not
necessarily pointwise (see for example \cite{Askey/Wainger/1965}).
Additional conditions, which a function has to satisfy in order to
guarantee that its Laguerre expansion converges pointwise, were discussed
by Szeg\"{o} \cite[Theorem 9.1.5 on p.\ 246]{Szegoe/1967} (see also
\cite[Appendix]{Filter/Steinborn/1980}).

In this article, exclusively the mathematical notation is used. A
different convention for Laguerre polynomials is frequently used in the
quantum mechanical literature. For example, Bethe and Salpeter \cite[Eq.\
(3.5)]{Bethe/Salpeter/1977} define \emph{associated Laguerre functions}
$\bigl[L_{n}^{m} (z)\bigr]_{\text{BS}}$ with $n, m \in \mathbb{N}_0$ via
the Rodrigues-type relationships
\begin{subequations}
  \label{AssLagFun_BS}
  \begin{align}
    \label{AssLagFun_BS_1}
    \bigl[L_{n}^{m} (z)\bigr]_{\text{BS}} & \; = \;
    \frac{\mathrm{d}^{m}}{\mathrm{d} z^{m}} \, \bigl[L_{n}
    (z)\bigr]_{\text{BS}} \, ,
    \\
    \label{AssLagFun_BS_2}
    \bigl[L_{n} (z)\bigr]_{\text{BS}} & \; = \; \mathrm{e}^z \,
    \frac{\mathrm{d}^{n}}{\mathrm{d} z^{n}} \, \bigl[ \mathrm{e}^{-z}
    z^{n} \bigr] \, .
  \end{align}
\end{subequations}
Comparison of (\ref{GLag_Rodrigues}) and (\ref{AssLagFun_BS_2}) implies:
\begin{equation}
  \label{GLagPol_2_GLagPol_BS}
L_{n}^{(m)} (z) \; = \;
\frac{(-1)^m}{(n+m)!} \, \bigl[L_{n+m}^{m} (z)\bigr]_{\text{BS}} \, .
\end{equation}
The convention of Bethe and Salpeter \cite{Bethe/Salpeter/1977} is also
used in the books by Condon and Shortley \cite[Eqs.\ (6) and (9) on p.\
115]{Condon/Shortley/1970} and by Condon and Odaba\c{s}i \cite[Eq.\ (2)
on p.\ 189]{Condon/Odabasi/1980} as well as in the numerous articles by
Guseinov and his coworkers.

In my opinion, the use of associated Laguerre functions defined by
(\ref{AssLagFun_BS}) is not recommendable. It follows from
(\ref{GLagPol_2_GLagPol_BS}) that these functions cannot express
generalized Laguerre polynomials $L_{n}^{(\alpha)}$ with nonintegral
superscripts $\alpha$. This is both artificial and unnecessary. For
example, the eigenfunctions $\Omega_{n, \ell}^{m} (\beta, \bm{r})$ of the
Hamiltonian $\beta^{-2} \nabla^2 - \beta^2 r^2$ of the three-dimensional
isotropic harmonic oscillator contain generalized Laguerre polynomials in
$r^{2}$ with half-integral superscripts (see for example \cite[Eq.\
(5.4)]{Weniger/1985} and references therein).

\typeout{==> Appendix: Semiconvergence}
\section{Semiconvergence}
\label{App:Semiconvergence}

If the sum of the leading terms of an infinite series seem to approach
the correct limit, it looks natural to conclude that increasing the
number of terms will improve the accuracy of the approximation, and that
ultimately the partial sums will converge to the correct
limit. Unfortunately, this assumption is overly optimistic and in spite
of its apparent plausibility not necessarily true.

Many series are known whose partial sums initially seem to converge. If
however, further terms are included, the accuracy decreases, and
ultimately the sequence of partial sums diverges. In the literature, this
phenomenon is well established and usually called \emph{semiconvergence}
(see for example \cite[p.\ 2, Footnote${}^{\dag}$]{Olver/1997a}). To the
best of my knowledge, this terminology was introduced by Stieltjes
\cite{Stieltjes/1886} already in 1886.

Semiconvergence is best known in connection with factorially divergent
asymptotic inverse power series for special functions. In Arfken's book
\cite[Chapter 5.10]{Arfken/1985}, one can find a comprehensive discussion
of the semiconvergence of the divergent asymptotic series of the
incomplete gamma function $\Gamma (a, z)$ \cite[Eq.\
(8.11.2)]{Olver/Lozier/Boisvert/Clark/2010} and its special case, the
asymptotic series for the exponential integral $E_{1} (z)$ \cite[Eq.\
(6.12.1)]{Olver/Lozier/Boisvert/Clark/2010}. In my opinion, Arfken's
treatment is well suited as a first introduction to this topic. Other
examples of semiconvergent series are the factorially divergent
asymptotic series for the complementary error function $\mathrm{erfc}
(z)$ \cite[Eq.\ (7.12.1)]{Olver/Lozier/Boisvert/Clark/2010}, and the
modified Bessel and Whittaker functions of the second kind $K_{\nu} (z)$
and $W_{\kappa, \mu} (z)$, respectively \cite[Eqs.\ (10.40.4) and
(13.19.3)]{Olver/Lozier/Boisvert/Clark/2010}.

If the argument of such a divergent asymptotic inverse power series is
sufficiently large, then the truncation of such a divergent series in the
vicinity of the minimal term can lead to excellent approximations to the
function it represents. Nevertheless, these partial sums diverge if
further terms beyond the minimal term are included (see for example
\cite[Figure 2.2 on p.\ 35]{Gil/Segura/Temme/2007}).

Accordingly, the accuracy, which can be obtained by truncating a
semiconvergent inverse power series in the vicinity of the minimal term,
depends crucially on the magnitude of the argument. If the argument is
large, excellent approximations can often be obtained. The situation is
not so good if the argument is small, because then the truncation of a
semiconvergent series can only provide relatively crude
approximations. But even for small arguments, it is often possible to
obtain very accurate approximations by using the partial sums of a
semiconvergent series as input data in nonlinear sequence transformations
\cite{Weniger/1996d,Weniger/Cizek/1990}.

Semiconvergent series occur also quite abundantly in quantum mechanical
perturbation expansions. For example, Ahlrichs \cite{Ahlrichs/1976}
showed that the total energy of interacting molecular systems $A$ and $B$
can be expressed by a semiconvergent series, the so called
$1/R$-expansion.

Other phenomena closely resembling semiconvergence are also
known. Baumel, Crocker, and Nuttall \cite{Baumel/Crocker/Nuttall/1975}
showed that in the case of scattering calculations with complex basis
functions low order approximations can produce good results although the
whole scheme ultimately diverges, and Gautschi \cite{Gautschi/1977}
observed initial apparent convergence to the wrong limit in the case of
continued fractions for Kummer functions.
\end{appendix}

\end{document}